\documentclass[11pt,oneside]{article}

\usepackage{epsfig,lscape}
\usepackage{amssymb,amsfonts,amsmath}
\usepackage{rotating}

\usepackage{amsthm}

\def\me{\mathrm e}

\def\dif{\mathrm d}

\def\var{\mathrm{var}}

\def\N{\mbox{N}}

\def\T{ {\mathrm{\scriptscriptstyle T}} }

\def\argmin{\mathrm{argmin}}

\def\myeta{ {g} }

\newenvironment{prf}
{\noindent \textbf{Proof.}}{\hfill $\Box$ \vspace{.1in}}

\newtheorem{lem}{Lemma}
\newtheorem{pro}{Proposition}
\newtheorem{cor}{Corollary}
\newtheorem{ass}{Assumption}

\theoremstyle{definition}

\theoremstyle{definition}

\begin{document}

\begin{titlepage}

\begin{center}
{\Large Regularized calibrated estimation of propensity scores with model misspecification and high-dimensional data}

\vspace{.1in} Zhiqiang Tan\footnotemark[1]

\vspace{.1in}
\today
\end{center}

\footnotetext[1]{Department of Statistics \& Biostatistics, Rutgers University. Address: 110 Frelinghuysen Road,
Piscataway, NJ 08854. E-mail: ztan@stat.rutgers.edu. The research was supported in part by PCORI grant ME-1511-32740.
The author thanks Cun-Hui Zhang for helpful discussions.}

\paragraph{Abstract.} Propensity score methods are widely used for estimating treatment effects from observational studies.
A popular approach is to estimate propensity scores by maximum likelihood based on logistic regression,
and then apply inverse probability weighted estimators or extensions to estimate treatment effects.
However, a challenging issue is that such inverse probability weighting methods including doubly robust methods
can perform poorly even when the logistic model appears adequate as examined by conventional techniques.
In addition, there is increasing difficulty to appropriately estimate propensity scores when dealing with a large number of covariates.
To address these issues, we study calibrated estimation as an alternative to maximum likelihood estimation for fitting logistic propensity score models.
We show that, with possible model misspecification, minimizing the expected calibration loss underlying the calibrated estimators
involves reducing both the expected likelihood loss and a measure of relative errors which controls the mean squared errors of inverse probability weighted estimators.
Furthermore, we propose a regularized calibrated estimator by minimizing the calibration loss with a Lasso penalty. We
develop a novel Fisher scoring descent algorithm for computing the proposed estimator,
and provide a high-dimensional analysis of the resulting inverse probability weighted estimators of population means, leveraging the control of relative errors
for calibrated estimation. We present a simulation study and an empirical application to demonstrate the advantages of the proposed methods compared with maximum likelihood and regularization.


\end{titlepage}

\section{Introduction} \label{sect:intro}

Statistical methods using propensity scores (Rosenbaum \& Rubin 1983) are extensively used for estimating treatment effects in causal inference with the potential outcome framework (Neyman 1923; Rubin 1974).
For each subject, potential outcomes are defined under different hypothetical treatments, but only one of them can be observed and the others are missing.
The propensity score is defined as the conditional probability of receiving a specific treatment given measured covariates (that is, possible confounding variables).
Similar methods using selection probabilities are also widely used in various related missing-data problems, such as regression analysis with missing outcomes or covariates (Robins et al.~1994; Tan 2011)
and with data combination (Graham et al.~2016).
There are several techniques for using propensity scores, including matching, stratification, and weighting (e.g., Imbens 2004).
Particularly, inverse probability weighting (IPW) is attractive and central to theory of semiparametric estimation with missing data (Tsiatis 2006; van der Laan \& Robins 2003).

One of the statistical challenges in applying propensity score methods for observational studies is that the propensity score is unknown
and need to be estimated from observed data.
A popular approach as demonstrated in Rosenbaum \& Rubin (1984) is
to fit a propensity score model (often logistic regression with main effects only), check covariate balance, and then modify and refit the propensity score model, using nonlinear terms and
interactions, until reasonable balance is achieved.
But this process can be work intensive and involve ad hoc choices for model refinement, and there is no formal mechanism to guarantee that covariates will eventually be balanced.
In addition, another statistical issue facing various IPW-based methods including doubly robust methods is that
these methods can perform poorly, due to instability to small propensity scores estimated for few treated subjects,
even when the propensity score model appears to be ``nearly correct" (e.g., Kang \& Schafer 2007).

To address the foregoing issues, calibrated estimation has been proposed as an alternative to maximum likelihood estimation for fitting propensity score models.
The basic idea is not to use maximum likelihood for parameter estimation, but a system of estimating equations, Eq.~(\ref{eq-CAL}) later, such that
the weighted averages of the covariates in the treated subsample are equal to the simple averages in the overall sample.
Subsequently, the fitted propensity scores can be used as usual in inverse probability weighted estimators or extensions.
While such ideas can be traced to Folsom (1991) in the survey literature,
calibrated estimators have been recently studied, sometimes independently under different names, from a number of perspectives.
In fact, Eq.~(\ref{eq-CAL}) can be formally deduced from Tan (2010, Eq.~17), with a working propensity score model being degenerate.
The same equations as (\ref{eq-CAL}) are also obtained in Graham et al.~(2012), Kim \& Haziza (2014), and Vermeulen \& Vansteelandt (2015)
to develop doubly robust estimators.
As shown in Section~\ref{sec:discussion}, the calibration equations for the untreated subsample, Eq.~(\ref{eq-CAL-0}) later, lead to the same estimator as in
entropy balancing (Hainmueller 2012).
Calibration equations (\ref{eq-CAL}) can also be seen from Chan et al.~(2016, Eq.~5) in a dual formulation,
for which our view is that a propensity score model is implicitly determined from the distance measure used.
The implied propensity scores may fall outside $(0,1)$, but the resulting estimators of population means
can still be shown to be consistent under suitable regularity conditions.

In spite of these developments, there remain at least two important questions. The previous works studied calibrated estimators mostly to the extent
of showing that the resulting estimators of population means are doubly robust, i.e., consistent if either a propensity score model or an outcome regression model
is correctly specified. However, doubly robust estimators may still perform poorly in
practical situations where both models tend to be misspecified (Kang \& Schafer 2007).
The first question is whether, with possible model misspecification, any advantage can be formally established for calibrated estimation, compared with maximum likelihood estimation,
when fitting propensity score models for inverse probability weighting, without additional conditions about outcome regression models.
In addition, calibrated estimation is previously analyzed with the number of covariates $p$ either fixed as the sample size $n$ increases
or growing slowly, e.g., $o(n^{1/11})$ in Chan et al.~(2016) under strong enough smoothness conditions.
The second question is how to extend and analyze calibrated estimation when the number of covariates is close to or greater than
the sample size.

In this article, we develop theory and methods to address the foregoing questions, with a logistic propensity score model.
First, we establish a simple relationship between the loss functions underlying the calibrated and maximum likelihood estimators.
From this result, we show that minimizing the expected calibration loss involves reducing both the expected likelihood loss and
a measure of {\it relative errors} of the target (or limiting) propensity score, which then controls the mean squared errors of the IPW estimators based on the target propensity score.
The relative error of a target propensity score is defined as the deviation of the ratio of the true and the target propensity scores from 1.
Such direct control of relative errors of propensity scores is not achieved by minimizing the expected likelihood loss alone.

Second, we propose a regularized calibrated estimator by minimizing the calibration loss with a Lasso penalty (Tibshirani 1992).
Using the Lasso penalty has an interesting consequence of relaxing calibration equations (\ref{eq-CAL}) to box constraints, that is,
the left hand side of (\ref{eq-CAL}) is bounded in the supremum norm by a constant, which is also the tuning parameter for the Lasso penalty.
We develop a novel algorithm for computing the proposed estimator, exploiting quadratic approximation, Fisher scoring (McCullagh \& Nelder 1989),
and the majorization-minimization technique (Wu \& Lange 2010).
We also provide a high-dimensional analysis of the regularized calibrated estimator and the
resulting IPW estimators of population means, allowing possible model misspecification.
Our slow-rate result shows that if the coefficients from the target propensity score are uniformly bounded, then
the squared difference between the IPW estimators based on the fitted and target propensity scores converges in probability to 0 at rate $|S| \sqrt{\log (p)/n}$
under simple conditions without a compatibility condition, where $|S|$ is the number of nonzero coefficients from the target propensity score.
This result is proved by leveraging the control of relative errors mentioned above for calibrated estimation, and hence would not be available for
regularized maximum likelihood estimation.
The rate of convergence can be improved to $|S| \log (p)/n$ under a compatibility condition.

The plan of the paper is as follows. We describe basic concepts from causal inference in Section~\ref{sec:background}, and then present theory and methods in Section~\ref{sec:proposed}, a simulation study in Section~\ref{sec:simulation}, and an empirical application in Section~\ref{sec:application}. We provide additional discussion, including comparison with related works, in Section~\ref{sec:discussion}.

\vspace{-.1in}
\section{Background: causal inference} \label{sec:background}

Suppose that 
the observed data consist
of independent and identically distributed observations $\{(Y_i, T_i, X_i): i=1,\ldots,n\}$ of $(Y,T,X)$, where
$Y$ is an outcome variable, $T$ is a treatment variable taking values 0 or 1, and $X=(X_{1},\ldots,X_{p})$ is a vector of measured
covariates. In the potential outcomes framework for causal inference (Neyman 1923; Rubin 1974), let
$(Y^0, Y^1)$ be potential outcomes that would be observed under treatment 0 or 1 respectively. By
consistency, assume that $Y$ is either $Y^0$ if $T=0$ or $Y^1$ if $T=1$, that is,
$Y= (1-T) Y^0 + T Y^1$.
There are two causal parameters commonly of interest: the average treatment effect (ATE), defined as
$E(Y^1- Y^0) = \mu^1 - \mu^0$ with $\mu^t= E(Y^t)$, and the average treatment effect on the treated (ATT),
defined as $E(Y^1 - Y^0 |T=1) = \nu^1 - \nu^0$ with $\nu^t = E(Y^t | T=1)$ for $t=0,1$.
For concreteness, we mainly discuss estimation of ATE until Section~\ref{sec:discussion} to discuss ATT.

Estimation of ATE is fundamentally a missing-data problem: only one potential outcome,
$Y^0_i$ or $Y^1_i$, is observed and the other one is missing for each subject $i$. For identification of
$(\mu^0,\mu^1)$ and ATE, we make the following two assumptions throughout:
\begin{itemize}\addtolength{\itemsep}{-.05in}
\item[(i)] Unconfoundedness: $T \perp Y^0 |X$ and $T \perp Y^1|X$, that is, $T$ and $Y^0$ and, respectively, $T$ and $Y^1$ are
conditionally independent given $X$ (Rubin 1976);

\item[(ii)] Overlap: $ 0 < \pi^*(x) < 1$ for all $x$, where $\pi^*(x)=P(T=1|X=x)$ is called the propensity score (PS) (Rosenbaum \& Rubin 1983).
\end{itemize}
Under these assumptions, $(\mu^0,\mu^1)$ and ATE are often estimated by imposing additional modeling (or dimension-reduction)
assumptions in two different ways.

One approach is to build a statistical model for the outcome regression (OR) function $m^*(t,X) = E(Y|T=t,X)$ in the form
\begin{align}
E(Y| T=t,X) = m(t,X; \alpha), \quad t=0,1, \label{model-OR}
\end{align}
where $m(t,x;\alpha)$ is a known function and $\alpha$ is a vector of unknown parameters.
Let $\hat\alpha_{\mbox{\tiny LS}}$ be an estimator of $\alpha$ by least squares or similar methods, and $\hat m_{\mbox{\tiny LS}}(t,X) = m(t,X;\hat\alpha_{\mbox{\tiny LS}})$.
If model (\ref{model-OR}) is correctly specified, then $\tilde E \{ \hat m_{\mbox{\tiny LS}}(t,X)\} = n^{-1} \sum_{i=1}^n \hat m_{\mbox{\tiny LS}}(t,X_i)$ is a consistent estimator of $\mu^t$ for $t=0,1$
under standard regularity conditions as $n\to\infty$ and the dimension of $\alpha$ is fixed. Throughout, $\tilde E()$ denotes the sample average.

Another approach, which is the main subject of our research, is to build a statistical model for the propensity score $\pi^*(X)=P(T=1|X)$ in the form
\begin{align}
P(T=1|X) = \pi(X;\gamma) = \Pi\{\gamma^\T f(X)\}, \label{model-PS}
\end{align}
where $\Pi()$ is an inverse link function, $f(x)$ is a vector of known functions, and $\gamma$ is a vector of unknown parameters.
Typically, logistic regression is used with $\pi(X;\gamma)= [1+\exp\{-\gamma^\T  f(X)\}]^{-1}$.
Let $\hat\gamma_{\mbox{\tiny ML}}$ be the maximum likelihood estimator of $\gamma$, which for logistic regression minimizes the average negative log-likelihood
\begin{align}
\ell_{\mbox{\tiny ML}}(\gamma) 
&= \tilde E \left[ \log\{ 1+\me^{ \gamma^\T f(X)} \}  -T\, \gamma^\T f(X) \right] \label{loss-ML}
\end{align}
or equivalently solves the score equation
\begin{align}
\tilde E \left[ \{T-\pi(X; \gamma)\} f(X) \right] = 0 . \label{eq-score}
\end{align}
Various methods have been proposed, using the fitted propensity score $\hat\pi_{\mbox{\tiny ML}}(X) = \pi( X; \hat \gamma_{\mbox{\tiny ML}})$, to estimate $(\mu^0,\mu^1)$ and ATE
(e.g., Imbens 2004). We focus on inverse probability weighting (IPW),
which is central to semiparametric theory of estimation in causal inference and missing-data problems (e.g., Tsiatis 2006).
Two IPW estimators for $\mu^1$ commonly used are
\begin{align*}
\hat \mu^1_{\mbox{\tiny IPW}} (\hat\pi_{\mbox{\tiny ML}}) = \tilde E\left\{ \frac{TY}{\hat\pi_{\mbox{\tiny ML}}(X)} \right\},\quad
\hat \mu^1_{\mbox{\tiny rIPW}} (\hat\pi_{\mbox{\tiny ML}})  = \hat \mu^1_{\mbox{\tiny IPW}}  / \tilde E\left\{ \frac{T}{\hat\pi_{\mbox{\tiny ML}}(X)} \right\}.
\end{align*}
Similarly, two IPW estimators for $\mu^0$ are
\begin{align*}
\hat \mu^0_{\mbox{\tiny IPW}} (\hat\pi_{\mbox{\tiny ML}}) = \tilde E\left\{ \frac{(1-T)Y}{1- \hat\pi_{\mbox{\tiny ML}}(X)} \right\},\quad
\hat \mu^0_{\mbox{\tiny rIPW}} (\hat\pi_{\mbox{\tiny ML}})  = \hat \mu^0_{\mbox{\tiny IPW}}  / \tilde E\left\{ \frac{1-T}{1-\hat\pi_{\mbox{\tiny ML}}(X)} \right\}.
\end{align*}
If model (\ref{model-PS}) is correctly specified, then the preceding IPW estimators are consistent under standard regularity conditions as $n\to\infty$ and the dimension of $\gamma$ is fixed.

The two approaches, OR and PS, rely on different modeling assumptions (see Tan 2007 for a comparison of their operating characteristics).
In addition, there are doubly robust (DR) methods using both OR and PS models, such that the resulting estimators of $(\mu^0,\mu^1)$ and ATE remain consistent if either of the two models is correctly specified.
See Kang \& Schafer (2007) and Tan (2010) for reviews and Section~\ref{sec:discussion} for further discussion.

\section{Theory and methods} \label{sec:proposed}

\subsection{Overview} \label{sec:overview}

A crucial aspect of propensity score methods for observational studies is that the propensity score, $\pi^*(X)=P(T=1|X)$, is unknown and need
to be estimated from data, often through a statistical model in the form (\ref{model-PS}).
A conventional method of estimation is to fit model (\ref{model-PS}) by maximum likelihood.
We study an alternative method of estimation for fitting propensity score model (\ref{model-PS}). The fitted propensity scores are to be used
for estimating $(\mu^0,\mu^1)$ and ATE by inverse probability weighting or related methods (including doubly robust methods) in the context
of causal inference or similar missing-data problems.

For concreteness, we assume that model (\ref{model-PS}) is logistic regression:
\begin{align}
P(T=1|X) = \pi(X;\gamma) =[1+\exp\{-\gamma^\T  f(X)\}]^{-1} , \label{logit-PS}
\end{align}
where $f(x) = \{1,f_1(x),\ldots, f_p(x)\}^\T$ is a vector of known functions including a constant and $\gamma = (\gamma_0,\gamma_1, \ldots,\gamma_p)^\T $ is a vector of unknown parameters.
Let $\hat\gamma^1_{\mbox{\tiny CAL}}$ be an estimator of $\gamma$ solving
\begin{align}
\tilde E \left[ \left\{\frac{T}{\pi(X;\gamma)}-1 \right\} f(X)\right] = 0. \label{eq-CAL}
\end{align}
The fitted propensity score is $\hat\pi^1_{\mbox{\tiny CAL}}(X) = \pi(X; \hat\gamma^1_{\mbox{\tiny CAL}})$. Then
$\mu^1$ can be estimated by $\hat\mu^1_{\mbox{\tiny IPW}}(\hat\pi^1_{\mbox{\tiny CAL}})$ or equivalently $\hat\mu^1_{\mbox{\tiny rIPW}}(\hat\pi^1_{\mbox{\tiny CAL}})$,
with $\hat\pi_{\mbox{\tiny ML}}(X)$ replaced by $\hat\pi^1_{\mbox{\tiny CAL}}(X)$.
The two IPW estimators are identical because $\tilde E \left\{ T/ \hat\pi^1_{\mbox{\tiny CAL}}(X) \right\}=1$ by (\ref{eq-CAL})
with a constant included in $f(X)$.

Similarly, let $\hat\gamma^0_{\mbox{\tiny CAL}}$ be an estimator of $\gamma$ solving
\begin{align}
\tilde E \left[ \left\{\frac{1-T}{1-\pi(X;\gamma)}-1 \right\} f(X)\right] = 0, \label{eq-CAL-0}
\end{align}
and let $\hat\pi^0_{\mbox{\tiny CAL}}(X) = \pi(X; \hat\gamma^0_{\mbox{\tiny CAL}})$. Then
$\mu^0$ can be estimated by $\hat\mu^0_{\mbox{\tiny IPW}}(\hat\pi^0_{\mbox{\tiny CAL}} )$ or equivalently $\hat\mu^0_{\mbox{\tiny rIPW}}(\hat\pi^0_{\mbox{\tiny CAL}} )$, with
$\hat\pi_{\mbox{\tiny ML}}(X)$ replaced by $\hat\pi^0_{\mbox{\tiny CAL}}(X)$,
where the equivalence of the two IPW estimators follows because $\tilde E \left[ (1- T)/\{1- \hat\pi^0_{\mbox{\tiny CAL}}(X) \} \right]=1$ by (\ref{eq-CAL-0})
with a constant included in $f(X)$.
See Section~\ref{sec:discussion} for remarks on the unusual fact that two different sets of fitted propensity scores, $\hat\pi^1_{\mbox{\tiny CAL}}(X)$ or $\hat\pi^0_{\mbox{\tiny CAL}}(X)$,
are used for estimating $\mu^1$ or $\mu^0$ respectively.

Estimating equations (\ref{eq-CAL}) and (\ref{eq-CAL-0}) and related ideas have been studied, sometimes independently (re)derived, in various contexts of causal inference, missing-data problems, and survey sampling
(e.g., Folsom 1991; Tan 2010; Hainmueller 2012; Graham et al.~2012; Imai \& Ratovic 2014; Kim \& Haziza 2014; Vermeulen \& Vansteelandt 2015; Chan et al.~2016).
See Section~\ref{sec:discussion} for further discussion.
To follow the survey literature where such calibration estimation appears to be first used,
Eq.~(\ref{eq-CAL}) is called calibration equations for the treated (i.e., treatment 1), because the inverse probability weighted average of $f(X_i)$ over the
treated group $\{i: T_i=1, i=1,\ldots,n\}$ is calibrated to the average of $f(X_i)$ over the entire sample including the
treated and untreated.
Similarly, Eq.~(\ref{eq-CAL-0}) is called calibration equations for the untreated (i.e., treatment 0).
The resulting estimators $\hat\gamma^1_{\mbox{\tiny CAL}}$ and $\hat\gamma^0_{\mbox{\tiny CAL}}$ are referred to as calibrated estimators of $\gamma$, in contrast with
the maximum likelihood estimator $\hat\gamma_{\mbox{\tiny ML}}$. The fitted values $\hat\pi^1_{\mbox{\tiny CAL}}(X)$ and $\hat\pi^0_{\mbox{\tiny CAL}}(X)$
are also called calibrated propensity scores.

We make two main contributions in this article.
First, the calibrated estimator $\hat\gamma^1_{\mbox{\tiny CAL}}$ can be equivalently defined as a minimizer of the loss function
\begin{align}
\ell_{\mbox{\tiny CAL}} (\gamma) &= \tilde E \left\{ T \me^{-\gamma^\T f(X)} + (1-T) \gamma^\T f(X) \right\}. \label{loss-CAL}
\end{align}
In fact, setting the gradient of $\ell_{\mbox{\tiny CAL}} (\gamma)$ to 0 is easily shown to yield calibration equation (\ref{eq-CAL}) with logistic $\pi(X;\gamma)$.
It can also be shown that $\ell_{\mbox{\tiny CAL}} (\gamma)$ is convex in $\gamma$, and is strictly convex and bounded from below under a certain non-separation condition (see Proposition~\ref{pro1}).
Previously, the loss function $\ell_{\mbox{\tiny CAL}}$ and related ones have been mainly used as a computational device (Tan 2010, Section 4.4; Graham et al.~2012; Vermeulen \& Vansteelandt 2015).

In Section~\ref{sec:cal}, we establish an interesting relationship between maximum likelihood and calibrated estimation in terms of their corresponding loss functions $\ell_{\mbox{\tiny ML}}$ and $\ell_{\mbox{\tiny CAL}}$,
beyond the apparent differences between the estimating equations (\ref{eq-score}) and (\ref{eq-CAL}).
This relationship provides a theoretical explanation for why calibrated propensity scores can be preferred over maximum likelihood fitted propensity scores for inverse probability weighting,
when propensity score model (\ref{logit-PS}) is possibly misspecified. Such a result has been lacking from previous works.

The second of our main contributions is to propose a regularized calibrated estimator of $\gamma$ in model (\ref{logit-PS})
and develop a computational algorithm and theoretical analysis, while allowing
that model (\ref{logit-PS}) may be misspecified and the dimension of the covariate vector $f(X)$ may be greater than the sample size $n$.
The new estimator, denoted by $\hat\gamma^1_{\mbox{\tiny RCAL}}$, is defined by minimizing the calibration loss $\ell_{\mbox{\tiny CAL}} (\gamma)$ with a Lasso penalty (Tibshirani 1992),
\begin{align}
\ell_{\mbox{\tiny RCAL}} (\gamma) &=\ell_{\mbox{\tiny CAL}} (\gamma) + \lambda \|\gamma_{1:p}\|_1, \label{reg-cal-loss}
\end{align}
where $\gamma_{1:p}= ( \gamma_1, \ldots,\gamma_p)^\T$ excluding $\gamma_0$, $\|\cdot\|_1$ denotes the $L_1$ norm such that $\|\gamma_{1:p}\|_1= \sum_{j=1}^p | \gamma_j |$, and $\lambda \ge 0$ is a tuning parameter.
By the Karush--Kuhn--Tucker condition for minimization of (\ref{reg-cal-loss}), the fitted propensity score, $\hat \pi^1_{\mbox{\tiny RCAL}}(X) = \pi(X; \hat\gamma^1_{\mbox{\tiny RCAL}})$, satisfies
\begin{align}
& \frac{1}{n} \sum_{i=1}^n \frac{T_i}{\hat \pi^1_{\mbox{\tiny RCAL}}(X_i) } = 1, \label{ineq-CAL-1} \\
& \frac{1}{n} \left| \sum_{i=1}^n \frac{T_i f_j(X_i)}{\hat \pi^1_{\mbox{\tiny RCAL}}(X_i) } - \sum_{i=1}^n f_j(X_i) \right| \le \lambda, \quad j=1,\ldots,p , \label{ineq-CAL-2}
\end{align}
where equality holds in (\ref{ineq-CAL-2}) for any $j$ such that the $j$th estimate $(\hat\gamma^1_{\mbox{\tiny RCAL}})_j$ is nonzero.
The inverse probability weights, $1/\hat \pi^1_{\mbox{\tiny RCAL}}(X_i)$ with $T_i=1$, still sum to the sample size $n$ by (\ref{ineq-CAL-1}), but
the weighted average of each covariate $f_j(X_i)$ over the treated group may differ from the overall average of $f_j(X_i)$ by no more than $\lambda$.
In other words, introducing the Lasso penalty to calibrated estimation leads to a relaxation of equalities (\ref{eq-CAL}) to box constraints (\ref{ineq-CAL-2}).

The Lasso method and generalizations have been extensively developed and used as a powerful tool for statistical learning in sparse, high-dimensional problems (e.g., Buhlmann \& van de Geer 2011).
For model (\ref{logit-PS}) as logistic regression, a Lasso penalized maximum likelihood estimator, denoted by $\hat\gamma_{\mbox{\tiny RML}}$, is obtained by minimizing
\begin{align}
\ell_{\mbox{\tiny RML}} (\gamma) &=\ell_{\mbox{\tiny ML}} (\gamma) + \lambda \|\gamma_{1:p}\|_1, \label{reg-ml-loss}
\end{align}
where $\ell_{\mbox{\tiny ML}} (\gamma)$ is the average negative log-likelihood in (\ref{loss-ML}).
Such Lasso penalized estimation has been studied in high-dimensional generalized linear models (including logistic regression)
by van de Geer (2008), Huang \& Zhang (2012), and Negahban et al.~(2012) among others.
However, existing results are mostly devoted to penalized maximum likelihood estimation and
are not directly applicable to the regularized calibrated estimator $\hat\gamma^1_{\mbox{\tiny RCAL}}$.
For example, the Hessian of the calibration loss $\ell_{\mbox{\tiny CAL}} (\gamma)$ depends on the response data $T_i$, but that
of $\ell_{\mbox{\tiny ML}} (\gamma)$ does not.
We provide a high-dimensional analysis of $\hat\gamma^1_{\mbox{\tiny RCAL}}$ and the resulting IPW estimator of $\mu^1$ under
simple technical conditions, while building on the previous works.

\subsection{Calibrated estimation} \label{sec:cal}

For model (\ref{logit-PS}), we compare the maximum likelihood estimator $\hat\gamma_{\mbox{\tiny ML}}$ and
the calibrated estimator $\hat\gamma^1_{\mbox{\tiny CAL}}$ and their loss functions $\ell_{\mbox{\tiny ML}} (\gamma)$ and $\ell_{\mbox{\tiny CAL}} (\gamma) $  in various ways.
First, the following result on convexity can be obtained similarly as
conditions (4) and (16) in Tan (2010) and, for $\ell_{\mbox{\tiny CAL}} (\gamma) $, directly from Vermeulen \& Vansteelandt (2015), Appendix D.

\begin{pro} \label{pro1}
The loss function $\ell_{\mbox{\tiny ML}} (\gamma) $ is convex in $\gamma$; it is strictly convex and bounded from below, and hence
has a unique minimizer $\hat\gamma_{\mbox{\tiny ML}}$, if and only if the set
\begin{align}
\Big\{ \gamma \not=0 : \gamma^\T f(X_i) \ge 0 \mbox{ if } T_i=1 \mbox{ and } \gamma^\T f(X_i) \le 0 \mbox{ if } T_i=0 \mbox{ for } i=1,\ldots,n \Big\} \mbox{ is empty}. \label{cond-ml}
\end{align}
The loss function $\ell_{\mbox{\tiny CAL}} (\gamma) $ is convex in $\gamma$; it is strictly convex and bounded from below, and hence
has a unique minimizer $\hat\gamma^1_{\mbox{\tiny CAL}}$, if and only if the set
\begin{align}
\Big\{ \gamma \not=0 : \gamma^\T f(X_i) \ge 0 \mbox{ if } T_i=1 \mbox{ for } i=1,\ldots,n \mbox{ and } \tilde E[ (1-T)\gamma^\T f(X)] \le 0 \Big\} \mbox{ is empty}. \label{cond-cal}
\end{align}
\end{pro}

As expected, condition (\ref{cond-ml}) requires that no linear predictor $\gamma^\T f(X)$ can separate the treated group $\{T_i=1\}$ and the untreated $\{T_i=0\}$.
In contrast, condition (\ref{cond-cal}) also amounts to some sort of non-separation of the two groups, but it is strictly more demanding than (\ref{cond-ml}):
it is possible that (\ref{cond-ml}) holds but (\ref{cond-cal}) fails, but not vice versa.
In other words, $\hat\gamma_{\mbox{\tiny ML}}$ may be well defined
but $\hat\gamma^1_{\mbox{\tiny CAL}}$ may not exist for certain datasets as found in our numerical study (see Table~\ref{table:non-conv} in the Supplementary Material).
This issue for calibrated estimation, however, can be effectively addressed by incorporating regularization, discussed in Section~\ref{sec:reg-cal}.

Next we study how the maximum likelihood and calibration loss functions $\ell_{\mbox{\tiny ML}} (\gamma)$ and $\ell_{\mbox{\tiny CAL}} (\gamma) $ are related to each other.
To allow for misspecification of model (\ref{logit-PS}), we write
$\ell_{\mbox{\tiny ML}} (\gamma) = \kappa_{\mbox{\tiny ML}} (\gamma^\T f)  $ and $\ell_{\mbox{\tiny CAL}} (\gamma) =\kappa_{\mbox{\tiny CAL}} (\gamma^\T f)$,
where for a function $\myeta(x)$,
\begin{align}
& \kappa_{\mbox{\tiny ML}} (\myeta) = \tilde E \left[ \log\left\{ 1+ \me^{\myeta(X)} \right\} - T \myeta(X) \right] , \label{kappa-ML}\\
& \kappa_{\mbox{\tiny CAL}} (\myeta) = \tilde E \left[ T \me^{-\myeta(X)} + (1-T) \myeta(X) \right]. \label{kappa-CAL}
\end{align}
Then $\kappa_{\mbox{\tiny ML}} (\myeta^*)$ and $\kappa_{\mbox{\tiny CAL}} (\myeta^*)$ are well defined for  the true log odds ratio
$\myeta^*(x) = \log[ \pi^*(x)/\{1-\pi^*(x)\} ]$, even when model (\ref{logit-PS}) is misspecified, that is, $\myeta^*(x)$ is not of the form $\gamma^\T f(x)$.
It can be easily shown that both $ \kappa_{\mbox{\tiny ML}} (\myeta) $ and $\kappa_{\mbox{\tiny CAL}} (\myeta)$ are convex in $\myeta$.
For two functions $\myeta(x)$ and $\myeta^\prime(x)$, consider the Bregman divergences associated with $\kappa_{\mbox{\tiny ML}}$ and $\kappa_{\mbox{\tiny CAL}}$ (Bregman 1967),
\begin{align*}
& D_{\mbox{\tiny ML}} (\myeta, \myeta^\prime) = \kappa_{\mbox{\tiny ML}} (\myeta) - \kappa_{\mbox{\tiny ML}} (\myeta^\prime) - \langle \nabla\kappa_{\mbox{\tiny ML}} (\myeta^\prime), \myeta - \myeta^\prime \rangle,\\
& D_{\mbox{\tiny CAL}} (\myeta, \myeta^\prime) = \kappa_{\mbox{\tiny CAL}} (\myeta) - \kappa_{\mbox{\tiny CAL}} (\myeta^\prime) - \langle \nabla\kappa_{\mbox{\tiny CAL}} (\myeta^\prime), \myeta - \myeta^\prime \rangle,
\end{align*}
where $\myeta$ is identified as a vector $(\myeta_1,\ldots,\myeta_n)$ with $\myeta_i = \myeta(X_i)$,
\begin{align*}
\langle \nabla\kappa_{\mbox{\tiny CAL}} (\myeta^\prime), \myeta - \myeta^\prime \rangle = n^{-1} \sum_{i=1}^n
\left[ \frac{\partial \{ T_i \me^{-\myeta^\prime_i} + (1-T_i) \myeta^\prime_i \} }{\partial \myeta^\prime_i}(\myeta_i-\myeta^\prime_i)\right] ,
\end{align*}
and $\langle \nabla\kappa_{\mbox{\tiny ML}} (\myeta^\prime), \myeta - \myeta^\prime \rangle$ is similarly defined.
For two probabilities $\rho\in(0,1)$ and $\rho^\prime\in (0,1)$, the Kullback--Liebler divergence is
\begin{align*}
L(\rho, \rho^\prime) = \rho^\prime \log(\rho/\rho^\prime) + (1-\rho^\prime) \log\{(1-\rho)/(1-\rho^\prime)\} \ge 0.
\end{align*}
In addition, let $K(\rho,\rho^\prime) = \rho^\prime /\rho - 1 - \log(\rho^\prime /\rho) \ge 0$, which is strictly convex in $\rho^\prime/\rho$ and has a minimum of 0 when $\rho^\prime/\rho=1$.

\begin{pro} \label{pro2}
(i) For any functions $\myeta(x)$ and $\myeta^\prime(x)$ and the corresponding functions $\pi(x) = \{1+ \me^{-\myeta(x)}\}^{-1}$ and $\pi^\prime(x) = \{1+ \me^{-\myeta^\prime(x)}\}^{-1}$, it holds that
\begin{align*}
& D_{\mbox{\tiny ML}} (\myeta, \myeta^\prime) = \tilde E \Big[  L\{ \pi(X), \pi^\prime(X)\} \Big], \\
& D_{\mbox{\tiny CAL}} (\myeta, \myeta^\prime) = \tilde E  \left( \frac{T}{\pi^\prime(X)} \Big[ K\{ \pi(X), \pi^\prime(X)\} + L\{ \pi(X), \pi^\prime(X)\} \Big] \right).
\end{align*}
(ii) As a result,
we have for any fixed value $\gamma$,
\begin{align}
& E \Big\{\ell_{\mbox{\tiny ML}} (\gamma) - \kappa_{\mbox{\tiny ML}} (\myeta^*) \Big\} = E \Big[ L\{ \pi(X;\gamma), \pi^*(X)\} \Big],  \label{exp-loss-ml} \\
& E \Big\{\ell_{\mbox{\tiny CAL}} (\gamma) - \kappa_{\mbox{\tiny CAL}} (\myeta^*) \Big\} =E \Big[ K\{ \pi(X;\gamma), \pi^*(X)\}  + L\{ \pi(X;\gamma), \pi^*(X)\} \Big]. \label{exp-loss-cal}
\end{align}
\end{pro}

\vspace{.1in}
There are interesting implications from Proposition~\ref{pro2}.
First, we briefly describe results from theory of estimation in misspecified models (White 1982; Manski 1988).
Under standard regularity conditions as $n\to \infty$ and $p$ is fixed,
the maximum likelihood estimator $\hat\gamma_{\mbox{\tiny ML}}$ can be shown to converge in probability to a target value
$\bar\gamma_{\mbox{\tiny ML}}$, which is defined as a minimizer of the expected loss $E\{\ell_{\mbox{\tiny ML}} (\gamma)\}$ or equivalently
the Kullback--Liebler divergence (\ref{exp-loss-ml}).
The target (or limiting) propensity score  $\pi(\cdot; \bar\gamma_{\mbox{\tiny ML}})$ is closest to the truth $\pi^*(\cdot)$
as measured by the Kullback--Liebler divergence.
Similarly, $\hat\gamma^1_{\mbox{\tiny CAL}}$ can be shown to converge in probability to a target value
$\bar\gamma^1_{\mbox{\tiny CAL}}$, which is defined as a minimizer of the expected loss $E\{\ell_{\mbox{\tiny CAL}} (\gamma)\}$ or equivalently the calibration divergence (\ref{exp-loss-cal}).
The target (or limiting) propensity score  $\pi(\cdot; \bar\gamma^1_{\mbox{\tiny CAL}})$ is closest to the truth $\pi^*(\cdot)$
as measured by the calibration divergence (\ref{exp-loss-cal}).
If model (\ref{logit-PS}) is correctly specified, then both $\bar\gamma_{\mbox{\tiny ML}}$ and $\bar\gamma^1_{\mbox{\tiny CAL}}$ coincide with
$\gamma^*$ such that $\pi(\cdot;\gamma^*) = \pi^*(\cdot)$. However, if model (\ref{logit-PS}) is misspecified,
then $\bar\gamma_{\mbox{\tiny ML}}$ and $\bar\gamma^1_{\mbox{\tiny CAL}}$ in general differ from each other.

To compare possibly misspecified propensity scores $\pi(\cdot;\gamma)$ used for inverse probability weighting,
consider the mean squared relative error
$$ \mbox{MSRE}(\gamma) = E \Big[ Q\{ \pi(X;\gamma), \pi^*(X)\} \Big] = E \left[ \left\{\frac{\pi^*(X)}{\pi(X;\gamma)}-1\right\}^2 \right], $$
where $Q(\rho,\rho^\prime)=(\rho^\prime/\rho-1)^2$ for two probabilities $\rho\in(0,1)$ and $\rho^\prime\in (0,1)$.
A justification for this measure of relative errors can be seen from the following bound on the bias of the IPW estimator based on $\pi(\cdot;\gamma)$ by the Cauchy--Schwartz inequality,
\begin{align}
\Big| E \left\{ \hat\mu^1_{\mbox{\tiny IPW}}(\gamma)\right\} - \mu^1 \Big|= \left| E \left[\left\{ \frac{\pi^*(X)}{\pi(X;\gamma)} -1 \right\} Y^1 \right]\right| \le
\sqrt{ \mbox{MSRE} (\gamma)} \sqrt{ E\{(Y^1)^2\} }. \label{bias-bound}
\end{align}
where $\hat\mu^1_{\mbox{\tiny IPW}}(\gamma) = \tilde E\{ TY /\pi(X;\gamma) \}$.
Similarly, the mean squared error of $\hat\mu^1_{\mbox{\tiny IPW}}(\gamma)$ can also be bounded in terms of $\mbox{MSRE}(\gamma)$ under additional conditions.

\begin{pro} \label{pro3}
Suppose that $E\{ (Y^1)^2 | X\} \le c$ and $\pi^*(X) \ge \delta$ almost surely for some constants $c >0$ and $\delta \in (0,1)$. Then
for any fixed value $\gamma$,
\begin{align}
E \left[ \left\{ \hat\mu^1_{\mbox{\tiny IPW}}(\gamma) - \mu^1 \right\}^2 \right] \le c\, \mbox{MSRE}(\gamma) + \frac{2}{n\delta} c \{1+\mbox{MSRE}(\gamma)\}.  \label{mse-bound}
\end{align}
\end{pro}

Combining equations (\ref{exp-loss-ml})--(\ref{exp-loss-cal}) and the definition of $\mbox{MSRE}(\gamma)$, we obtain a formal explanation for why
the limiting propensity score  $\pi(\cdot; \bar\gamma^1_{\mbox{\tiny CAL}})$ can be preferred over $\pi(\cdot; \bar\gamma_{\mbox{\tiny ML}})$ for
achieving small relative errors with possible model misspecification. The argument is as follows, depending particularly on the presence of the function $K\{ \pi(X;\gamma), \pi^*(X)\}$ in (\ref{exp-loss-cal}):
\begin{align*}
\mbox{minimizing (\ref{exp-loss-cal})} \Longrightarrow \mbox{ reducing } E \Big[ K\{ \pi(X;\gamma), \pi^*(X)\} \Big]
\Longrightarrow \mbox{ controling } E \Big[ Q\{ \pi(X;\gamma), \pi^*(X)\} \Big] .
\end{align*}
That is,
minimization of the calibration divergence (\ref{exp-loss-cal}) results in small
$E [ K\{ \pi(X;\gamma), \pi^*(X)\}]$, which in turn leads to
a small mean squared relative error $E [ Q\{ \pi(X;\gamma), \pi^*(X)\}]$.
The first step is immediate because $E [ K\{ \pi(X;\gamma), \pi^*(X)\}]$ is no greater than (\ref{exp-loss-cal}).
The second step can be justified by the following proposition,
which shows that $E [ Q\{ \pi(X;\gamma), \pi^*(X)\}]$ is upper-bounded by $E [ K\{ \pi(X;\gamma), \pi^*(X)\}]$
up to a factor depending on the supremum of $\pi^*(X) /\pi(X;\gamma)$, at most that of $\pi^{-1}(X;\gamma)$.
In contrast, minimization of the Kullback--Liebler
divergence (\ref{exp-loss-ml}) does not seem to present a similar mechanism for controling relative errors.
See Figure~\ref{fig:obj} in the Supplementary Material for a numerical illustration of the behavior of the functions $L(\rho,\rho^\prime)$, $K(\rho,\rho^\prime)$, and $Q(\rho,\rho^\prime)$.

\begin{pro} \label{pro4}
(i) For a constant $a \in (0,1/2]$, if any two probabilities $\rho\in(0,1)$ and $\rho^\prime\in (0,1)$ satisfy $\rho \ge a \rho^\prime$, then
\begin{align*}
Q(\rho,\rho^\prime) \le \frac{5}{3\, a} K(\rho, \rho^\prime).
\end{align*}
By comparison,
$\sup_{\rho \ge a \rho^\prime} \{ Q(\rho, \rho^\prime) / L(\rho, \rho^\prime) \} = \infty$ for any constant $a>0$.\\
(ii) For a fixed value $\gamma$, suppose that $\pi(X;\gamma) \ge a\, \pi^*(X) $ almost surely for some constant $a \in (0,1/2]$. Then
\begin{align*}
E \Big[ Q\{ \pi(X;\gamma), \pi^*(X)\} \Big]  \le \frac{5}{3\, a} E \Big[ K\{ \pi(X;\gamma), \pi^*(X)\} \Big].
\end{align*}
\end{pro}

\begin{figure}
\caption{\small Limiting propensity scores (left) based on $\bar\gamma_{\mbox{\tiny ML}}$, $\bar\gamma^1_{\mbox{\tiny CAL}}$ and
 $\bar\gamma_{\mbox{\tiny BAL}}$, and the ratios (right) of the true propensity scores over those when a propensity score model is misspecified.} \label{fig:normal2} \vspace{.1in}
\begin{tabular}{c}
\includegraphics[width=6in, height=2.5in]{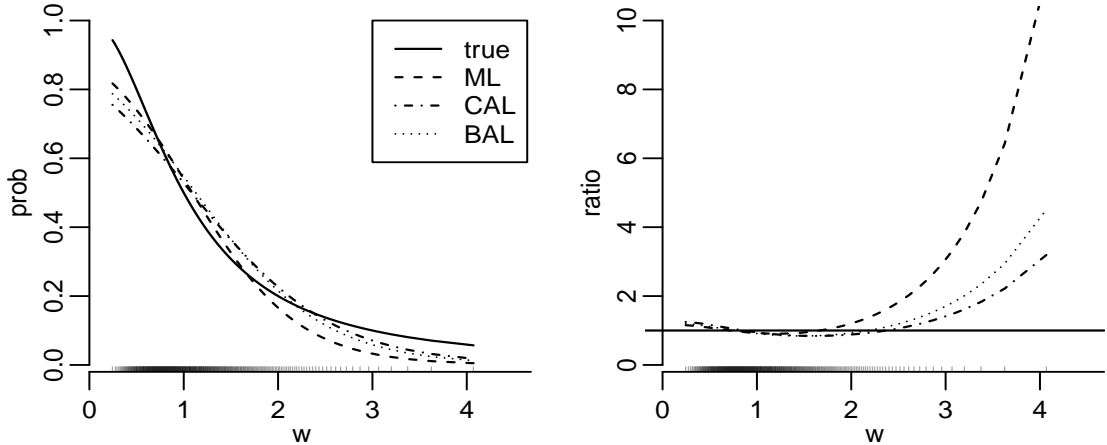} \vspace{-.15in}
\end{tabular}
\end{figure}

For illustration, consider a simple setting adapted from the simulation study in Section~\ref{sec:simulation}. Let $W=f_1(X) = \exp(X/2)$ with $X \sim \N(0,1)$ and $\pi^*(X) =\{1+\exp(X)\}^{-1}$.
The propensity score model $\pi(X;\gamma) = \{1+\exp(\gamma_0 + \gamma_1 W)\}^{-1}$ is misspecified, perhaps in a mild manner.
Figure~\ref{fig:normal2} shows the limiting propensity scores $\pi(\cdot; \bar\gamma_{\mbox{\tiny ML}})$, $\pi(\cdot; \bar\gamma^1_{\mbox{\tiny CAL}})$,
and $\pi(\cdot; \bar\gamma_{\mbox{\tiny BAL}})$ (Imai \& Ratkovic 2014), conditionally on
$n=400$ design points $(W_1,\ldots,W_n)$, where $W_i = \exp(X_i/2)$ and $X_i$ is the $i/401$ quantile of $\N(0,1)$ for $i=1,\ldots,n$.
The values $\bar\gamma_{\mbox{\tiny ML}}$, $\bar\gamma^1_{\mbox{\tiny CAL}}$, and $\bar\gamma_{\mbox{\tiny BAL}}$ are computed by minimizing respectively
$\ell_{\mbox{\tiny ML}}(\gamma)$, $\ell_{\mbox{\tiny CAL}}(\gamma)$, $\ell_{\mbox{\tiny BAL}}(\gamma)$ in (\ref{loss-ML}), (\ref{loss-CAL}), and (\ref{loss-BAL})
with $T_i$ replaced by $\pi^*(X_i)$. If judged by pointwise absolute errors, that is, $|\pi(\cdot; \bar\gamma_{\mbox{\tiny ML}})-\pi^*(\cdot)|$, etc., the three propensity scores are comparable and
reasonably capture the main trend of the true propensity scores.
However, substantial differences emerge, when the propensity scores are compared by pointwise relative errors, that is, $|\pi^*(\cdot)/\pi(\cdot; \bar\gamma_{\mbox{\tiny ML}})-1|$, etc.
The calibrated propensity scores are the most accurate, the maximum likelihood propensity scores are the least, and the balancing propensity scores are in-between,
especially in the right tail of $W$ where the true propensity scores are small. If a true propensity score $0.05$ is estimated by, for example, $0.005$, then
the relative error is large even though the absolute error appears small. As suggested by (\ref{bias-bound})--(\ref{mse-bound}), it is relative errors rather than absolute errors that are
relevant for evaluating propensity scores used for inverse probability weighting.

\subsection{Regularized calibrated estimation} \label{sec:reg-cal}

We turn to the regularized calibrated method. There are two motivations for incorporating regularization into calibrated estimation:
(i) to deal with the situation where $\hat\gamma^1_{\mbox{\tiny CAL}}$ may not exist because the calibration loss (\ref{loss-CAL}) may not admit a finite minimizer (see Proposition~\ref{pro1}),
and (ii) to improve statistical estimation when the dimension of covariate vector $f(X)$ is close to or greater than the sample size.
In particular, we study the Lasso penalized calibrated estimator $\hat\gamma^1_{\mbox{\tiny RCAL}}$ defined by minimizing the Lasso penalized loss (\ref{reg-cal-loss}).
As discussed in Sectoin~\ref{sec:overview}, using Lasso has a convenient interpretation of
relaxing the calibration equations (\ref{eq-CAL}) to inequalities (\ref{ineq-CAL-2}), in addition to the theoretical properties to be shown below.

\subsubsection{Computation} \label{sec:comput}

We present a Fisher scoring descent
algorithm for computing the estimator $\hat\gamma^1_{\mbox{\tiny RCAL}}$, that is, minimizing $\ell_{\mbox{\tiny RCAL}}(\gamma)$ in (\ref{reg-cal-loss}) for any fixed choice of $\lambda$.
The basic idea of the algorithm is to iteratively form a quadratic approximation to the calibration loss $\ell_{\mbox{\tiny CAL}}(\gamma)$ in (\ref{loss-CAL}) and
solve a Lasso penalized, weighted least squares problem, similarly as existing algorithms for Lasso penalized (maximum likelihood based) logistic regression (e.g., Friedman et al.~2010).
However, a suitable quadratic approximation is obtained only after an additional step, which is, in general, known as Fisher scoring.
In fact, Fisher scoring is previously used to derive the iterative reweighted least squares (IRLS) for fitting
generalized linear models with non-canonical links, for example, probit regression (McCullagh \& Nelder 1989).

The quadratic approximation directly from a Taylor expansion of $\ell_{\mbox{\tiny CAL}}(\gamma)$ about current estimates (denoted by $\tilde\gamma$) is
\begin{align}
\ell_{\mbox{\tiny CAL, Q1}}(\gamma; \tilde\gamma) & = \ell_{\mbox{\tiny CAL}}(\tilde\gamma)+\tilde E \left[ \left\{-T \me^{-f^\T(X) \tilde \gamma} + 1-T\right\} f^\T(X) (\gamma-\tilde\gamma) + \right. \nonumber \\
& \quad \left. \frac{1}{2} (\gamma-\tilde\gamma)^\T f^\T(X)
\left\{ T\me^{-f^T(X)\tilde\gamma}\right\} f(X) (\gamma-\tilde\gamma)  \right]. \label{QA-1}
\end{align}
As suggested from the quadratic term, it is tempting to recast (\ref{QA-1}) as a weighted least squares objective function with weights $T_i \exp\{ -f^\T(X_i)\tilde\gamma\}$ for $i=1,\ldots,n$.
But this would then imply that the linear term could depend only on $\{X_i: T_i=1, i=1,\ldots,n\}$, which is a contradiction.
Instead, we replace $T_i \exp\{ -f^\T(X_i)\tilde\gamma\}$ by its expectation $[1 + \exp\{ f^\T(X_i) \tilde\gamma\}]^{-1}$ under (\ref{logit-PS}) with parameter $\tilde\gamma$,
and obtain
\begin{align}
\ell_{\mbox{\tiny CAL, Q2}}(\gamma; \tilde\gamma) & = \ell_{\mbox{\tiny CAL}}(\tilde\gamma)+ \tilde E \left[ \left\{-T \me^{-f^\T(X) \tilde \gamma} + 1-T\right\} f^\T(X) (\gamma-\tilde\gamma) +\right.\nonumber \\
& \quad \left. \frac{1}{2} (\gamma-\tilde\gamma)^\T f^\T(X)
\left\{ 1 + \me^{f^\T(X_i) \tilde\gamma} \right\}^{-1} f(X) (\gamma-\tilde\gamma)  \right], \label{QA-2}
\end{align}
which is easily shown to be a weighted least squares objective function with covariate vector $f(X_i)$ and
working response and weights respectively
\begin{align}
\tilde T_i 
& = f^\T(X_i)\tilde\gamma + \frac{T_i - \pi(X_i;\tilde\gamma)}{\pi(X_i;\tilde\gamma)\{1-\pi(X_i;\tilde\gamma)\}} , \label{working-response} \\
w_i 
& = 1- \pi(X_i;\tilde\gamma) . \label{working-weight}
\end{align}
By comparison, in the IRLS algorithm for fitting logistic regression by maximum likelihood, the working response is the same as (\ref{working-response})
but the weight is $\pi(X_i; \tilde\gamma) \{1-\pi(X_i; \tilde\gamma)\}$.
Therefore, observations are weighted more with $\pi(X_i;\tilde\gamma)$ closer to $1/2$ for maximum likelihood estimation, but
with $\pi(X_i;\tilde\gamma)$ closer to 0 for calibrated estimation by (\ref{working-weight}).

To reduce computational cost, we also exploit the use of the majorization-minimization (MM) technique (Wu \& Lange 2010), similarly as in existing algorithms
for logistic regression. In particular, a majorizing function
of (\ref{QA-2}) at current estimates $\tilde\gamma$ is, by the quadratic lower bound principle (Bohning \& Lindsay 1988), the quadratic function obtained
by replacing the Hessian $\tilde E[ f^\T(X) \{1- \pi(X;\tilde\gamma)\} f(X)]$ by
$ \tilde E[ f^\T(X)  f(X)]$ in (\ref{QA-2}).
The resulting quadratic function of $\gamma$, denoted by $\ell_{\mbox{\tiny CAL, Q3}}(\gamma; \tilde\gamma)$, can be shown to be a weighted least squares objective function with
working response and weights
\begin{align*}
\tilde T_i  = f^\T(X_i)\tilde\gamma + \frac{T_i}{\pi(X_i;\tilde\gamma)}-1 , \quad w_i =1 .
\end{align*}
A complication from Fisher scoring, i.e., transition from (\ref{QA-1}) to (\ref{QA-2}) is that, unlike a direct majorization of the quadratic approximation from a Taylor expansion,
the function $\ell_{\mbox{\tiny CAL, Q3}}(\gamma; \tilde\gamma)$ may not
be a majorizing function of $\ell_{\mbox{\tiny CAL}}(\gamma)$ and hence minimization of $\ell_{\mbox{\tiny CAL, Q3}}(\gamma; \tilde\gamma)+ \lambda \|\gamma_{1:p}\|_1$
may not lead to a decrease of the objective function $\ell_{\mbox{\tiny RCAL}}(\gamma) = \ell_{\mbox{\tiny CAL}}(\gamma)+ \lambda \|\gamma_{1:p}\|_1$ from
the current value $\ell_{\mbox{\tiny RCAL}}(\tilde\gamma)$, as otherwise would be achieved by the MM technique.
However, the descent property, when occasionally violated, can be restored by incorporating a backtracking line search, because
the direction found from minimizing $\ell_{\mbox{\tiny CAL, Q3}}(\gamma; \tilde\gamma)+ \lambda \|\gamma_{1:p}\|_1$ must be a descent direction
for the objective function $\ell_{\mbox{\tiny RCAL}}(\gamma)$.

\begin{pro} \label{pro5}
Let $\tilde\gamma^{(1)} \not= \tilde\gamma$ be a minimizer of $\ell_{\mbox{\tiny CAL, Q2}}(\gamma; \tilde\gamma)+ \lambda \|\gamma_{1:p}\|_1$
or alternatively $\ell_{\mbox{\tiny CAL, Q3}}(\gamma; \tilde\gamma)+ \lambda \|\gamma_{1:p}\|_1$
and $\tilde\gamma^{(t)} = (1-t) \tilde\gamma + t \tilde\gamma^{(1)}$ for $0 \le t \le 1$.
Then any subgradient of $\ell_{\mbox{\tiny CAL}}(\tilde\gamma^{(t)})+ \lambda \|\gamma^{(t)}_{1:p}\|_1$ at $t=0$ is negative.
\end{pro}

\vspace{.05in}
Combining the preceding discussion leads to the following algorithm.

\vspace{.05in}
\noindent {\bf Algorithm 1}.\; {\it
Fisher scoring descent algorithm for minimizing (\ref{reg-cal-loss}):
\begin{itemize} \addtolength{\itemsep}{-.15in}
\item[(i)] Set an initial value $\gamma^{(0)}$.
\item[(ii)] Repeat the following updates for $k=1,2,\ldots$ until convergence to obtain $\hat\gamma^1_{\mbox{\tiny CAL}}$:
\begin{itemize}\addtolength{\itemsep}{-.1in}
\item[(ii1)] Compute $\gamma^{(k-1/2)} = \argmin_\gamma \, \ell_{\mbox{\tiny CAL, Q2}}(\gamma; \gamma^{(k-1)})+ \lambda \|\gamma_{1:p}\|_1$ or alternatively\\
$\gamma^{(k-1/2)} = \argmin_\gamma \, \ell_{\mbox{\tiny CAL, Q3}}(\gamma; \gamma^{(k-1)})+ \lambda \|\gamma_{1:p}\|_1$.

\item[(ii2)] If $\ell_{\mbox{\tiny RCAL}}( \gamma^{(k-1/2)} ) < \ell_{\mbox{\tiny RCAL}}( \gamma^{(k-1)} )$, then set $\gamma^{(k)}  = \gamma^{(k-1/2)} $;
otherwise set $\gamma^{(k)}  = (1-t) \gamma^{(k-1)} + t \gamma^{(k-1/2)}$ for some $0<t<1$, through a backtracking line search,
such that $\ell_{\mbox{\tiny RCAL}}( \gamma^{(k)} ) < \ell_{\mbox{\tiny RCAL}}( \gamma^{(k-1)} )$.
\end{itemize}
\end{itemize}
}

Various algorithms, for example, coordinate descent as in Friedman et al.~(2010) can be used for solving the least-squares Lasso problem in Step (ii2).
Our numerical implementation employs the simple surrogate function $\ell_{\mbox{\tiny CAL, Q3}}(\gamma; \tilde\gamma)$ and
then a variation of the active set algorithm in Osborne et al.~(2000), which enjoys a finite termination property.
We need to compute only once and save the QR decompostion of the Gram matrix defined from the vectors $\{f_j(X_1),\ldots,f_j(X_n)\}$ for the active coordinates $\gamma_j$
in the active set algorithm.
Computer codes will be made publicly available (currently submitted as a supplementary file).

\subsubsection{High-dimensional analysis}

We provide a high-dimensional analysis of the regularized calibrated estimator $\hat\gamma^1_{\mbox{\tiny RCAL}}$ and
the resulting IPW estimator of $\mu^1$,
allowing for misspecification of model (\ref{logit-PS}).
In fact, we obtain a general result with possible model misspecification on convergence of Lasso penalized M-estimators,
including $\hat\gamma^1_{\mbox{\tiny RCAL}}$ and $\hat\gamma_{\mbox{\tiny RML}}$, in the high-dimensional setting where the number of covariates $p$
is close to or greater than the sample size $n$.
See Appendix I in Supplementary Material. Such general results
can also be useful in other applications.

As discussed in Section~\ref{sec:cal}, for calibrated estimation with the loss $\ell_{\mbox{\tiny CAL}}(\gamma)$,
the target value of $\gamma$, denoted as $\bar \gamma^1_{\mbox{\tiny CAL}}$, is defined as a minimizer of the expected calibration loss
\begin{align*}
E \{ \ell_{\mbox{\tiny CAL}} (\gamma) \} &= E \left\{ T \me^{-\gamma^\T f(X)} + (1-T) \gamma^\T f(X) \right\}.
\end{align*}
The resulting approximation of $\myeta^*$ is $\bar\myeta^1_{\mbox{\tiny CAL}} = (\bar\gamma^1_{\mbox{\tiny CAL}} )^\T f$,
in general different from $\myeta^*$ in the presence of model misspecification.
For our theoretical analysis of $\hat\gamma^1_{\mbox{\tiny RCAL}}$, the tuning parameter in the Lasso penalized loss (\ref{reg-cal-loss}) is specified as $\lambda = A_0 \lambda_0$, with
a constant $A_0 >1$ and
\begin{align*}
\lambda_0 = O(1)\sqrt{ \log\{(1+p)/\epsilon\}/n},
\end{align*}
where $O(1)$ is a constant depending only on $(B_0,C_0)$ from the conditions (i) and (ii) of Proposition~\ref{pro-RCAL},
and $0 <\epsilon<1$ is a tail probability for the error bound. For example, taking $\epsilon=1/(1+p)$ gives $\lambda_0 =O(1)\sqrt{ 2 \log(1+p)/n}$, a familiar rate in high-dimensional analysis.

Our main result, Proposition~\ref{pro-RCAL}, establishes the convergence of  $\hat\gamma^1_{\mbox{\tiny RCAL}}$ to $\bar \gamma^1_{\mbox{\tiny CAL}}$
in the $L_1$ norm $\|\hat\gamma^1_{\mbox{\tiny RCAL}} -\bar\gamma^1_{\mbox{\tiny CAL}} \|_1$ and
the symmetrized Bregman divergence between $\hat\myeta^1_{\mbox{\tiny RCAL}} = (\hat\gamma^1_{\mbox{\tiny RCAL}} )^\T f$ and $\bar\myeta^1_{\mbox{\tiny CAL}} = (\bar\gamma^1_{\mbox{\tiny CAL}} )^\T f$.
In fact, convergence is obtained in terms of
$D^\dag_{\mbox{\tiny CAL}} ( \hat \myeta^1_{\mbox{\tiny RCAL}}, \bar \myeta^1_{\mbox{\tiny CAL}} )$,
where for two functions $\myeta= \gamma^\T f$ and $\myeta^\prime = \gamma^{\prime \T} f$,
\begin{align*}
D^\dag_{\mbox{\tiny CAL}}  (  \myeta , \myeta^\prime ) = D_{\mbox{\tiny CAL}} (  \myeta , \myeta^\prime ) + D_{\mbox{\tiny CAL}} ( \myeta^\prime , \myeta )
+(A_0-1) \lambda_0 \|\gamma - \gamma^\prime \|_1 .
\end{align*}
See Appendix I in the Supplementary Material for a discussion of the technical conditions imposed and a comparison with related results in high-dimensional analysis
including Buhlmann \& van de Geer (2011), Huang \& Zhang (2012), and Negahban et al.~(2012).

\begin{pro} \label{pro-RCAL}
Suppose that (i) $\bar \myeta^1_{\mbox{\tiny CAL}}(X) \ge B_0$ for a constant $B_0 \in \mathbb R$, that is,
$\pi(X; \bar\gamma^1_{\mbox{\tiny CAL}})$ is bounded from below by $( 1+\me^{-B_0})^{-1}$,
(ii) Assumption~\ref{ass-compat} in Appendix I holds with some subset $S \subset \{0,1,\ldots,p\}$ containing 0 and constants $\nu_0>0$ and $\xi_0>1$,
where $\psi(T, \myeta) = T\me^{-\myeta} + (1-T)\myeta$,
(iii) Assumption~\ref{ass-covariate} in Appendix I holds: $\max_{j=0,1,\ldots,p} |f_j(X)| \le C_0$ for a constant $C_0>0$,  and (iv)
$|S|\lambda_0 \le \eta_0$ for a sufficiently small constant $\eta_0 >0$, as derived from Assumption~\ref{ass-rate} in Appendix I.\
Then for a sufficiently large constant $A_0$ depending only on $(B_0,C_0)$, we have
with probability at least $1-4\epsilon$,
\begin{align}
D^\dag_{\mbox{\tiny CAL}} ( \hat \myeta^1_{\mbox{\tiny RCAL}}, \bar \myeta^1_{\mbox{\tiny CAL}} )
\le  O(1) \left\{ \lambda_0\sum_{j\not\in S} |\bar \gamma^1_{\mbox{\tiny CAL},j}| + |S| \lambda_0^2 \right\}, \label{pro-RCAL-eq}
\end{align}
where $O(1)$ depends only on $(A_0, B_0, \xi_0, \nu_0, C_0, \eta_0)$.
\end{pro}

From Proposition~\ref{pro-RCAL}, the following slow and fast rates can be deduced.
In spite of their names, the two rates are of distinct interest, being valid under different assumptions.
Taking $S = \{0\}$ leads to a slow rate, of order $\lambda_0\sum_{j=1}^p |\bar \gamma^1_{\mbox{\tiny CAL},j}| $,
where the corresponding compatibility assumption is explicitly satisfied under mild conditions.

\begin{cor} \label{cor-RCAL-slow}
Suppose that conditions (i), (iii), and (iv) in Proposition~\ref{pro-RCAL} hold with $|S|=1$ and that either no linear combination of $f_1(X), \ldots, f_p(X)$ is close to being a constant
or the $\psi_2$-weighted $L_2$ norms of $f_1(X), \ldots, f_p(X)$ are bounded away from above by 1, as defined in (\ref{cor-slow-rate-eq1}) and (\ref{cor-slow-rate-eq2}) of Appendix I.\
Then for a sufficiently large constant $A_0$ depending only on $(B_0,C_0)$, we have
with probability at least $1-4\epsilon$,
\begin{align}
D^\dag_{\mbox{\tiny CAL}} ( \hat \myeta^1_{\mbox{\tiny RCAL}}, \bar \myeta^1_{\mbox{\tiny CAL}} )
\le  O(1) \lambda_0\sum_{j=1}^p |\bar \gamma^1_{\mbox{\tiny CAL},j}| , \label{cor-RCAL-slow-eq}
\end{align}
where $O(1)$ depends only on $(A_0, B_0, C_0, \eta_0)$ and $\eta_3$ or $\eta_4$ from (\ref{cor-slow-rate-eq1}) or (\ref{cor-slow-rate-eq2}).
\end{cor}

Taking $S = \{0\} \cup \{j: \bar\gamma^1_{\mbox{\tiny CAL},j} \not= 0, j=1,\ldots,p\}$ yields a fast rate, of order $|S| \lambda_0^2$, albeit under
a compatibility condition on the linear dependency between $f_1(X),\ldots,f_p(X)$,
which may be violated when the number of covariates, $p$, is large.

\begin{cor} \label{cor-RCAL-fast}
Suppose that conditions (i)--(iv) in Proposition~\ref{pro-RCAL} hold with $S = \{0\} \cup \{j: \bar\gamma^1_{\mbox{\tiny CAL},j} \not= 0, j=1,\ldots,p\}$.
Then for a sufficiently large constant $A_0$ depending only on $(B_0,C_0)$, we have
with probability at least $1-4\epsilon$,
\begin{align}
D^\dag_{\mbox{\tiny CAL}} ( \hat \myeta^1_{\mbox{\tiny RCAL}}, \bar \myeta^1_{\mbox{\tiny CAL}} )
\le  O(1) |S| \lambda_0^2,  \label{cor-RCAL-fast-eq}
\end{align}
where $O(1)$ depends only on $(A_0, B_0, \xi_0, \nu_0, C_0, \eta_0)$.
\end{cor}

We now examine implications of the preceding results together with those in Section~\ref{sec:cal} on IPW estimation.
Denote $\hat\pi^1_{\mbox{\tiny RCAL}}(X) =\pi(X; \hat\gamma^1_{\mbox{\tiny RCAL}})$, the fitted propensity score based on $\hat\gamma^1_{\mbox{\tiny RCAL}}$.
Consider the resulting IPW estimator in two equivalent forms due to (\ref{ineq-CAL-1}),
\begin{align*}
\hat \mu^1_{\mbox{\tiny IPW }} ( \hat\pi^1_{\mbox{\tiny RCAL}})= \hat \mu^1_{\mbox{\tiny rIPW }} ( \hat\pi^1_{\mbox{\tiny RCAL}}) = \tilde E\left\{ \frac{TY}{\hat\pi^1_{\mbox{\tiny RCAL}}(X)} \right\} .
\end{align*}
Then a high-probability bound can be obtained on the difference between $\hat \mu^1_{\mbox{\tiny IPW }} ( \hat\pi^1_{\mbox{\tiny RCAL}})$
and the limiting version $\hat \mu^1_{\mbox{\tiny IPW }} ( \bar\pi^1_{\mbox{\tiny CAL}})$ with
$\bar\pi^1_{\mbox{\tiny CAL}}(X) =\pi(X; \bar\gamma^1_{\mbox{\tiny CAL}})$.

\begin{cor} \label{cor-diff}
(i) Suppose that the conditions in Corollary~\ref{cor-RCAL-slow} hold and that
$\sum_{j=1}^p |\bar \gamma^1_{\mbox{\tiny CAL},j}| \le M_1$ for a constant $M_1>0$.
Then for a sufficiently large constant $A_0$ depending only on $(B_0,C_0)$, we have
with probability at least $1-4\epsilon$,
\begin{align}
\left| \hat \mu^1_{\mbox{\tiny IPW }} ( \hat\pi^1_{\mbox{\tiny RCAL}}) - \hat \mu^1_{\mbox{\tiny IPW }} ( \bar\pi^1_{\mbox{\tiny CAL}}) \right|^2
\le  O(1)\lambda_0 \tilde E \left\{ \frac{TY^2}{\bar\pi^1_{\mbox{\tiny CAL}}(X)} \right\}, \label{cor-diff-eq1}
\end{align}
where $O(1)$ depends only on $(A_0, B_0, C_0, \eta_0, M_1)$ and $\eta_3$ or $\eta_4$ from (\ref{cor-slow-rate-eq1}) or (\ref{cor-slow-rate-eq2}).\\
(ii) Suppose that the conditions in Corollary~\ref{cor-RCAL-fast} hold.
Then for a sufficiently large constant $A_0$ depending only on $(B_0,C_0)$, we have
with probability at least $1-4\epsilon$,
\begin{align}
\left| \hat \mu^1_{\mbox{\tiny IPW }} ( \hat\pi^1_{\mbox{\tiny RCAL}}) - \hat \mu^1_{\mbox{\tiny IPW }} ( \bar\pi^1_{\mbox{\tiny CAL}}) \right|^2
\le  O(1) |S| \lambda_0^2 \tilde E \left\{ \frac{TY^2}{\bar\pi^1_{\mbox{\tiny CAL}}(X)} \right\}, \label{cor-diff-eq2}
\end{align}
where $O(1)$ depends only on $(A_0, B_0, \xi_0, \nu_0,  C_0, \eta_0)$.
\end{cor}

A remarkable aspect of Corollary~\ref{cor-diff} is that as $\lambda_0 \to 0$, the difference between $\hat \mu^1_{\mbox{\tiny IPW }} ( \hat\pi^1_{\mbox{\tiny RCAL}})$
and $\hat \mu^1_{\mbox{\tiny IPW }} ( \bar\pi^1_{\mbox{\tiny CAL}})$ is shown to converge in probability to 0,
even when the $L_1$ norm $\|\hat\gamma^1_{\mbox{\tiny RCAL}} -\bar\gamma^1_{\mbox{\tiny CAL}} \|_1$ may not converge to 0.
As a special case, it can be shown that under the conditions in either (i) or (ii),
$\hat \mu^1_{\mbox{\tiny IPW }} ( \hat\pi^1_{\mbox{\tiny RCAL}})$ converges in probability to $\mu^1$ as $\lambda_0\to 0$, if model (\ref{logit-PS}) is correctly specified and $E(Y^2) < \infty$.
In fact, $\|\hat\gamma^1_{\mbox{\tiny RCAL}} -\bar\gamma^1_{\mbox{\tiny CAL}} \|_1$ is, in general, only bounded from above in probability,
under the conditions for the slow rate in Corollary~\ref{cor-RCAL-slow}.
The situation with the fast rate in Corollary~\ref{cor-RCAL-fast} is similar, but technically subtler:
$\|\hat\gamma^1_{\mbox{\tiny RCAL}} -\bar\gamma^1_{\mbox{\tiny CAL}} \|_1$ is generally of order $|S| \lambda_0$, which
is only required to be sufficiently small (no greater than some constant $\eta_0$) but need not be arbitrarily close to 0.
As seen from our proofs in the Supplementary Material, these results are demonstrated with key steps depending on the properties
of the calibration loss $\ell_{\mbox{\tiny CAL}}$ in Propositions \ref{pro2} and \ref{pro4}.
By comparison, similar rates of convergence as in Proposition~\ref{pro-RCAL} and Corollaries~\ref{cor-RCAL-slow} and \ref{cor-RCAL-fast}
can be obtained for the Lasso penalized maximum likelihood estimator $\hat \gamma_{\mbox{\tiny RML}}$ under comparable conditions
(Buhlmann \& van de Geer 2011; Huang \& Zhang 2012). See also Theorem~\ref{pro-reg} in Appendix I.\ However,
a similar result as Corollary~\ref{cor-diff} 
would not be available for
the IPW estimator based on $ \hat\gamma_{\mbox{\tiny RML}}$ without additional conditions.

Finally, although Corollary~\ref{cor-diff} deals with convergence of $\hat \mu^1_{\mbox{\tiny IPW }} ( \hat\pi^1_{\mbox{\tiny RCAL}})$
to $\hat \mu^1_{\mbox{\tiny IPW }} ( \bar\pi^1_{\mbox{\tiny CAL}})$, which may differ from the parameter of interest $\mu^1$,
we point out that Corollary~\ref{cor-diff} and the results in Section~\ref{sec:cal} are complementary in providing support for the use
of $\hat\pi^1_{\mbox{\tiny RCAL}}$ for IPW estimation of $\mu^1$.
The argument is based on the triangle inequality:
\begin{align*}
| \hat \mu^1_{\mbox{\tiny IPW }} ( \hat\pi^1_{\mbox{\tiny RCAL}}) - \mu^1 |
\le |\hat \mu^1_{\mbox{\tiny IPW }} ( \hat\pi^1_{\mbox{\tiny RCAL}}) - \hat \mu^1_{\mbox{\tiny IPW }} ( \bar\pi^1_{\mbox{\tiny CAL}})| +
|  \hat \mu^1_{\mbox{\tiny IPW }} ( \bar\pi^1_{\mbox{\tiny CAL}}) - \mu^1|.
\end{align*}
On one hand, as discussed through Propositions~\ref{pro2}--\ref{pro4}, the use of the calibration loss facilitates achieving a small second term,
$|\hat \mu^1_{\mbox{\tiny IPW }} ( \bar\pi^1_{\mbox{\tiny CAL}}) - \mu^1|$, in the presence of model misspecification.
On the other hand, specific properties of the calibration loss makes it possible
to achieve sharper rates of convergence of the first term,
$|\hat \mu^1_{\mbox{\tiny IPW }} ( \hat\pi^1_{\mbox{\tiny RCAL}}) - \hat \mu^1_{\mbox{\tiny IPW }} ( \bar\pi^1_{\mbox{\tiny CAL}})|$,
than based on maximum likelihood, when combined with Lasso penalization in high-dimensional settings.

\section{Simulation study} \label{sec:simulation}

We present a simulation study extending the design of Kang \& Schafer (2007) to high-dimensional, sparse settings.
For $p \ge 4$, let $X=(X_1,\ldots,X_p)^\T$ be independent, standard normal covariates,
and $T$ be a binary variable such that
\begin{align}
P(T=1 | X ) = \pi^*(X) = [ 1+ \exp\{ X_1 - 0.5 X_2 + 0.25 X_3 + 0.1 X_4 \} ]^{-1}, \label{simulation-PS}
\end{align}
depending only on the four covariates $(X_1,X_2,X_3,X_4)$.
Consider two specifications of logistic model (\ref{logit-PS}) with the following regressors:
\begin{itemize}\addtolength{\itemsep}{-.05in}
\item[(i)] $f_j(X)= X_j$ for $j=1,\ldots,p$.

\item[(ii)] $f_j(X)$ is a standardized version of $W_j$ with sample mean 0 and sample variance 1, where
$W_1 = \exp(0.5 X_1)$, $W_2 = 10 + \{ 1+ \exp(X_1) \}^{-1} X_2$, $W_3 = (0.04 X_1 X_3 + 0.6)^3$, $W_4 = (X_2 + X_4 + 20)^2$,
and, if $p >4$, $W_j = X_j$ for $j=5,\ldots,p$.
\end{itemize}
Then model (\ref{logit-PS}) is correctly specified in the scenario (i), but is misspecified in the scenario (ii).
For $p=4$, Kang \& Schafer (2007) showed that model (\ref{logit-PS}) in the scenario (ii), although misspecified,
appears adequate as examined by conventional techniques for logistic regression.
In addition, Kang \& Schafer (2007) constructed an outcome variable $Y^1 = 210 + 13.7 (2X_1 + X_2 + X_3 + X_4) + \varepsilon$ with $\varepsilon|(T,X) \sim \N(0,1)$,
and considered a linear model of $Y^1$ given $X$, which can be correctly specified with regressors as in scenario (i)
or misspecified with regressors as in scenario (ii) above.
The linear model of $Y^1$ given $X$ in the misspecified case can also been shown as ``nearly correct" by standard techniques for linear regression.
This simulation setting with $p=4$ has since been widely used to study various estimators for $\mu^1 = E(Y^1)$ with observed data $\{(T_i Y^1_i, T_i, X_i): i=1,\ldots,n\}$.
See, for example, Tan (2010), Imai \& Ratkovic (2014), Vermeulen \& Vansteelandt (2015), and Chan et al.~(2016).

We compare IPW estimators in the ratio form $\hat\mu^1_{\mbox{\tiny rIPW}}(\hat\pi)$, which are numbered as follows.
\begin{itemize}\addtolength{\itemsep}{-.05in}
\item[(1)] $\hat\pi$ is replaced by the true propensity score $\pi^*$.

\item[(2)] $\hat\pi = \tilde E(T)$, obtained from model (\ref{logit-PS}) with only the intercept $f\equiv 1$.

\item[(3)] $\hat\pi=\hat\pi_{\mbox{\tiny ML}}$ obtained by maximum likelihood, i.e., minimizing (\ref{loss-ML}).

\item[(4)] $\hat\pi=\hat\pi_{\mbox{\tiny RML}}$ obtained by Lasso penalized maximum likelihood, i.e., minimizing (\ref{reg-ml-loss}).

\item[(5)] $\hat\pi=\hat\pi^1_{\mbox{\tiny CAL}}$ obtained by calibrated estimation, i.e., minimizing (\ref{loss-CAL}).

\item[(6)] $\hat\pi=\hat\pi^1_{\mbox{\tiny RCAL}}$ obtained by regularized calibrated estimation, i.e., minimizing (\ref{reg-cal-loss}).
\end{itemize}
The functions (\ref{loss-ML}) and (\ref{loss-CAL}) are minimized using a trust-region algorithm in the R package $\texttt{trust}$ (Geyer 2014),
and (\ref{reg-cal-loss}) and (\ref{reg-ml-loss}) are minimized using the Fisher-scoring descent algorithm described in Section~\ref{sec:comput}.
The tuning parameter $\lambda$ in (\ref{reg-cal-loss}) or (\ref{reg-ml-loss}) is determined using 5-fold cross validation
based on the corresponding loss function. For $k=1,\ldots,5$, let
$\mathcal I_k$ be a random subsample of size $n/5$ from $\{1,2,\ldots,n\}$.
For a loss function $\ell (\gamma)$, for example $\ell_{\mbox{\tiny CAL}}(\gamma)$ in (\ref{loss-CAL}), denote by $\ell (\gamma; \mathcal I)$ the loss function
obtained when the sample average $\tilde E()$ is computed over only the subsample $\mathcal I$.
The 5-fold cross-validation criterion is defined as
\begin{align*}
\mbox{CV}_5 (\lambda) = \frac{1}{k}\sum_{k=1}^5 \ell ( \hat\gamma_\lambda^{(k)}; \mathcal I_k ),
\end{align*}
where $\hat\gamma^{(k)}_\lambda$ is a minimizer of the penalized loss $\ell(\gamma; \mathcal I^c_k) + \lambda \| \gamma_{1:p} \|_1$ over the subsample $\mathcal I^c_k$ of size $4n/5$,
i.e., the complement to $\mathcal I_k$.
Then $\lambda$ is selected by minimizing $\mbox{CV}_5(\lambda)$ over the discrete set $\{ \lambda_0 / 2^j: j=0,1,\ldots,10\}$,
where for $\hat\pi_0=\tilde E(T)$, the value $\lambda_0$ is computed as
\begin{align*}
\lambda_0 = \max_{j=1,\ldots,p} \left| \tilde E \{ (T -\hat\pi_0) f_j(X) \} \right|
\end{align*}
when the likelihood loss (\ref{loss-ML}) is used, or
\begin{align*}
\lambda_0 = \max_{j=1,\ldots,p} \left| \tilde E \{ (T/\hat\pi_0 -1) f_j(X) \} \right|
\end{align*}
when the calibration loss (\ref{loss-CAL}) is used. It can be shown that in either case, the penalized loss
$\ell(\gamma) + \lambda \| \gamma_{1:p} \|_1$ over the original sample has a minimum at $\gamma_{1:p} =0$ for all $\lambda \ge \lambda_0$.

The performance of an IPW estimator $\hat\mu^1_{\mbox{\tiny rIPW}}(\hat\pi)$ is affected by not only the closeness of $\hat\pi$ to $\pi^*$ but also
the outcome regression function $m^*_1(X)=E(Y^1| X)$ and the error $\varepsilon = Y^1 - m^*_1(X)$. See Section~\ref{sec:discussion} for
a related discussion about double robustness. Under unconfoundedness, it can be shown
via conditioning on $\{(T_i,X_i): i=1,\ldots,n\}$ that
\begin{align*}
&  E\{\hat\mu^1_{\mbox{\tiny rIPW}}(\hat\pi)\}  = E \{ \hat\mu^1_{\mbox{\tiny rIPW}}(\hat\pi; m_1^*) \}, \\
& \var \{\hat\mu^1_{\mbox{\tiny rIPW}}(\hat\pi)\} = \var \{ \hat\mu^1_{\mbox{\tiny rIPW}}(\hat\pi; m_1^*) \} + \var  \{ \hat\mu^1_{\mbox{\tiny rIPW}}(\hat\pi; \varepsilon) \},
\end{align*}
where $ \hat\mu^1_{\mbox{\tiny rIPW}}(\hat\pi; h)  = \tilde E \{ T h(X) / \hat\pi(X)\} / \tilde E \{ T / \hat\pi(X)\}$ for a function $h(X)$
and  $ \hat\mu^1_{\mbox{\tiny rIPW}}(\hat\pi; \varepsilon)  = \tilde E \{ T \varepsilon / \hat\pi(X)\} / \tilde E \{ T / \hat\pi(X)\}$.
As a result, the mean squared error $E[ \{ \hat\mu^1_{\mbox{\tiny rIPW}}(\hat\pi) - \mu^1\}^2]$, can be decomposed as
$\mbox{MSE} \{ \hat\mu^1_{\mbox{\tiny rIPW}}(\hat\pi; m^*_1) \} + \var\{\hat\mu^1_{\mbox{\tiny rIPW}}(\hat\pi; \varepsilon)\}$, where
\begin{align*}
\mbox{MSE} \{ \hat\mu^1_{\mbox{\tiny rIPW}}(\hat\pi; h) \} = E \left( \left[  \hat\mu^1_{\mbox{\tiny rIPW}}(\hat\pi; h) - E\{h(X)\} \right]^2 \right) .
\end{align*}
We consider a number of configurations for $m^*_1(X)$ including
\begin{itemize}\addtolength{\itemsep}{-.05in}
\item[](lin1): $h(X) = X_1 + 0.5 X_2 + 0.5 X_3 + 0.5 X_4$,

\item[](lin2): $h(X) = X_1 + 2 X_2 + 2 X_3 + 2 X_4$,

\item[](quad1): $h(X) = \sum_{j=1}^4 \{ \max (X_j,0) \}^2$,

\item[](quad2): $h(X) = \sum_{j=1}^4 \{ \max (-X_j,0) \}^2$,

\item[](exp): $h(X) = \sum_{j=1}^4 \exp(X_j/2)$.
\end{itemize}
The first configuration, lin1, corresponds to that used in Kang \& Schafer (2007), up to a linear transformation.
But the relative order of $\mbox{MSE}\{\hat\mu^1_{\mbox{\tiny rIPW}}(\hat\pi; m_1^*)\}$ from different estimators $\hat\pi$ remains the same under linear transformations of $m^*_1(X)$.

\begin{figure}[t!]
\caption{\small Root mean squared errors of $ \hat\mu^1_{\mbox{\tiny rIPW}}(\hat\pi; h) $
and $ \hat\mu^1_{\mbox{\tiny rIPW}} (\hat\pi; \varepsilon) $ for the estimators $\hat\pi$ labeled 1--6 when logistic model (\ref{logit-PS}) is correctly specified,
with $p = 4$ ($\triangle$), $20$ ($+$), $50$ ($\bullet$), $100$ ($\times$), or $200$ ($\nabla$) and
$n=200$ (left), $400$ (middle), or $800$ (right).
The estimators $\hat\pi_{\mbox{\tiny ML}}$ and $\hat\pi^1_{\mbox{\tiny CAL}}$ (3 and 5) are computed only for $p=4$, $20$, and $50$.
The results for $\hat\pi^1_{\mbox{\tiny CAL}}$ should be interpreted with caution for $(p,n)=(20,200)$ and $(50,\le 400)$ due to non-convergence found in 30--99\% of 1000 repeated simulations (see Table~\ref{table:non-conv}).}
\label{fig:est-cor} \vspace{.15in}
\begin{tabular}{c}
\includegraphics[width=6.2in, height=5in]{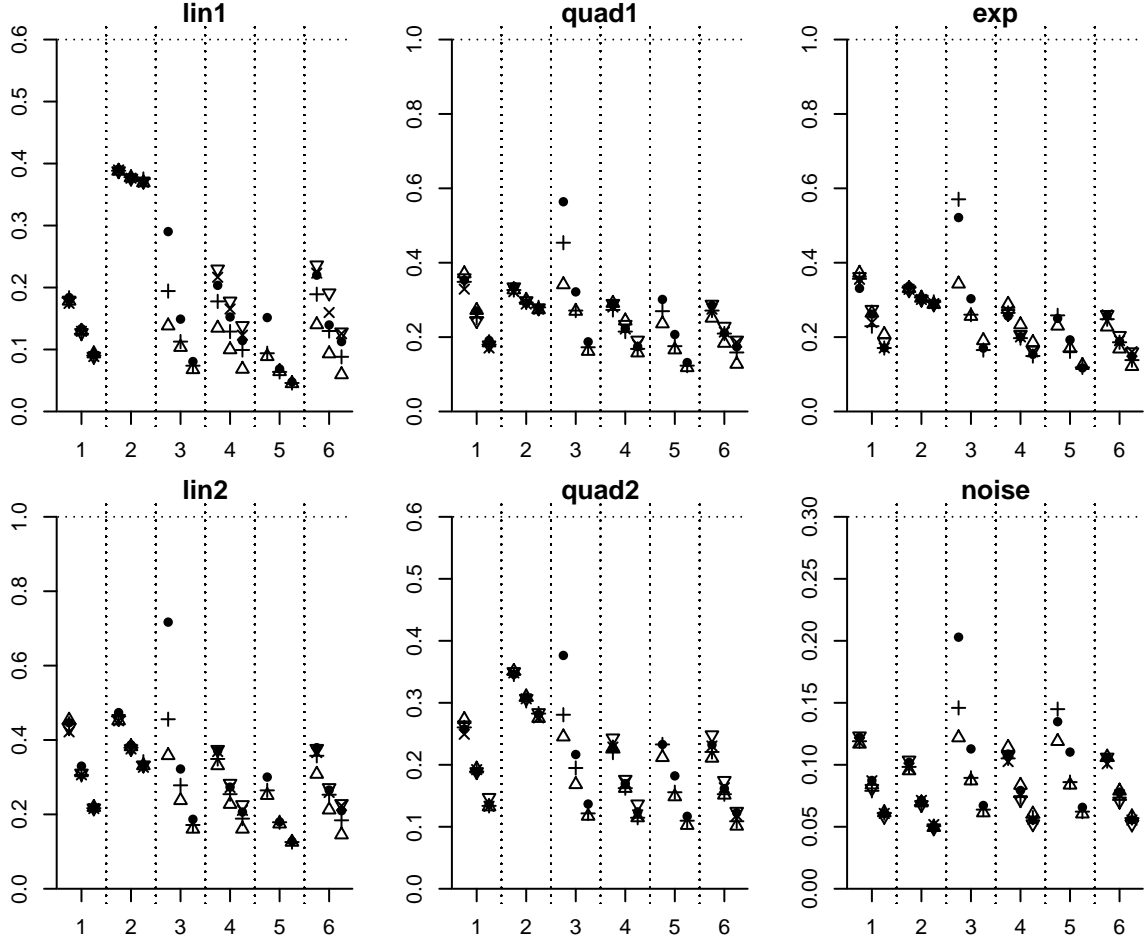} \vspace{-.25in}
\end{tabular}
\end{figure}

\begin{figure} [t!]
\caption{\small Root mean squared errors of $ \hat\mu^1_{\mbox{\tiny rIPW}}(\hat\pi; h) $
and $ \hat\mu^1_{\mbox{\tiny rIPW}} (\hat\pi; \varepsilon) $, plotted similarly as in Figure~\ref{fig:est-cor},
for the estimators $\hat\pi$ labeled 1--6 when logistic model (\ref{logit-PS}) is misspecified.
See the notes provided in Figure~\ref{fig:est-cor}. The values are censored within the upper limit of $y$-axis (dotted line). } \label{fig:est} \vspace{.15in}
\begin{tabular}{c}
\includegraphics[width=6.2in, height=5in]{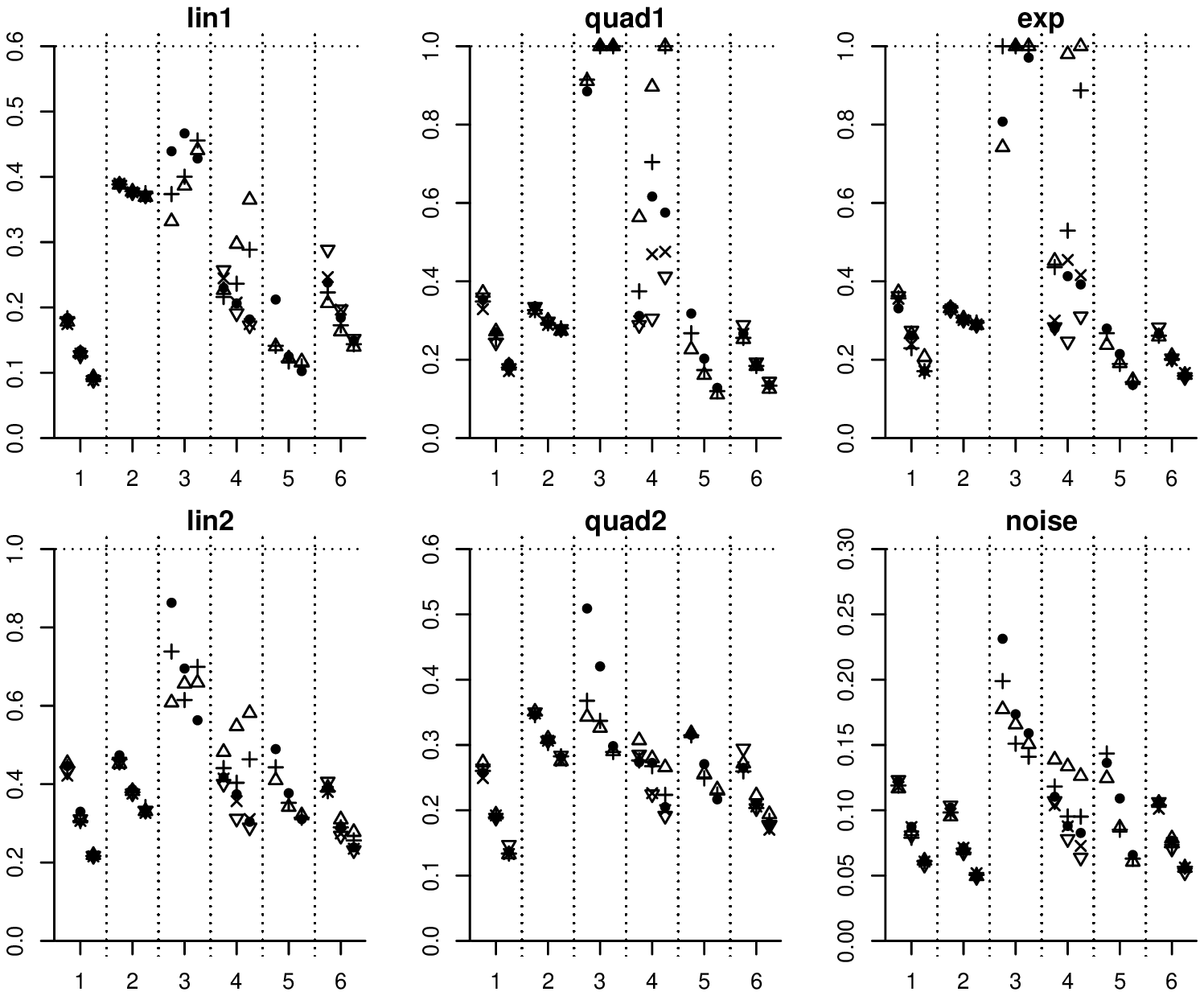} \vspace{-.15in}
\end{tabular}
\end{figure}

For model (\ref{logit-PS}) correctly specified or misspecified,
Figure~\ref{fig:est-cor} or \ref{fig:est} respectively
shows Monte Carlo estimates of $\mbox{MSE}^{1/2} \{ \hat\mu^1_{\mbox{\tiny rIPW}}(\hat\pi; h) \}$ with five choices of $h(X)$ and $\var^{1/2}\{\hat\mu^1_{\mbox{\tiny rIPW}} (\hat\pi; \varepsilon)\}$
with $\varepsilon \sim \N(0,1)$, for six estimators $\hat\pi$ (labeled 1--6 above) from 1000 repeated simulations with $n=200, 400, 800$ and $p=4, 20, 50, 100, 200$.
See Tables~\ref{table:est-lin1}--\ref{table:est-eps} in the Supplementary Material for numerical values.
The non-penalized estimators  $\hat\pi_{\mbox{\tiny ML}}$ and $\hat\pi^1_{\mbox{\tiny CAL}}$ are computed only for $p$ from $4$ to $50$.
For $(p,n)=(20,200)$ or $(50,\le 400)$, the estimator $\hat\pi^1_{\mbox{\tiny CAL}}$ is obtained with non-convergence declared by the R package $\texttt{trust}$ in a considerable fraction of simulations,
indicating that the loss function $\ell_{\mbox{\tiny CAL}}(\gamma)$ may not have a finite minumum (see Table~\ref{table:non-conv}).

We provide comments about comparison of the estimators in Figures~\ref{fig:est-cor} and \ref{fig:est}.
\begin{itemize}\addtolength{\itemsep}{-.05in}
\item For all the choices of $(n,p)$ and $h(X)$ studied, the estimator $\hat\pi^1_{\mbox{\tiny RCAL}}$  yields similar or smaller mean squared errors than $\hat\pi_{\mbox{\tiny RML}}$,
whether model (\ref{logit-PS}) is correctly specified or misspecified.
The advantage of $\hat\pi^1_{\mbox{\tiny RCAL}}$ is substantial in the case of misspecified model (\ref{logit-PS}).

\item For relatively small $p \le 50$, the estimator $\hat\pi^1_{\mbox{\tiny CAL}}$, in spite of the non-convergence issue mentioned above, consistently leads to smaller mean squared errors than $\hat\pi_{\mbox{\tiny ML}}$
whether model (\ref{logit-PS}) is correctly specified or misspecified. In the case of misspecified model (\ref{logit-PS}),
the performance of $\hat\pi_{\mbox{\tiny ML}}$ deteriorates substantially, particularly for estimation associated with configurations ``quad2" and ``exp" for $h(X)$.
A possible explanation is that $h(X)$ in these cases quickly increases as $(X_1,X_3,X_4)$ become large, which by definition (\ref{simulation-PS}) is the region where
the propensity scores $\pi^*(X)$ becomes small. Even a small discrepancy (especially under-estimation) between $\hat\pi$ and $\pi^*$ for the few observations with $T=1$ in this region can
yield large errors for the estimates $\hat\mu^1_{\mbox{\tiny rIPW}} (\hat\pi; h)$.

\item The regularized estimator $\hat\pi_{\mbox{\tiny RML}}$ or $\hat\pi^1_{\mbox{\tiny RCAL}}$ yields smaller or slightly larger means squared errors than
the corresponding non-regularized estimator $\hat\pi_{\mbox{\tiny ML}}$ or $\hat\pi^1_{\mbox{\tiny CAL}}$ with $p \le 50$, except for $\hat\pi^1_{\mbox{\tiny RCAL}}$
versus $\hat\pi^1_{\mbox{\tiny CAL}}$ in the configuration ``lin1" for $h(X)$. This exception can be explained as follows, due to several coincidental factors:
$X_1$ is the dominating component in the ``lin1" function $h(X)$, with $\var(X_1) > \var\{ 0.5(X_2+X_3+X_4)\}$.
At the same time, $X_1$ is also the dominating direction in determining the magnitude of the fitted propensity score $\hat\pi$,
either by definition (\ref{simulation-PS}) when model (\ref{logit-PS}) is correctly specified
or by the particular construction of $(W_1,W_2,W_3,W_4)$ when model (\ref{logit-PS}) is misspecified.
In fact, $X_1$ is the most important direction of propensity scores that can be recovered with misspecified model (\ref{logit-PS})
because $X_1$ and $W_1$ are monotone transformations of each other.
Regularization tends to introduce some bias into $\hat\pi$ along the direction of $X_1$, which then increases errors in $\hat\mu^1_{\mbox{\tiny rIPW}} (\hat\pi; h)$
with the ``lin1" function $h(X)$. This phenomenon seems related to the bias of Lasso in the presence of strong signals (e.g., Zhang \& Zhang 2012).
\end{itemize}

\begin{figure}
\caption{\small Root mean squared errors of global measures $\tilde\kappa_{\mbox{\tiny ML}} (\hat\myeta)$,
$\tilde\kappa_{\mbox{\tiny CAL}} (\hat\myeta) $, $\mbox{mse}(\hat\pi)$, and $\mbox{msre}(\hat\pi)$
for the estimators $\hat\pi$ labeled 1--6
when logistic model (\ref{logit-PS}) is misspecified.
See the notes provided in Figure~\ref{fig:est-cor}. The values are censored within the upper limit of $y$-axis (dotted line). } \label{fig:div} \vspace{.15in}
\begin{tabular}{c}
\includegraphics[width=6.2in, height=4in]{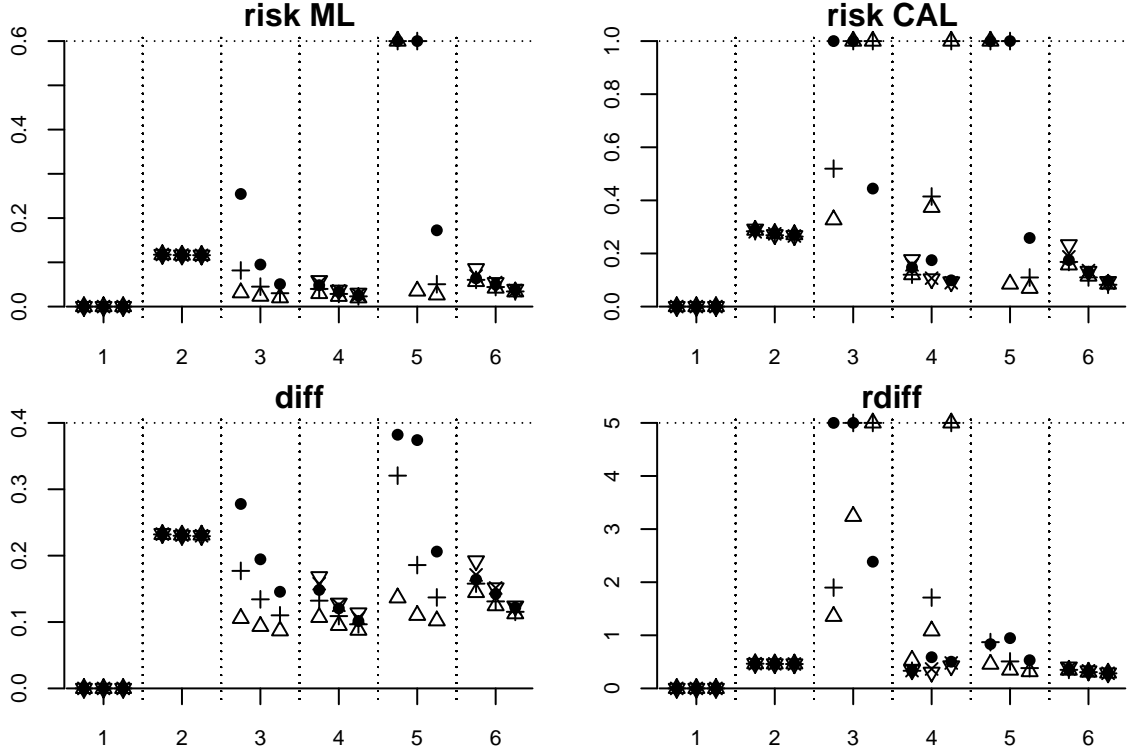} \vspace{-.15in}
\end{tabular}
\end{figure}

As seen from the preceding discussion, comparison of $\mbox{MSE} \{\hat\mu^1_{\mbox{\tiny rIPW}} (\hat\pi; h) \}$  between different estimators $\hat\pi$
may depend on the choice of $h(X)$. Alternatively, we consider several global measures of closeness of $\hat\pi$ to $\pi^*$ as follows.
Denote $\hat\myeta = \log\{ \hat\pi/(1-\hat\pi)\}$.
The expected likelihood loss (i.e., likelihood risk) $E \{\kappa_{\mbox{\tiny ML}} (\hat\myeta)\}$ achieved by $\hat \myeta$ is estimated as
\begin{align*}
\tilde\kappa_{\mbox{\tiny ML}} (\hat\myeta) = \tilde E \left[ \log\left\{ 1+ \me^{\hat\myeta(X)} \right\} - \pi^*(X) \hat\myeta(X) \right].
\end{align*}
The expected calibration loss (i.e., calibration risk) is $E \{\kappa_{\mbox{\tiny CAL}} (\myeta)\} = E[ \pi^*(X) \me^{-\myeta(X)} + \{1-\pi^*(X)\} \myeta(X)]$, and
its value achieved by $\hat\myeta$ is estimated as
\begin{align*}
\tilde\kappa_{\mbox{\tiny CAL}} (\hat\myeta) = \tilde E \left[ T \left\{ \me^{-\hat\myeta(X)}  - \me^{-\myeta^*(X)} \hat\myeta(X) \right\}\right].
\end{align*}
We find $\tilde\kappa_{\mbox{\tiny CAL}} (\hat\myeta)$ a more relevant measure than the direct estimate $\tilde E[ \pi^*(X) \me^{-\hat\myeta(X)} + \{1-\pi^*(X)\} \hat\myeta(X)]$,
because $\tilde\kappa_{\mbox{\tiny CAL}} (\hat\myeta)$ only involves comparison of $\hat\pi(X_i)$ and $\pi^*(X_i)$ for $\{i: T_i=1, i=1,\ldots,n\}$,
by which the performance of $\hat\mu^1_{\mbox{\tiny rIPW}} (\hat\pi; h)$ is mainly affected.
For similar reasons, the mean squared error $E [\{ \hat\pi(X) - \pi^*(X) \}^2]$ is estimated as
$$
\mbox{mse} (\hat\pi) = \tilde E \left[\frac{T}{\pi^*(X)}  \left\{ \hat\pi(X) - \pi^*(X) \right\}^2 \right],
$$
and the mean squared relative error is estimated as
$$
\mbox{msre} (\hat\pi) = \tilde E \left[ \frac{T}{\pi^*(X)} \left\{ \frac{\pi^*(X)}{\hat\pi(X)}-1 \right\}^2 \right].
$$
For misspecified model (\ref{logit-PS}), Figure~\ref{fig:div} presents Monte Carlo estimates of the four ``root mean squared errors",
$E^{1/2} [ \{ \tilde\kappa_{\mbox{\tiny ML}} (\hat\myeta) - \tilde\kappa_{\mbox{\tiny ML}} (\myeta^*) \}^2 ] $,
$E^{1/2} [ \{ \tilde\kappa_{\mbox{\tiny CAL}} (\hat\myeta) - \tilde\kappa_{\mbox{\tiny CAL}} (\myeta^*) \}^2 ] $,
$E^{1/2} \{ \mbox{mse}^2 (\hat\pi) \}$,
and $E^{1/2} \{ \mbox{msre}^2 (\hat\pi) \}$, referred to as riskML, riskCAL,
diff, and rdiff respectively.
See Figure~\ref{fig:div-cor} in the Supplementary Material for the results with correctly specified model (\ref{logit-PS}).

The following remarks can be drawn from Figure~\ref{fig:div}.
\begin{itemize}\addtolength{\itemsep}{-.05in}
\item The estimators $\hat\pi_{\mbox{\tiny RML}}$ and $\hat\pi^1_{\mbox{\tiny RCAL}}$ lead to similar mean squared errors of
the excess likelihood risk $\tilde\kappa_{\mbox{\tiny ML}} (\hat\myeta) - \tilde\kappa_{\mbox{\tiny ML}} (\myeta^*)$,
which are in general smaller than the corresponding mean squared errors from the non-penalized estimators $\hat\pi_{\mbox{\tiny ML}}$ and $\hat\pi^1_{\mbox{\tiny CAL}}$.

\item The estimator $\hat\pi^1_{\mbox{\tiny RCAL}}$ consistently yields similar or smaller mean squared errors of
the excess calibration risk $\tilde\kappa_{\mbox{\tiny CAL}} (\hat\myeta) - \tilde\kappa_{\mbox{\tiny CAL}} (\myeta^*)$ than
both non-penalized estimators $\hat\pi_{\mbox{\tiny ML}}$ and $\hat\pi^1_{\mbox{\tiny CAL}}$ and the penalized likelihood estimator $\hat\pi_{\mbox{\tiny RML}}$.

\item The estimator $\hat\pi_{\mbox{\tiny RML}}$ leads to slightly smaller mean squared errors of $\mbox{mse} (\hat\pi)$, a measure of absolute errors,
than $\hat\pi^1_{\mbox{\tiny RCAL}}$. However,
$\hat\pi^1_{\mbox{\tiny RCAL}}$ yields smaller, sometimes substantially smaller, mean squared errors of $\mbox{msre} (\hat\pi)$, a measure of relative errors,
than $\hat\pi_{\mbox{\tiny RML}}$.
\end{itemize}
These results provide empirical support for the rationale of regularized calibrated estimation: minimizing the expected calibration loss, through regularization,
leads to controling relative errors of the fitted propensity scores
as well as reducing the Kullback--Liebler divergence.

In the Supplementary Material, we provide various additional results, including the number of samples with non-convergence for $\hat\gamma_{\mbox{\tiny ML}}$
and $\hat\gamma^1_{\mbox{\tiny CAL}}$, the average numbers of nonzero coefficients obtained in $\hat\gamma_{\mbox{\tiny RML}}$
and $\hat\gamma^1_{\mbox{\tiny RCAL}}$,
and the root mean squared errors of the differences $ \hat\mu^1_{\mbox{\tiny rIPW}}(\hat\pi^1; h) -\hat\mu^0_{\mbox{\tiny rIPW}}(\hat\pi^0; h) $
and $ \hat\mu^1_{\mbox{\tiny rIPW}} (\hat\pi^1; \varepsilon) -\hat\mu^0_{\mbox{\tiny rIPW}} (\hat\pi^0; \varepsilon) $,
which are relevant for estimation of ATE $=\mu^1-\mu^0$.
The estimators $\hat\mu^0_{\mbox{\tiny rIPW}}(\hat\pi^0; h) $ and $\hat\mu^0_{\mbox{\tiny rIPW}}(\hat\pi^0; \varepsilon) $ are defined as
$\hat\mu^1_{\mbox{\tiny rIPW}}(\hat\pi^1; h) $ and $\hat\mu^1_{\mbox{\tiny rIPW}}(\hat\pi^1; \varepsilon) $ respectively,
but with $T$ replaced by $1-T$ and $\hat\pi^1$ by $1-\hat\pi^0$.
The fitted propensity scores $\hat\pi^1$ and $\hat\pi^0$ are the same when  maximum likelihood is used, but
separately computed for calibrated estimation and regularization.
See Section~\ref{sec:discussion} for a discussion of $\hat\pi^0_{\mbox{\tiny RCAL}}$ and ATE estimation.

\section{Application to a medical study} \label{sec:application}

We provide an empirical application of the proposed methods to a medical study in Connors et al.~(1996) on the effects of right heart catheterization (RHC).
The observational study was of interest at the time when many physicians believed that the RHC procedure led to better patient outcomes, but
the benefit of RHC had not been demonstrated in any randomized clinical trials.
The study of Connors et al.~(1996) included $n=5735$ critically ill patients  admitted  to  the  intensive  care  units  of  5  medical  centers.
For each patient, the data consist of treatment status $T$ ($=1$ if RHC was used
within  24  hours  of  admission  and  0  otherwise),  health  outcome $Y$
(survival  time  up  to  30  days),  and  a list of 75 covariates $X$ (including dummy variables from multi-valued factors),
specified by medical specialists in critical care.
For previous analyses using propensity scores, logistic regression was employed
either with main effects only (e.g., Hirano \& Imbens 2002; Vermeulen \& Vansteelandt 2015)
or with interaction terms manually added (Tan 2006) in the approach of Rosenbaum \& Rubin (1984).

To capture possible dependency beyond main effects, we consider a logistic propensity score model (\ref{logit-PS})
with the vector $f(X)$ including all main effects and two-way interactions of $X$ except those with the fractions of nonzero values less than 46 (i.e., 0.8\% of the sample size 5735).
The dimension of $f(X)$ is $p=1855$, excluding the constant. All variables in $f(X)$ are standardized with sample means 0 and sample variances 1.
We apply the methods of regularized maximum likelihood (RML) and regularized calibrated (RCAL) estimation similarly as in the simulation study,
with the Lasso tuning parameter $\lambda$ attempted in a finer set $\{\lambda_0 / 2^{j/4}: j=0,1,\ldots,24\}$,
where $\lambda_0$ is the value leading to a zero solution $\gamma_1=\cdots=\gamma_p=0$.

\begin{figure}[t!]
\caption{\small (i) Standardized differences $\mbox{CAL}^1(\hat \pi; f_j)$ over index $j$ for the estimators $\hat\pi=\tilde E(T)$, $\hat\pi_{\mbox{\tiny RML}}$ and $\hat\pi^1_{\mbox{\tiny RCAL}}$
with $\lambda$ selected from cross validation
(upper row and lower left). A vertical line is placed at the end of indices for 71 main effects.
Two horizontal lines are placed at the maximum absolute standardized differences in two directions.
Marks ($\times$) are plotted at the indices $j$ corresponding to 188 nonzero estimates of $\gamma_j$ for $\hat\pi_{\mbox{\tiny RML}}$
and 32 nonzero estimates of $\gamma_j$ for $\hat\pi^1_{\mbox{\tiny RCAL}}$.
(ii) The fitted propensity scores $\{\hat\pi_{\mbox{\tiny RML}}(X_i), \hat\pi^1_{\mbox{\tiny RCAL}}(X_i)\}$ in the treated sample $\{i:T_i=1,i=1,\ldots,n\}$ (lower right).
} \label{fig:balance} \vspace{.15in}
\begin{tabular}{c}
\includegraphics[width=6.2in, height=4in]{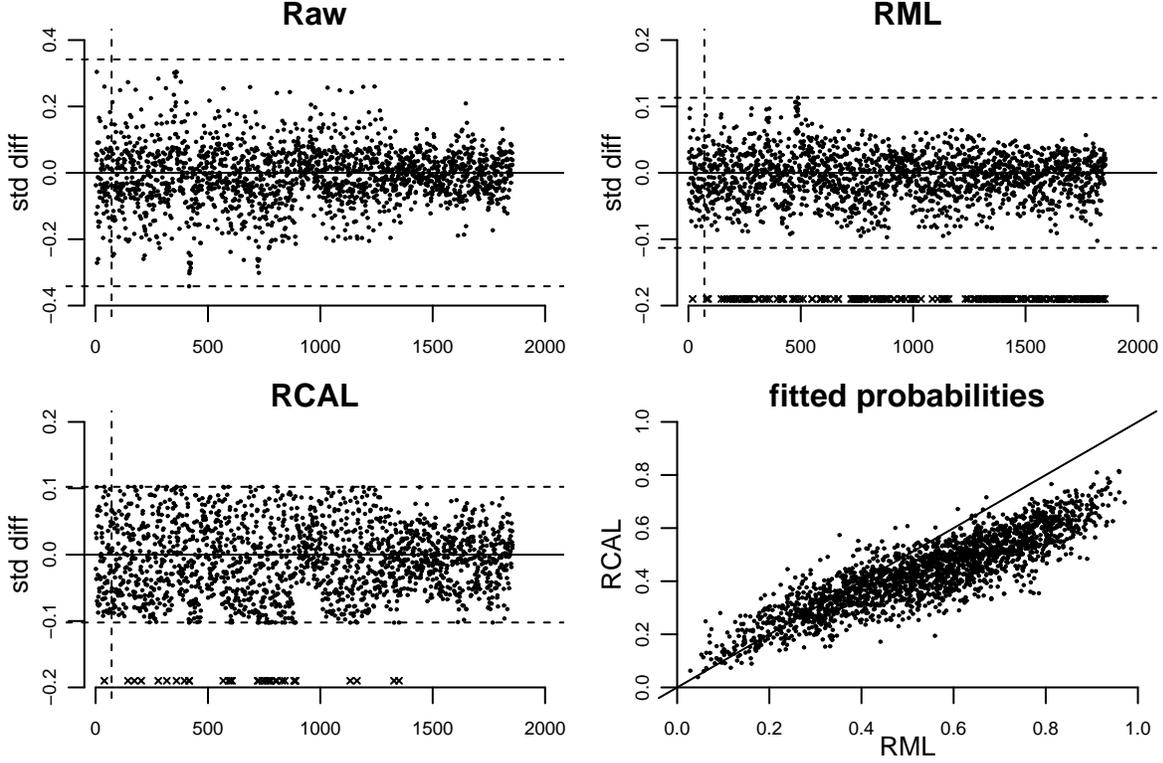} \vspace{-.15in}
\end{tabular}
\end{figure}

To measure the effect of calibration in the treated sample for a function $h(X)$ using a propensity score estimate $\hat\pi$, we use the standardized calibration difference
\begin{align*}
\mbox{CAL}^1(\hat\pi; h) = \frac{\hat\mu^1_{\mbox{\tiny rIPW}}(\hat\pi; h) - \tilde E \{h(X)\}}{ \sqrt{ \widetilde \var \{h(X)\}} },
\end{align*}
where $\tilde E()$ and $\widetilde\var()$ denote the sample sample and variance and $\hat\mu^1_{\mbox{\tiny rIPW}}(\hat\pi; h)$ is defined as
$\hat\mu^1_{\mbox{\tiny rIPW}}(\hat\pi)$ with $Y$ replaced by $h(X)$.
For $f_j(X)$ standardized with sample mean 0 and sample variance 1, $\mbox{CAL}^1(\hat\pi; f_j)$ reduces to $\hat\mu^1_{\mbox{\tiny rIPW}}(\hat\pi; f_j)$.
See for example Austin \& Stuart (2015, Section 4.1.1) for a related statistic based on $\hat\mu^1_{\mbox{\tiny rIPW}}(\hat\pi; h) -\hat\mu^0_{\mbox{\tiny rIPW}}(\hat\pi; h)$
for balance checking.
Figure~\ref{fig:balance} presents the standardized calibration differences for all the variables $f_j(X)$
and the fitted propensity scores in the treated sample, obtained from the regularized estimators $\hat\pi_{\mbox{\tiny RML}}$ and $\hat\pi^1_{\mbox{\tiny RCAL}}$, with the tuning parameter $\lambda$ selected by 5-fold cross validation as in Section~\ref{sec:simulation}.

Several interesting remarks can be drawn from Figure~\ref{fig:balance}. The maximum absolute standardized differences are reduced from $35\%$ to about 10\% ($.113$ and $.102$ respectively) based on
the estimators $\hat\pi_{\mbox{\tiny RML}}$ and $\hat\pi^1_{\mbox{\tiny RCAL}}$.
But the latter estimator $\hat\pi^1_{\mbox{\tiny RCAL}}$ is obtained with a much smaller number ($32$ versus $188$) of nonzero estimates of coefficients $\gamma_j$.
The corresponding standardized differences for these 32 nonzero coefficients precisely attain the maximum absolute value, $.102$,
which is also the tuning parameter $\lambda$ used for the Lasso penalty by Eq.~(\ref{ineq-CAL-2}).
The fitted propensity scores  $\hat\pi^1_{\mbox{\tiny RCAL}}(X_i)$ in the treated are consistently larger (or smaller) than  $\hat\pi_{\mbox{\tiny RML}}(X_i)$ when close to 0 (or 1).
As a result, the inverse probability weights $1/\hat\pi^1_{\mbox{\tiny RCAL}}(X_i)$ tend to be less variable than $1/\hat\pi_{\mbox{\tiny RML}}(X_i)$, which is also confirmed in the following discussion.

\begin{figure}
\caption{\small Maximum absolute standardized differences, $\max_j |\mbox{CAL}^1(\hat \pi; f_j)|$, against the numbers of nonzero estimates of $(\gamma_1,\ldots,\gamma_p)$ (left)
and the relative variances of the inverse probability weights in the treated sample (right) as the tuning parameter $\lambda$ varies for the Lasso penalty. Vertical
lines are placed at the values corresponding to $\lambda$ selected by cross validation.} \label{fig:tune1} \vspace{.15in}
\begin{tabular}{c}
\includegraphics[width=6.2in, height=2in]{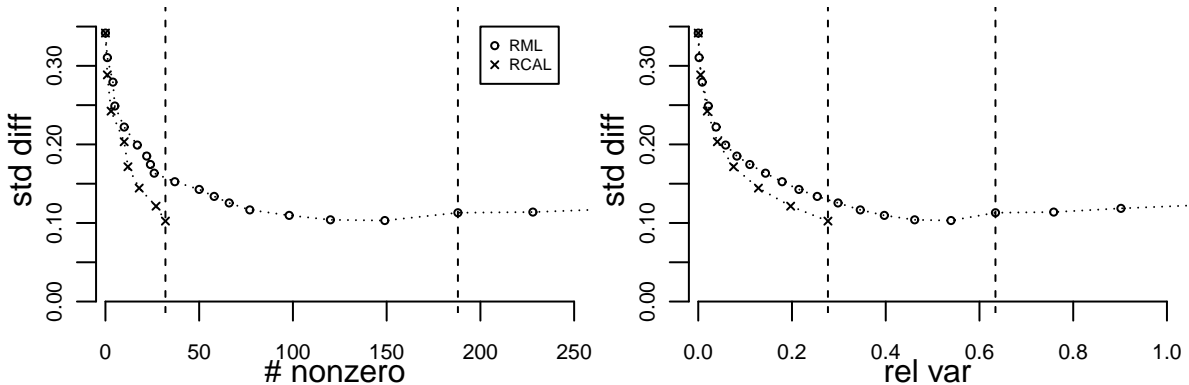} \vspace{-.15in}
\end{tabular}
\end{figure}

Figure~\ref{fig:tune1} shows how the maximum absolute standardized differences are related to the numbers of nonzero estimates of $\gamma_j$
and the relative variances of the inverse probability weights in the treated sample as the tuning parameter $\lambda$ varies.
For a set of weights $\{w_i: T_i=1, i=1,\ldots,n\}$, the relative variance is defined as
$ \sum_{i: T_i=1} (w_i - \bar w)^2 / \{(n_1-1)\bar w^2\}$,
where $\bar w = \sum_{i: T_i=1} w_i/n_1$ and $n_1/n = \tilde E(T)$.
See Liu (2001, Section 2.5.3) for a discussion about use of the relative variance to measure the efficiency of a weighted sample.
As seen from Figure~\ref{fig:tune1}, in the process of reducing the standardized differences, the estimator $\hat\pi^1_{\mbox{\tiny RCAL}}$
is associated with a much smaller number of nonzero coefficients $\gamma_j$ (greater sparsity) and smaller relative variance of the inverse probability weights (greater efficiency)
than $\hat\pi_{\mbox{\tiny RML}}$. These results demonstrate advantages of regularized calibrated estimation in high-dimensional settings.

In the Supplementary Material, we present additional results, including the results in parallel to Figure~\ref{fig:tune1} for the fitted propensity score $\hat\pi^0_{\mbox{\tiny RCAL}}$
in the untreated sample, and the estimates of $\mu^1$, $\mu^0$, and the ATE for the 30-day survival (i.e., $Y \ge 30$).

\vspace{-.1in}
\section{Additional discussion} \label{sec:discussion}

\vspace{-.05in}
\noindent{\bf Dual formulation.} We point out that the regularized calibrated estimator $\hat\gamma^1_{\mbox{\tiny RCAL}}$
 can also be derived in a dual formulation.
Denote $w = \{w_i >1: T_i=1, i=1,\ldots, n\}$. For fixed $\lambda \ge 0$, consider the following optimization problem:
\begin{align}
& \mbox{minimize} \quad \ell^D_{\mbox{\tiny CAL}}( w) = \sum_{1\le i \le n:\, T_i=1} \left\{ (w_i-1) \log (w_i-1) - (w_i-1) \right\} \label{dual-obj} \\
& \mbox{subjec to} \sum_{1\le i \le n:\, T_i=1} w_i = n, \label{dual-constr1} \\
& \hspace{.6in} \left| \sum_{1\le i \le n:\, T_i=1} w_i f_j(X_i) - \sum_{i=1}^n f_j(X_i) \right| \le \lambda, \quad j=1,\ldots,p. \label{dual-constr2}
\end{align}
It can be shown directly via the Karush-–Kuhn–-Tucker condition that if $\hat\gamma^1_{\mbox{\tiny RCAL}}$ minimizes the penalized loss
$\ell_{\mbox{\tiny RCAL}}(\gamma)$ in (\ref{reg-cal-loss}), then the inverse probability weights
$$
 \hat w_i = \{\hat \pi^1_{\mbox{\tiny RCAL}}(X_i)\}^{-1} = 1 + \me^{- (\hat\gamma^1_{\mbox{\tiny RCAL}})^\T f(X_i)}, \quad 1\le i\le n \mbox{ with } T_i=1,
$$
are a solution to the optimization problem (\ref{dual-obj})--(\ref{dual-constr2}).
In the case of exact calibration ($\lambda=0$), the program (\ref{dual-obj})--(\ref{dual-constr2}) can be
obtained from Chan et al.~(2016) with the particular distance measure $\ell^D_{\mbox{\tiny CAL}}( w)$. See also Zubizarreta (2015) for a related method.
Similarly, for the regularized likelihood estimator $\hat\gamma_{\mbox{\tiny RML}}$ minimizing $\ell_{\mbox{\tiny RML}}(\gamma)$ in (\ref{reg-ml-loss}),
it can also be shown that the fitted propensity scores, $\hat \pi_i = \hat \pi_{\mbox{\tiny RML}}(X_i)$ for  $i=1,\ldots,n$, solve the following optimization problem
with $\pi = \{0< \pi_i <1: i=1,\ldots,n\}$:
\begin{align}
& \mbox{minimize} \quad \ell^D_{\mbox{\tiny ML}}(\pi) = \sum_{i=1}^n \left\{ (1-\pi_i) \log (1-\pi_i) + \pi_i \log(\pi_i) \right\} \label{dual2-obj} \\
& \mbox{subjec to} \sum_{i=1}^n (T_i- \pi_i)=0, \label{dual2-constr1} \\
& \hspace{.6in} \left| \sum_{i=1}^n (T_i - \pi_i) f_j(X_i) \right| \le \lambda, \quad j=1,\ldots,p. \label{dual2-constr2}
\end{align}
See Dudik et al.~(2007) for general results relating box constraints such as (\ref{dual-constr1})--(\ref{dual-constr2}) and (\ref{dual2-constr1})--(\ref{dual2-constr2}) to Lasso penalty
in a different context.
These formulations shed light on differences between maximum likelihood and calibration estimation, which deal with propensity scores in the probability scale
or, respectively, the scale of inverse probability weights.

We distinguish two types of calibration estimators that can be derived using unit-specific weights in a dual formulation, which usually involves exact constraints in previous works.
The first type is survey calibration (Deville \& Sarndal 1992),
where calibration weights are constructed by minimizing a distance measure to the design weights (i.e., inverse of inclusion probabilities) subject to calibration equations.
Similar ideas are used in Tan (2010, 2013) to derive improved doubly robust estimators, through adjusting inverse of fitted propensity scores to achieve calibration constraints, possibly
depending on a fitted outcome regression function.
The second type of calibrated estimators, such as $\hat\gamma^1_{\mbox{\tiny CAL}}$ for $\gamma$ or $\hat\mu^1(\hat\pi^1_{\mbox{\tiny CAL}})$ for $\mu^1$,
are typically derived to deal with non-response or missing data, in a similar manner as the first type with uniform design weights.
But there is a subtle difference. The survey calibration weights (Deville \& Sarndal 1992)
are expected to deviate from design weights by $O_p( n^{-1/2})$ and mainly used to reduce asymptotic variances of the resulting estimators of population quantities.
The calibration weights of the second type can be viewed as the inverse of fitted response probabilities or propensity scores from an implied model (by the choice
of a distance measure)
and are expected to behave as $O_p(1)$ to achieve bias reduction.

\vspace{.1in}
\noindent{\bf Estimation of ATE.} Our theory and methods are presented mainly on estimation of $\mu^1$,
but they can be directly extended for estimating $\mu^0$ and hence ATE, that is, $\mu^1-\mu^0$.
As mentioned in Section~\ref{sec:overview}, for IPW estimation of $\mu^0$ with model (\ref{logit-PS}),
the calibrated estimator of $\gamma$, denoted by $\hat\gamma^0_{\mbox{\tiny CAL}}$,
is defined as a solution to equation (\ref{eq-CAL-0}).
By exchanging $T$ with $1-T$ and $\gamma$ with $-\gamma$ in (\ref{loss-CAL}), the corresponding loss function minimized by $\hat\gamma^0_{\mbox{\tiny CAL}}$
is
\begin{align*}
\ell^0_{\mbox{\tiny CAL}} (\gamma) &= \tilde E \left\{ (1-T) \me^{\gamma^\T f(X)} - T \gamma^\T f(X) \right\}.
\end{align*}
For fixed $\lambda \ge 0$, the regularized calibrated estimator $\hat\gamma^0_{\mbox{\tiny RCAL}}$ is defined as a minimizer of
\begin{align*}
\ell^0_{\mbox{\tiny RCAL}} (\gamma) &=  \ell^0_{\mbox{\tiny CAL}} (\gamma) + \lambda \|\gamma_{1:p} \|_1 .
\end{align*}
The fitted propensity score, $\hat\pi^0_{\mbox{\tiny RCAL}}(X) = \pi(X; \hat\gamma^0_{\mbox{\tiny RCAL}})$, then satisfies equations (\ref{ineq-CAL-1})--(\ref{ineq-CAL-2}) with
$T_i$ replaced by $1-T_i$ and $\hat\pi^1_{\mbox{\tiny RCAL}}(X_i)$ replaced by $1-\hat\pi^0_{\mbox{\tiny RCAL}}(X_i)$.
The resulting IPW estimator of $\mu^0$ is $\hat\mu^0_{\mbox{\tiny IPW}} ( \hat \pi^0_{\mbox{\tiny RCAL}})=\hat\mu^0_{\mbox{\tiny rIPW}} ( \hat \pi^0_{\mbox{\tiny RCAL}})$,
and that of ATE is $\hat\mu^1_{\mbox{\tiny IPW}} ( \hat \pi^1_{\mbox{\tiny RCAL}}) -\hat\mu^0_{\mbox{\tiny IPW}} ( \hat \pi^0_{\mbox{\tiny RCAL}})$.

An interesting aspect of our approach is that two different estimators of the propensity score are used when
estimating $\mu^0$ and $\mu^1$.
The estimators  $\hat\gamma^0_{\mbox{\tiny RCAL}}$ and $\hat\gamma^1_{\mbox{\tiny RCAL}}$
may in general have different asymptotic limits when the propensity score model (\ref{logit-PS}) is misspecified,
even though their asymptotic limits coincide when model(\ref{logit-PS}) is correctly specified. Such
possible differences should not be of concern:
the IPW estimators $\hat\mu^0_{\mbox{\tiny IPW}} ( \hat \pi)$ and
$\hat\mu^1_{\mbox{\tiny IPW}} ( \hat \pi)$ are decoupled, involving two disjoint subsets of fitted propensity scores on
the untreated $\{i:T_i=0\}$ and the treated $\{i: T_i=1\}$ respectively.
It seems reasonable, especially in the case of potential model misspecification, to estimate propensity scores and construct inverse
probability weights separately for the treated or the untreated. Furthermore,
whether substantial differences exist between these separately fitted propensity scores
can be used for diagnosis of the validity of model (\ref{logit-PS}). See Chan et al.~(2016), Section 2.3, for a related discussion.

\vspace{.1in}
\noindent{\bf Calibration or balancing.} It is interesting to compare calibrated propensity scores
with covariate balancing propensity scores in Imai \& Ratkovic (2014). For model (\ref{model-PS}),
the covariate-balancing estimator of $\gamma$, denoted by $\hat\gamma_{\mbox{\tiny BAL}}$, is defined as a solution to
\begin{align}
\tilde E \left[ \left\{ \frac{T}{\pi(X;\gamma)} - \frac{1-T}{1-\pi(X;\gamma)} \right\} f(X) \right] = 0 . \label{eq-BAL}
\end{align}
The same fitted propensity score $\hat\pi_{\mbox{\tiny BAL}}(X) =\pi(X; \hat\gamma_{\mbox{\tiny BAL}})$ can used in
$\hat\mu^1_{\mbox{\tiny IPW}}( \hat\pi_{\mbox{\tiny BAL}})$ or alternatively $\hat \mu^1_{\mbox{\tiny rIPW}}(\hat\pi_{\mbox{\tiny BAL}})$ for estimating $\mu^1$
and in $\hat\mu^0_{\mbox{\tiny IPW}}( \hat\pi_{\mbox{\tiny BAL}})$ or $\hat \mu^0_{\mbox{\tiny rIPW}}(\hat\pi_{\mbox{\tiny BAL}})$ for estimating $\mu^0$.
Eq.~(\ref{eq-BAL}) amounts to finding a single value $\hat\gamma_{\mbox{\tiny BAL}}$ such that the left hand sides of (\ref{eq-CAL}) and (\ref{eq-CAL-0})
are equal, although they may each deviate from 0.
For calibrated estimation, (\ref{eq-CAL}) and (\ref{eq-CAL-0}) are satisfied separately by two estimators  $\hat\gamma^1_{\mbox{\tiny CAL}}$ and $\hat\gamma^0_{\mbox{\tiny CAL}}$.
An advantage of using the calibration equations (\ref{eq-CAL}) and (\ref{eq-CAL-0}) is that for each $t=0$ or 1,
$\hat\mu^t_{\mbox{\tiny IPW}}( \hat\pi^t_{\mbox{\tiny CAL}})$, but not $\hat\mu^t_{\mbox{\tiny IPW}}( \hat\pi_{\mbox{\tiny BAL}})$,
is doubly robust, i.e., remains consistent if either propensity score model (\ref{model-PS}) or a linear outcome model is correct,
$E(Y^t |X) = \alpha^\T_t f(X)$ for a coefficient vector $\alpha_t$ (Graham et al.~2012).
We also point out that with logistic model (\ref{logit-PS}), $\hat\gamma_{\mbox{\tiny BAL}}$ can be obtained by minimizing the loss function
\begin{align}
\ell_{\mbox{\tiny BAL}}(\gamma) = \ell_{\mbox{\tiny CAL}}(\gamma) + \ell^0_{\mbox{\tiny CAL}}(\gamma), \label{loss-BAL}
\end{align}
which is still convex in $\gamma$. Our results developed for calibrated estimation and regularization
can be adapted to $\hat\gamma_{\mbox{\tiny BAL}}$ and its regularized version.
See Figure~\ref{fig:normal2} for a comparison of limiting propensity scores in a simple example with model misspecification.

\vspace{.1in}
\noindent{\bf Estimation of ATT.} There is a simple extension of our approach to estimation of ATT,
that is, $\nu^1-\nu^0$ as defined in Section~\ref{sec:background}.
The parameter $\nu^1 = E(Y^1 | T=1)$ can be directly estimated by $\tilde E (TY) / \tilde E(T)$.
Two standard IPW estimators for $\nu^0$ are
\begin{align*}
\hat \nu^0_{\mbox{\tiny IPW}} (\hat\pi_{\mbox{\tiny ML}}) = \tilde E\left\{ \frac{(1-T)\hat\pi_{\mbox{\tiny ML}}(X) Y}{1- \hat\pi_{\mbox{\tiny ML}}(X)} \right\}/\tilde E (T)
\end{align*}
and $\hat \nu^0_{\mbox{\tiny rIPW}} (\hat\pi_{\mbox{\tiny ML}})$, defined as $\hat \nu^0_{\mbox{\tiny IPW}} (\hat\pi_{\mbox{\tiny ML}})$
but with $\tilde E(T)$ replaced by
$\tilde E [(1-T)\hat\pi_{\mbox{\tiny ML}}(X) / \{1- \hat\pi_{\mbox{\tiny ML}}(X)\}]$,
where $\hat\pi_{\mbox{\tiny ML}}(X)$ is the maximum likelihood fitted propensity score.
To derive a calibrated estimator of $\gamma$, consider the following set of calibration equations
\begin{align}
\tilde E \left[ \left\{\frac{(1-T)\pi(X;\gamma)}{1-\pi(X;\gamma)}-T \right\} f(X)\right] = 0 . \label{eq-CAL-0-ATT}
\end{align}
Equation (\ref{eq-CAL-0-ATT}) is used in Imai \& Ratkovic (2014) as balancing equations for propensity score estimation for estimating ATT.
We point out two simple results, which, although as straightforward as shown below, do not seem to be discussed before.
\begin{itemize}\addtolength{\itemsep}{-.05in}
\item[(i)] Equation (\ref{eq-CAL-0-ATT}) is equivalent to calibration equations (\ref{eq-CAL-0}) for $\hat\gamma^0_{\mbox{\tiny CAL}}$ when estimating $\mu^0$.
This follows from the simple identity:
$$
\frac{(1-T)\pi(X;\gamma)}{1-\pi(X;\gamma)}- T = \frac{1-T}{1-\pi(X;\gamma)}-1 .
$$
Therefore, the same set of fitted propensity scores, for example $\hat\pi^0_{\mbox{\tiny RCAL}}(X_i)$ based on the regularized estimator $\hat\gamma^0_{\mbox{\tiny RCAL}}$, can be used
for estimating $\mu^0$  by $\hat\mu^0_{\mbox{\tiny IPW}} ( \hat \pi^0_{\mbox{\tiny RCAL}})$ and
for estimating $\nu^0$ by  $\hat\nu^0_{\mbox{\tiny IPW}} ( \hat \pi^0_{\mbox{\tiny RCAL}})
=\hat\nu^0_{\mbox{\tiny rIPW}} ( \hat \pi^0_{\mbox{\tiny RCAL}})$ due to similar equation as (\ref{ineq-CAL-1}).

\item[(ii)] With logistic model (\ref{logit-PS}), the IPW estimator $\hat\nu^0_{\mbox{\tiny IPW}} ( \hat \pi^0_{\mbox{\tiny CAL}})
=\hat\nu^0_{\mbox{\tiny rIPW}} ( \hat \pi^0_{\mbox{\tiny CAL}})$ is identical to the estimator of $\nu^0$ by entropy balancing (Hainmueller 2012).
In fact, $\hat\nu^0_{\mbox{\tiny rIPW}} ( \hat \pi^0_{\mbox{\tiny CAL}})$ can be written as
$\sum_{i: T_i=0} \hat w_i Y_i$, where for $\hat\gamma=\hat\gamma^0_{\mbox{\tiny CAL}}$, $\hat\gamma_{1:p} = (\hat\gamma_1,\ldots,\hat\gamma_p)^\T$, and $f_{1:p}=(f_1,\ldots,f_p)^\T$,
$$
\hat w_i
=\frac{\exp\{-\hat\gamma_{1:p}^\T f_{1:p}(X_i)\}}{\sum_{i^\prime: T_{i^\prime}=0} \exp\{-\hat\gamma_{1:p}^\T f_{1:p}(X_{i^\prime})\} } .
$$
Equation (\ref{eq-CAL-0-ATT}) for $\hat \gamma^0_{\mbox{\tiny CAL}}$ then implies that
for $j=1,\ldots,p$,
$$
\sum_{i: T_i=0} \hat w_i f_j(X_i) = \frac{\tilde E \left\{\frac{(1-T)\pi(X;\hat\gamma)}{1-\pi(X;\gamma)}f_j(X)\right\}} { \tilde E \left\{ \frac{(1-T)\pi(X;\hat\gamma)}{1-\pi(X;\hat\gamma)}\right\}}
= \frac{\tilde E \left\{Tf_j(X)\right\}} { \tilde E (T)}
=\frac{1}{n_1} \sum_{i: T_i=1} f_j(X_i) ,
$$
where $n_1/n=\tilde E(T)$. These equations together with $\sum_{i: T_i=0} \hat w_i=1$ are the same as in entropy balancing.
From this connection, our regularized method also extends entropy balancing to allow box constraints similar to (\ref{ineq-CAL-2}).
\end{itemize}

\vspace{.1in}
\noindent{\bf Doubly robust estimation.}
Our development is mainly focused estimation of propensity scores to improve IPW estimation of population means with missing data.
The new methods for propensity score estimation can be
adapted in various manners to explicitly achieve double robustness.
As mentioned earlier, the estimator $\hat\mu^1 (\hat\pi^1_{\mbox{\tiny CAL}})$ itself is known to be doubly robust with respect to propensity score model (\ref{logit-PS})
and a linear outcome model, $E(Y^1 |X) = \alpha_1^\T f(X)$ for a coefficient vector $\alpha_1$ (Graham et al.~2012).
For a general outcome model, let $\hat m_1(X)$ be a fitted outcome regression function, by maximum quasi-likelihood or similar methods.
One approach is to directly use the calibrated propensity scores $\hat\pi^1_{\mbox{\tiny CAL}}$ or $\hat\pi^1_{\mbox{\tiny RCAL}}$,
and the fitted outcome regression function $\hat m_1(X)$ as inputs to existing doubly robust
estimators, for example, the augmented IPW estimator (Robins et al.~1994)
or the calibrated likelihood estimator (Tan 2010).
Another approach is to incorporate the fitted
outcome regression function $\hat m_1(X)$ in $f(X)$ and then enforce the corresponding calibration equation to exactly
hold, that is, redefine $f=(1, \hat m_1, f_1,\ldots, f_p)^\T$ and $\gamma= (\gamma_{00},\gamma_{01},\gamma_1,\ldots,\gamma_p)^\T$
and leave $(\gamma_{00},\gamma_{01})$ non-penalized in our regularized method. This topic can be investigated in future research.

\vspace{.1in}
\noindent{\bf Related works on high-dimensional causal inference.}
There is a growing literature on causal inference in high-dimensional settings.
For ATE estimation, Farrell (2015) and Belloni et al.~(2017) studied the augmented IPW estimator, with both the propensity score and the outcome regression function estimated using Lasso or related methods.
Their focus is to obtain valid confidence intervals when both the propensity score and outcome regression models
are correctly specified, but approximately sparse.
For ATT estimation, Athey et al.~(2016) studied a hybrid method combining
penalized estimation of a linear outcome model and construction of balancing weights similar as in Zubizarreta (2015),
and also obtained valid confidence intervals when the linear outcome model is correctly specified.
These works differ from our development focused on IPW estimation with possibly misspecified propensity score models, but without any outcome regression model,
in high-dimensional settings. On the other hand, results from these works can be useful
in developing confidence intervals for our approach.

\vspace{.25in}
\centerline{\bf\Large References}

\begin{description}\addtolength{\itemsep}{-.1in}

\item Athey, S., Imbens, G.W., and Wager, S. (2016) ``Approximate residual balancing: De-biased inference of average treatment effects in high dimensions," arXiv:1604.07125.

\item Austin, P.C. and Stuart, E.A. (2015) ``Moving towards best practice when using
inverse probability of treatment
weighting (IPTW) using the propensity
score to estimate causal treatment effects
in observational studies," {\em Statistics in Medicine}, 34, 3661--3679.

\item Belloni, A., Chernozhukov, V., Fernandez-Val, I., and Hansen, C. (2017) "Program evaluation and causal inference
with high-dimensional data," {\em Econometrica}, 85, 233--298.

\item Bohning, D. and Lindsay, B.G. (1988) ``Monotonicity of
quadratic approximation algorithms," {\em  Annals of the Institute of Statistical Mathematics}, 40, 641--663.

\item Bregman, L.M. (1967) ``The relaxation method of finding the common point of convex sets and its application
to the solution of problems in convex programming," {\em USSR Computational Mathematics and Mathematical Physics}, 7, 200–-217.

\item Buhlmann, P. and van de Geer, S. (2011) {\em Statistics for High-Dimensional Data: Methods, Theory and Applications}, New York: Springer.

\item Chan, K.C.G., Yam, S.C.P., and Zhang, Z. (2016)
``Globally efficient non-parametric inference of average treatment effects by empirical balancing
calibration weighting," {\em Journal of the Royal Statistical Society}, Ser. B,  78, 673--700.

\item Connors, A.F., Speroff, T., Dawson, N.V., et al. (1996) ``The effectiveness
of right heart catheterization in the initial care of critically ill patients,"
{\em Journal of the American Medical Association}, 276, 889--897.

\item Deville, J.C. and Sarndal, C.E. (1992) ``Calibration estimators in survey sampling," {\em Journal of the American
Statistical Association}, 87, 376--382.

\item Dudik, M., Phillips, S.J., and Schapire, R.E. (2007) ``Maximum entropy density estimation with generalized regularization
and an application to species distribution modeling," {\em Journal of Machine Learning Research}, 8, 1217--1260.

\item Farrell, M.H. (2015) ``Robust inference on average treatment effects with possibly more covariates
than observations." {\em Journal of Econometrics}, 189, 1--23.

\item Folsom, R.E. (1991) ``Exponential and logistic weight adjustments for sampling and nonresponse error reduction,"
{\em Proceedings of the American Statistical Association}, Social Statistics Section, 197--202.


\item Friedman, J., Hastie, T., and Tibshirani, R. (2010) ``Regularization paths for generalized linear models via coordinate descent,"
{\em Journal of Statistical Software}, 33, 1--22.

\item Geyer, C.J. (2014) ``Trust: Trust Region Optimization," R package 0.1-6, https://cran.r-project.org/web/packages/trust/index.html.

\item Graham, B.S., de Xavier Pinto, C.C., and Egel, D. (2012) ``Inverse probability
tilting for moment condition models with missing data," {\em Review of Economic Studies}, 79, 1053--1079.

\item Graham, B.S., de Xavier Pinto, C.C., and Egel, D. (2016) ``Efficient estimation
of data combination models by the method of auxiliary-to-study tilting (AST),"
{\em Journal of Business and Economic Statistics}, 34, 288--301

\item Hainmueller, J. (2012) ``Entropy balancing for causal effects: Multivariate reweighting method
to produce balanced samples in observational studies," {\em Political Analysis}, 20, 25--46.

\item Hirano, K., and Imbens,  G.W. (2002) ``Estimation of causal effects using
propensity score weighting: An application to data on right heart catheterization,"
{\em Health Services and Outcomes Research Methodology}, 2, 259--278.


\item Huang, J. and Zhang, C.-H. (2012) ``Estimation and selection via absolute penalized convex minimization
and its multistage adaptive applications," {\em Journal of Machine Learning Research}, 13, 1839--1864.

\item Imai, K. and Ratkovic, M. (2014) ``Covariate balancing propensity score," {\em Journal of the Royal
Statistical Society}, Ser. B, 76, 243--263.

\item Imbens, G.W. (2004) ``Nonparametric estimation of average treatment effects
under exogeneity: A review," {\em Review of Economics and Statistics}, 86, 4--29.

\item Kang, J.D.Y. and Schafer, J.L. (2007) ``Demystifying double robustness: A comparison of alternative strategies for estimating a population mean from incomplete data" (with discussion), {\em Statistical Science}, 523--539.

\item Kim, J.K. and Haziza, D. (2014) ``Doubly robust inference with missing data in survey sampling," {\em Statistica Sinica}, 24, 375--394.

\item Liu, J. S. (2001) {\em Monte Carlo Strategies in Scientific Computing}, New York: Springer.

\item Manski, C.F. (1988) {\em Analog Estimation Methods in Econometrics}, New York:
Chapman \& Hall.

\item McCullagh, P. and Nelder, J. (1989) {\em Generalized Linear Models} (2nd edition), New York: Chapman \& Hall.

\item Negahban, S.N., Ravikumar, P., Wainwright, M.J., and Yu, B. (2012) ``A unified
framework for high-dimensional analysis of M-estimators with decomposable
regularizers," {\em Statistical Science}, 27, 538--557.

\item Neyman, J. (1923) ``On the application of probability theory to agricultural
experiments: Essay on principles, Section 9," translated in {\em Statistical Science}, 1990, 5, 465--480.

\item Osborne, M., Presnell, B., and Turlach, B. (2000) ``A new approach to variable selection in least
squares problems." {\em IMA Journal of Numerical Analysis}, 20, 389--404.

\item Robins, J.M., Rotnitzky, A., and Zhao, L.P. (1994) ``Estimation of regression
coefficients when some regressors are not always observed," {\em Journal of the American
Statistical Association}, 89, 846--866.

\item Rosenbaum, P.R. and Rubin, D.B. (1983) ``The central role of the propensity score in observational studies for causal
effects," {\em Biometrika}, 70, 41-–55.

\item Rosenbaum, P.R. and Rubin, D.B. (1984) ``Reducing bias in observational studies using subclassification on the
propensity score," {\em Journal of the American Statistical Association}, 79, 516--524.

\item Rubin, D.B. (1976) ``Inference and missing data," {\em Biometrika}, 63, 581--590.

\item Tan, Z. (2006) ``A distributional approach for causal inference using propensity scores,"
{\em Journal of the American Statistical Association}, 101, 1619–-1637.

\item Tan, Z. (2010) ``Bounded, efficient, and doubly robust estimation with inverse weighting," {\em Biometrika}, 97, 661--682.

\item Tan, Z. (2011) ``Efficient restricted estimators for conditional mean models with missing data," {\em Biometrika}, 98, 663--684.

\item Tan, Z. (2013) ``Simple design-efficient calibration estimators for rejective and high-entropy sampling," {\em Biometrika}, 100, 399--415.

\item Tibshirani, R. (1996) ``Regression shrinkage and selection via the Lasso," {\em Journal of the Royal Statistical Society}, Ser. B, 58, 267--288.

\item Tsiatis, A.A. (2006) {\em Semiparametric Theory and Missing Data}, New York: Springer.

\item van de Geer, S. (2008) ``High-dimensional generalized linear models and the Lasso." {\em Annals of Statistics}, 36, 614--645.

\item van der Laan, M.J. and Robins, J.M. (2003) {\em Unified Methods for Censored Longitudinal Data and Causality}, New York: Springer.

\item Vermeulen. K. and Vansteelandt, S. (2015) ``Bias-reduced doubly
robust estimation," {\em Journal of the American Statistical Association}, 110, 1024--1036.

\item White, H. (1982) ``Maximum Likelihood Estimation of Misspecified Models," {\em Econometrica}, 50, 1--25.

\item Wu, T.T. and Lange, K. (2010) ``The MM alternative to EM," {\em Statistical Science}, 25, 492--505.

\item Zhang, C.-H. and Zhang, T. (2012) ``A general theory of concave regularization for high-dimensional sparse estimation problems,"
{\em Statistical Science}, 27, 576--593.

\item Zubizarreta, J.R. (2015)  ``Stable weights that balance covariates for estimation with incomplete outcome
data," {\em Journal of the American Statistical Association}, 110, 910--922.

\end{description}


\clearpage

\setcounter{page}{1}

\setcounter{section}{0}
\setcounter{equation}{0}

\setcounter{figure}{0}

\setcounter{pro}{0}
\renewcommand{\thepro}{S\arabic{pro}}

\setcounter{cor}{0}
\renewcommand{\thecor}{S\arabic{cor}}

\renewcommand{\theequation}{S\arabic{equation}}
\renewcommand{\thesection}{\Roman{section}}

\renewcommand\thefigure{S\arabic{figure}}
\renewcommand\thetable{S\arabic{table}}

\begin{center}
{\Large Supplementary Material for ``Regularized calibrated estimation of propensity scores with model misspecification and high-dimensional data"}

\vspace{.1in} Zhiqiang Tan
\end{center}
\vspace{.1in}

The Supplementary Material contains Appendices I--V.

\section{On Lasso penalized M-estimation}

Let $\{ (T_i,X_i) : i=1,\ldots,n\}$ be independent and identically distributed observations of $(T,X)$.
Suppose that a function of interest, $\myeta^*(x)$, is approximated as $\gamma^\T f(x)$, where
$f(x)=(1,f_1,\ldots,$ $f_p)^\T$ is a vector of known functions and
$\gamma=(\gamma_0,\gamma_1,\ldots,\gamma_p)^\T$ is a vector of unknown coefficients.
In general, $\myeta^*(x)$ may fall outside the linear subspace $\{ \gamma^\T f(x): \gamma\in \mathbb R^{1+p} \}$.
Let $\kappa(\myeta)$ be a loss function, defined in the form
\begin{align*}
\kappa( \myeta) = \tilde E \left[ \psi \{ T, \myeta(X)\} \right]
\end{align*}
for some function $\psi(t,u)$, assumed to be convex and twice-differentiable in $u$.
Denote $\psi_1(t,u) = \partial \psi(t,u)/\partial u$ and $\psi_2(t,u) = \partial^2 \psi(t,u)/\partial u^2$.
Let $\ell(\gamma) = \kappa( \gamma^\T f)$ be the loss function induced on $\gamma$.
Then $\kappa(g)$ is convex in $g$, and $\ell(\gamma)$ is convex in $\gamma$.

Consider a regularized estimator
\begin{align*}
\hat \gamma = \argmin_\gamma \left\{ \ell(\gamma ) + \lambda R(\gamma) \right\},
\end{align*}
where $R(\gamma)= \| \gamma_{1:p} \|_1 = \sum_{j=1}^p |\gamma_j|$, that is, a Lasso penalty on $\gamma$ except $\gamma_0$,
and $\lambda \ge 0$ is a tuning parameter. The resulting estimator of $\myeta^*$ is
then $\hat\myeta = \hat\gamma^\T f$.

The target linear approximation of $\myeta^*$ is defined as $\bar \myeta= \bar\gamma^\T f$, where $\bar \gamma$ is a minimizer of the theoretical loss $E [ \psi\{T, \gamma^\T f(X)\}]$, which is also convex in $\gamma$.
Setting the gradient of the theoretical loss to 0 shows that $\bar \gamma$ satisfies $ E [ \psi_1\{T, \bar \gamma^\T f(X)\} f(X)] = 0$ under mild conditions.
Asymptotic theory has been established about convergence of $\hat\gamma$ (typically non-penalized) to $\bar \gamma$ at the $n^{-1/2}$ rate
in the classical setting where $p$ is much smaller than the sample size $n$ (e.g., White 1982; Manski 1988).
It is desired to develop corresponding theory in the high-dimensional setting where $p$
can be close to or greater than the sample size $n$.

For our theoretical analysis, the tuning parameter is specified as $\lambda = A_0 \lambda_0$, where $A_0>1$ is a constant and
\begin{align*}
\lambda_0 = \max \left\{ \sqrt{8(D_0^2+D_1^2)}, 4 C_0^2 C_1 \right\} \sqrt{ \log\{(1+p)/\epsilon\}/n},
\end{align*}
depending on a tail probability $0 <\epsilon<1$ for the error bound
and the constants $(D_0,D_1)$ from the sub-gaussian Assumption~\ref{ass-mean0} and $(C_0,C_1)$ from Assumptions~\ref{ass-covariate} and \ref{ass-hessian2}
on the boundedness of $f(X)$ and $\psi_2\{T,\bar \myeta(X)\}$, to be discussed below.

Our analysis involves a number of assumptions. The first is a sub-gaussian condition on the ``score" variables
defined as $Z_j =\psi_1 \{T, \bar \myeta(X)\} f_j(X)$. In the presence of model misspecification,
the ``noise" variable $\psi_1 \{T, \bar \myeta(X)\}$ may not have mean 0 conditionally on $X$. Nevertheless,
Assumption~\ref{ass-mean0} is easily shown to hold if $\psi_1 \{T, \bar \myeta(X)\}$ is sub-gaussian and $f_j(X)$, $j=0,1,\ldots,p$, are uniformly bounded (that is, Assumption~\ref{ass-covariate} below).

\begin{ass} \label{ass-mean0}
Let $Z_j =\psi_1 \{T, \bar \myeta(X)\} f_j(X)$. Assume that $E(Z_j)=0$ for $j=0,1,\ldots,p$, and $(Z_0,Z_1,\ldots,Z_p)$ are uniformly sub-gaussian: $\max_{i=1,\ldots,p} D_0^2  E \{\exp(Z_j^2/D_0^2) - 1 \} \le D_1^2$
for some constants $(D_0,D_1)$.
\end{ass}

The second assumption is a theoretical compatibility condition. Similar conditions are commonly used in high-dimensional analysis (Buhlmann \& van de Geer 2011).
Our assumption is formulated with a subset $S$ required to contain 0, as a way to deal with the fact that $\gamma_0$ is not penalized.
Denote the Hessian of $\ell(\gamma) = \kappa(\gamma^\T f)$ as
$$
\tilde \Sigma_\gamma = \tilde E [ f(X) \psi_2 \{T, \gamma^\T f (X)\} f^\T(X) ] .
$$
The corresponding population matrix is
$$
\Sigma_\gamma = E [ f(X) \psi_2 \{T, \gamma^\T f (X)\} f^\T(X) ] .
$$

\begin{ass} \label{ass-compat}
For certain subset $S \subset \{0,1,\ldots,p\}$ containing 0 and constants $\nu_0 >0$ and $\xi_0>1$, assume that
\begin{align}
\nu_0^2  \left(\sum_{j\in S} |b_j| \right)^2 \le |S| \left( b^\T \Sigma_{\bar\gamma} b  \right) \label{ass-compat-eq1}
\end{align}
for any vector $b=(b_0,b_1,\ldots,b_p)^\T \in \mathbb R^{1+p} $ satisfying
\begin{align}
\sum_{j\not\in S} |b_j| \le \xi_0 \sum_{j\in S} |b_j| . \label{ass-compat-eq2}
\end{align}
\end{ass}

By the Cauchy--Schwartz inequality, Assumption~\ref{ass-compat} is implied by (hence weaker than) a restricted eigenvalue condition (Bickel et al.~2009) such that
$\nu_0^2 (\sum_{j\in S} b_j^2) \le b^\T \Sigma_{\bar\gamma} b$
for any vector $b=(b_0,b_1,\ldots,b_p)^\T $ satisfying (\ref{ass-compat-eq2}).
Denote the population Gram matrix as $\Sigma^0 = E \{ f(X) f^\T(X) \}$.
Assumption~\ref{ass-compat} can also be justified from a simpler compatibility condition (e.g., Buhlmann \& van de Geer 2011):
$\nu_0^2 (\sum_{j\in S} b_j^2) \le b^\T \Sigma^0 b$
for any vector $b=(b_0,b_1,\ldots,b_p)^\T $ satisfying (\ref{ass-compat-eq2}),
in conjunction with the assumption that $E [ \psi_2\{T,\bar\myeta(X)\} |X] \ge c $ for a constant $c>0$.
In the context of Proposition~\ref{pro-RCAL}, the latter assumption can be easily shown to be valid
when $g^*(X) \ge B^*_0$ and $\bar \myeta^1_{\mbox{\tiny CAL}}(X) \le B_1$,
that is, $\pi^*(X) \ge (1+\me^{-B^*_0})^{-1}$ and
$\bar \pi^1_{\mbox{\tiny CAL}}(X) \le (1+\me^{-B_1})^{-1}$, for some constants $B_0^* >0$
and $B_1>0$.

The following Assumptions~\ref{ass-covariate} and \ref{ass-hessian} are mainly used in Lemma~\ref{lem-hessian} to bound the curvature of a symmetrized Bregman divergence
associated with a non-quadratic loss function.
Assumptions~\ref{ass-covariate} and \ref{ass-hessian2} are involved in showing that, with a high probability,
the empirical Hessian $\tilde\Sigma_{\bar\gamma}$ is close to the theoretical Hessian $\Sigma_{\bar\gamma}$
and hence an empirical compatibility condition can be derived from the theoretical compatibility condition (see Lemma~\ref{lem-compat}).

\begin{ass} \label{ass-covariate}
Assume that $\sup_{j=0,1,\ldots,p} |f_j(X) | \le C_0$ for a constant $C_0 >0$.
\end{ass}

\begin{ass} \label{ass-hessian2}
Assume that $\psi_2 \{T, \bar \myeta(X)\} \le C_1$ for a constant $C_1 >0$.
\end{ass}

\begin{ass} \label{ass-hessian}
Assume that for any $t$ and $(u,u^\prime)$,
\begin{align*}
\psi_2(t,u) \le \psi_2(t,u^\prime) \me^{C_2 |u-u^\prime|},
\end{align*}
where $C_2>0$ is a constant depending only on $\psi_2()$.
\end{ass}

The last assumption requires that $|S| \lambda_0$ be sufficiently small, and is used to facilitate both the derivation of the empirical compatibility condition (Lemma~\ref{lem-compat}) and
the localized analysis with a non-quadratic loss function (Lemma~\ref{lem-orac-ineq}).

\begin{ass} \label{ass-rate}
Assume that (i) $(1+\xi_0)^2 \nu_0^{-2} |S| \lambda_0 \le \eta_1$ for a constant $0 < \eta_1 <1$, and
(ii) $C_0 C_2 \xi_2(1-\eta_1)^{-1}  \nu_0^{-2} |S| \lambda_0 \le \eta_2$ for a constant $0 <\eta_2 <1$, where $\xi_2 = (\xi_0+1)(A_0-1)$.
\end{ass}

From the preceding assumptions, we provide a general result about the convergence of $\hat\gamma$ to $\bar \gamma$ in the $\|\cdot\|_1$ norm
and the symmetrized Bregman divergence between $\hat g$ and $\bar \myeta$.
For two functions $\myeta$ and $\myeta^\prime$, the Bregman divergence associated with $\kappa$ is
\begin{align*}
D (\myeta, \myeta^\prime) = \kappa(\myeta) - \kappa(\myeta^\prime) - \langle \nabla\kappa(\myeta^\prime), \myeta-\myeta^\prime \rangle,
\end{align*}
where
$\langle \nabla \kappa(\myeta ), h \rangle = \lim_{u\to 0}\{\kappa ( \myeta  + uh ) - \kappa(\myeta)\}/u$.
If $\myeta=\gamma^\T f$ and $\myeta^\prime = {\gamma^\prime}^\T f$, then
$D(\myeta, \myeta^\prime) = \ell(\gamma) - \ell(\gamma^\prime) - (\gamma-\gamma^\prime)^\T \tilde E[ \psi_1\{T, {\gamma^\prime}^\T f(X)\} f(X)] $.

\begin{pro} \label{pro-reg}
Suppose that Assumptions~\ref{ass-mean0}--\ref{ass-rate} hold. Then for $A_0 > (\xi_0 +1)/(\xi_0-1)$, we have
with probability at least $1-4\epsilon$,
\begin{align}
D( \hat \myeta, \bar \myeta ) + D( \bar \myeta, \hat\myeta) +(A_0-1) \lambda_0 \|\hat\gamma -\bar\gamma \|_1
\le  2 \xi_1^{-1} A_0 \lambda_0\sum_{j\not\in S} |\bar \gamma_j| + \xi_2^2 \nu_1^{-2}  |S| \lambda_0^2, \label{pro-reg-eq}
\end{align}
where $\xi_1 =1-2A_0/\{(\xi_0+1)(A_0-1)\} \in (0,1]$, $\xi_2 = (\xi_0+1)(A_0-1)$, and
$\nu_1 = \nu_0 (1-\eta_1)(1-\eta_2)$.
\end{pro}

Various implications can be deduced from Proposition~\ref{pro-reg}. Taking $S=\{ 0 \}$ leads to a slow rate, of order $ \lambda_0\sum_{j=1}^p|\bar \gamma_j|$,
where the compatibility condition holds under mild conditions: either no linear combination of $f_1(X), \ldots, f_p(X)$ is close to being a constant,
or the $L_2$ norms of $f_1(X), \ldots, f_p(X)$, weighted by $\psi_2\{T,\bar \myeta(X)\}$, are bounded away from above by 1.

\begin{cor} \label{cor-slow-rate}
Suppose that either (i) 
for a constant $0<\eta_3<1$,
\begin{align}
E^{\otimes 2} [ \psi_2\{T,\bar \myeta(X)\} f_{1:p}(X) ] \le \eta_3^2 E [\psi_2\{T,\bar \myeta(X)\} ] E[\psi_2\{T,\bar \myeta(X)\} f^{\otimes 2}_{1:p}(X)  ], \label{cor-slow-rate-eq1}
\end{align}
where $f_{1:p}=(f_1,\ldots,f_p)^\T$ and $b^{\otimes 2} = b b^\T$, or (ii) for a constant $0<\eta_4 <1$,
\begin{align}
\max_{j=1,\ldots, p} E [ \psi_2\{T,\bar \myeta(X)\} f_j^2(X) ] \le \eta_4^2 E[ \psi_2\{T,\bar \myeta(X)\} ] . \label{cor-slow-rate-eq2}
\end{align}
Then Assumption~\ref{ass-compat} is satisfied with $S=\{0\}$ and some constants $\nu_0>0$ and $\xi_0 >1$ depending only on $\eta_3$ or $\eta_4$.
If, in addition, Assumptions~\ref{ass-mean0}, \ref{ass-covariate}, \ref{ass-hessian2}, \ref{ass-hessian}, and \ref{ass-rate} hold with $|S|=1$,
then (\ref{pro-reg-eq}) holds with probability at least $1-4\epsilon$.
\end{cor}

Taking $S = \{0\} \cup \{j: \bar\gamma_j \not= 0, j=1,\ldots,p\}$ yields a fast rate, of order $|S| \lambda_0^2$.


\begin{cor} \label{cor-fast-rate}
Suppose that Assumptions~\ref{ass-mean0}--\ref{ass-rate} hold with $S = \{0\} \cup \{j: \bar\gamma_j \not= 0, j=1,\ldots,p\}$. Then for $A_0 > (\xi_0 +1)/(\xi_0-1)$, we have
with probability at least $1-4\epsilon$,
\begin{align*}
D( \hat \myeta, \bar \myeta ) + D( \bar \myeta, \hat\myeta) +(A_0-1) \lambda_0 \|\hat\gamma -\bar\gamma \|_1
\le  \xi_2^2 \nu_1^{-2}  |S| \lambda_0^2,
\end{align*}
where $\xi_2$ and $\nu_1$ are as in Proposition~\ref{pro-reg}.
\end{cor}

The following result provides a bound relating $D(\hat \myeta, \myeta^*)$ to $D(\bar \myeta, \myeta^*)$, which compare the predictor $\hat\myeta$ and the oracle $\bar \myeta$ respectively
with the truth $\myeta^*$. This result, with leading coefficient one for $D(\bar \myeta, \myeta^*)$, is distinct from previous results, for example, Buhlmann \& van de Geer (2011, Theorem 6.4).
See Zhang \& Zhang (2012, Section 3.2) for a related discussion.

\begin{cor} \label{cor-comp-truth}
In addition to Assumptions~\ref{ass-mean0}--\ref{ass-rate}, suppose that Assumption~\ref{ass-mean0} also holds with $Z_j$ replaced by $Z_j^*=\psi_1 \{T, \myeta^*(X)\} f_j(X)$ for $j=0,1,\ldots,p$.
Then for $A_0 > (\xi_0 +1)/(\xi_0-1)$, we have
with probability at least $1-6\epsilon$,
\begin{align*}
D( \hat \myeta, \myeta^* ) + D( \bar \myeta,  \hat \myeta )
\le  D( \bar \myeta, \myeta^* )  + \frac{A_0+1}{A_0-1} \Delta (\bar \myeta, S),
\end{align*}
where $\Delta(\bar \myeta, S)$ denotes the right hand side of (\ref{pro-reg-eq}) and $\xi_1$, $\xi_2 $, and $\nu_1 $ are as in Proposition~\ref{pro-reg}.
\end{cor}

Finally, we provide additional comments on how our results are related to previous works.
Our approach mainly builds on techniques developed in Huang \& Zhang (2012) and Zhang \& Zhang (2012) for high-dimensional analysis, including the use of Bregman divergences and
the derivation of basic inequalities exploiting the convexity of loss functions. Our analysis, however,
provides explicit assumptions, notably Assumption~\ref{ass-mean0} in terms of the target value $\bar \gamma$,
and yields direct results on the convergence of $\hat\gamma$ to $\bar \gamma$.
Such convergence is also focused on in classical theory of estimation with misspecified models (e.g., Manski 1988).
From this perspective, our analysis also differs from van de Geer (2008) and Buhlmann \& van de Geer (2011),
where the main results are oracle inequalities comparing
the closeness of the predictor $\hat g$ to $g^*$ with that of the oracle $\bar g$ to $g^*$,
but in a different manner than Corollary~\ref{cor-comp-truth} as discussed above.
Negahban et al.~(2012) provided high-dimensional analysis of regularized M-estimators in general settings, including
Lasso penalized maximum likelihood estimators. But their analysis involves the stronger assumption that
the variables $f_1(X),\ldots,f_p(X)$, are jointly sub-gaussian.

\newpage
\section{Additional numerical illustration}

\begin{figure}[h]
\caption{\small Behavior of functions $L(\rho \pm .01, \rho)$, $K(\rho \pm .01, \rho)$, and $Q(\rho \pm .01, \rho)$.} \label{fig:obj} \vspace{.1in}
\begin{tabular}{c}
\includegraphics[width=6in, height=2.5in]{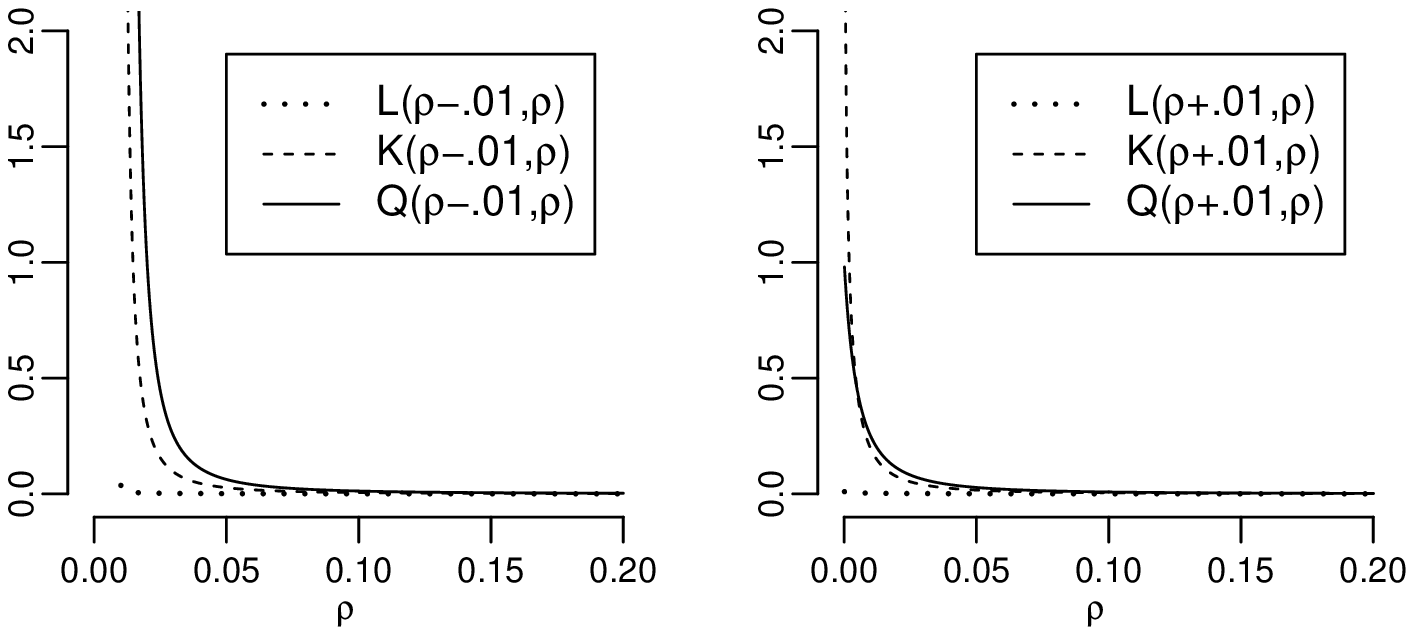} \vspace{-.15in}
\end{tabular}
\end{figure}

Figure~\ref{fig:obj} illustrates how the functions $L(\rho^\prime, \rho)$, $K(\rho^\prime, \rho)$, and $Q(\rho^\prime, \rho)$
are related to each other, with $\rho^\prime = \rho\pm .01$, that is, $\rho^\prime$ is somewhat close to $\rho$.
The Kullback--Liebler divergence $L(\rho \pm .01, \rho)$ is close to 0. But the relative error $Q(\rho^\prime, \rho) = (\rho/\rho^\prime-1)^2$
with $\rho^\prime=\rho \pm .01$ can be very large, particularly when $\rho^\prime=\rho-.01$ and $\rho$ is close to $.01+$.
This shows that an absolute error of $.01$ can still lead to a large relative error.
The function $K(\rho \pm .01, \rho)$ can be seen to upper-bound $Q(\rho \pm .01, \rho)$ up to a constant
depending on how close $\rho$ is to $.01+$, that is, how large $\rho/(\rho-.01)$ is.
This is the main point in Proposition~\ref{pro4}(i).

\vspace{.2in}

\newpage
\section{Additional simulation results}

We present additional results from the simulation study in Section \ref{sec:simulation}.

Table~\ref{table:non-conv} shows the number of samples from 1000 simulations, with non-convergence declared by the R package $\texttt{trust}$ when computing the non-penalized
estimators $\hat\gamma_{\mbox{\tiny ML}}$ and $\hat\gamma^1_{\mbox{\tiny CAL}}$.
In fact, convergence is obtained for $\hat\gamma_{\mbox{\tiny ML}}$ in all simulations.
But non-convergence is found for $\hat\gamma^1_{\mbox{\tiny CAL}}$ in a considerable fraction of simulations when $(p,n)=(20,200)$ or $(50, \le 400)$.

Table~\ref{table:non-zero} shows the average numbers of all nonzero coefficients and truly nonzero ones (i.e., only associated with the first 4 covariates),
for the regularized estimators $\hat\gamma_{\mbox{\tiny RML}}$ and $\hat\gamma^1_{\mbox{\tiny RCAL}}$.
In either case, the average numbers from $\hat\gamma^1_{\mbox{\tiny RCAL}}$ are consistently lower than from $\hat\gamma_{\mbox{\tiny RML}}$.

Tables~\ref{table:est-lin1}--\ref{table:est-eps} present the root mean squared errors of $ \hat\mu^1_{\mbox{\tiny rIPW}}(\hat\pi; h) $ with
5 configurations ``lin1", ``lin2", ``quad1", ``quad2", and ``exp" for $h(X)$ and $ \hat\mu^1_{\mbox{\tiny rIPW}}(\hat\pi; \varepsilon) $, which
are plotted in Figures~\ref{fig:est-cor} and \ref{fig:est} to facilitate visual comparison.

\begin{figure}
\caption{\small Root mean squared errors of the differences $ \hat\mu^1_{\mbox{\tiny rIPW}}(\hat\pi^1; h) -\hat\mu^0_{\mbox{\tiny rIPW}}(\hat\pi^0; h) $
and $ \hat\mu^1_{\mbox{\tiny rIPW}} (\hat\pi^1; \varepsilon) -\hat\mu^0_{\mbox{\tiny rIPW}} (\hat\pi^0; \varepsilon) $, plotted similarly as in Figure~\ref{fig:est-cor},
for the estimators $\hat\pi$ labeled 1--6 when logistic model (\ref{logit-PS}) is correctly specified.} \label{fig:diff-cor} \vspace{.2in}
\begin{tabular}{c}
\includegraphics[width=6.2in, height=5in]{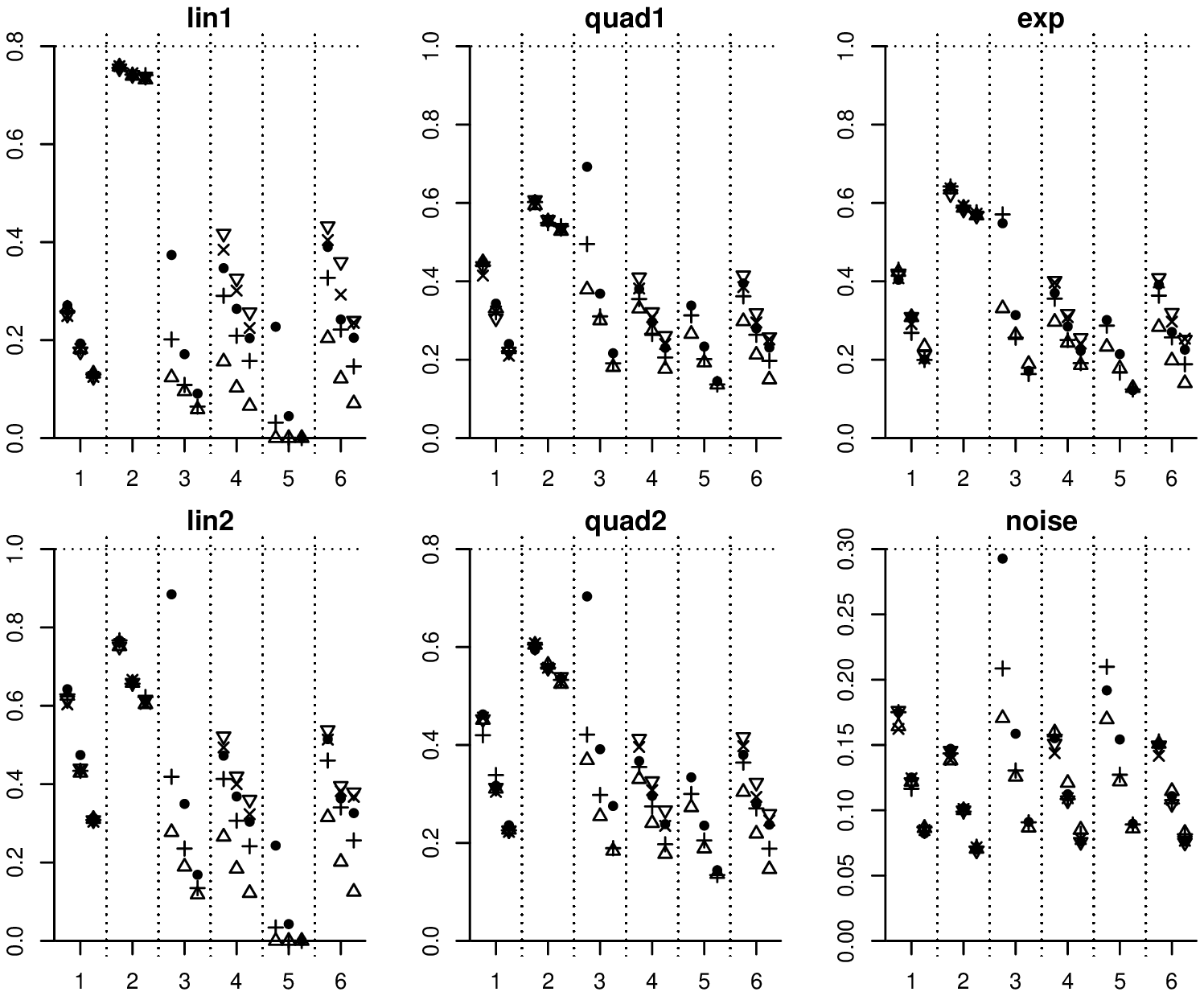} 
\end{tabular}
\end{figure}

\begin{figure}
\caption{\small Root mean squared errors of the differences $ \hat\mu^1_{\mbox{\tiny rIPW}}(\hat\pi^1; h) -\hat\mu^0_{\mbox{\tiny rIPW}}(\hat\pi^0; h) $
and $ \hat\mu^1_{\mbox{\tiny rIPW}} (\hat\pi^1; \varepsilon) -\hat\mu^0_{\mbox{\tiny rIPW}} (\hat\pi^0; \varepsilon) $, plotted similarly as in Figure~\ref{fig:est-cor},
for the estimators $\hat\pi$ labeled 1--6 when logistic model (\ref{logit-PS}) is misspecified.} \label{fig:diff} \vspace{.2in}
\begin{tabular}{c}
\includegraphics[width=6.2in, height=5in]{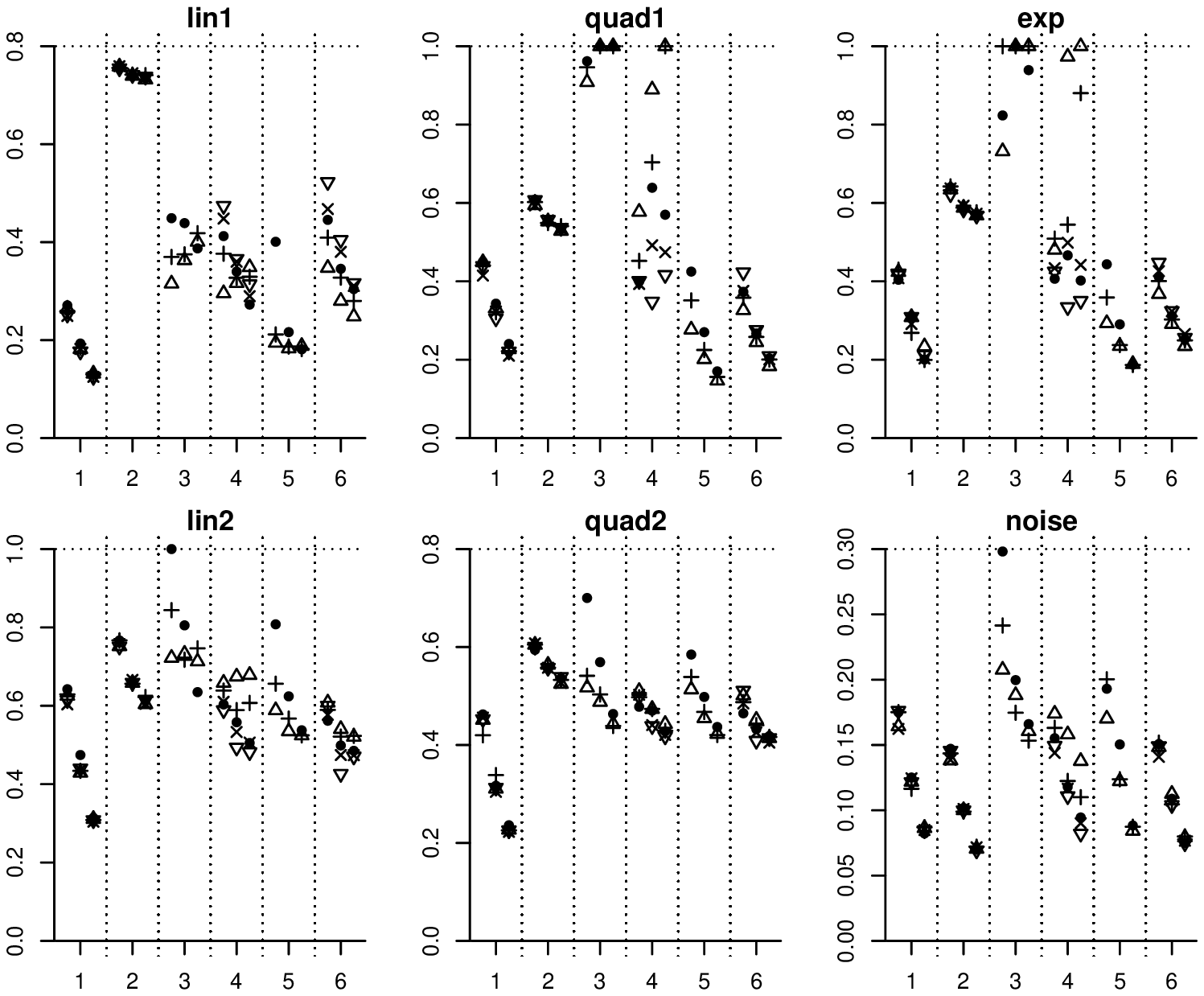} 
\end{tabular}
\end{figure}

Figures~\ref{fig:diff-cor} and \ref{fig:diff} show the root mean squared errors of the differences $ \hat\mu^1_{\mbox{\tiny rIPW}}(\hat\pi^1; h) -\hat\mu^0_{\mbox{\tiny rIPW}}(\hat\pi^0; h) $
and $ \hat\mu^1_{\mbox{\tiny rIPW}} (\hat\pi^1; \varepsilon) -\hat\mu^0_{\mbox{\tiny rIPW}} (\hat\pi^0; \varepsilon) $,
when logistic model (\ref{logit-PS}) is correctly specified or misspecified.
The fitted propensity scores $\hat\pi^1$ and $\hat\pi^0$ are the same when maximum likelihood is used, but
separately computed for calibrated estimation and regularization.
In particular, the estimator $\hat\gamma^0_{\mbox{\tiny RCAL}}$ corresponding to
$\hat\pi^0_{\mbox{\tiny RCAL}}$ is computed with the tuning parameter $\lambda$ determined by 5-fold cross validation, similarly as
but separately from $\hat\gamma^1_{\mbox{\tiny RCAL}}$.
The relative performances of the estimators are similar to those in Figures~\ref{fig:est-cor} and \ref{fig:est}.

\begin{figure}
\caption{\small Root mean squared errors of global measures $\tilde\kappa_{\mbox{\tiny ML}} (\hat\myeta)$,
$\tilde\kappa_{\mbox{\tiny CAL}} (\hat\myeta) $, $\mbox{mse}(\hat\pi)$, and $\mbox{msre}(\hat\pi)$, plotted similarly as in Figure~\ref{fig:div},
for the estimators $\hat\pi$ labeled 1--6
when logistic model (\ref{logit-PS}) is correctly specified. } \label{fig:div-cor} \vspace{.2in}
\begin{tabular}{c}
\includegraphics[width=6.2in, height=4in]{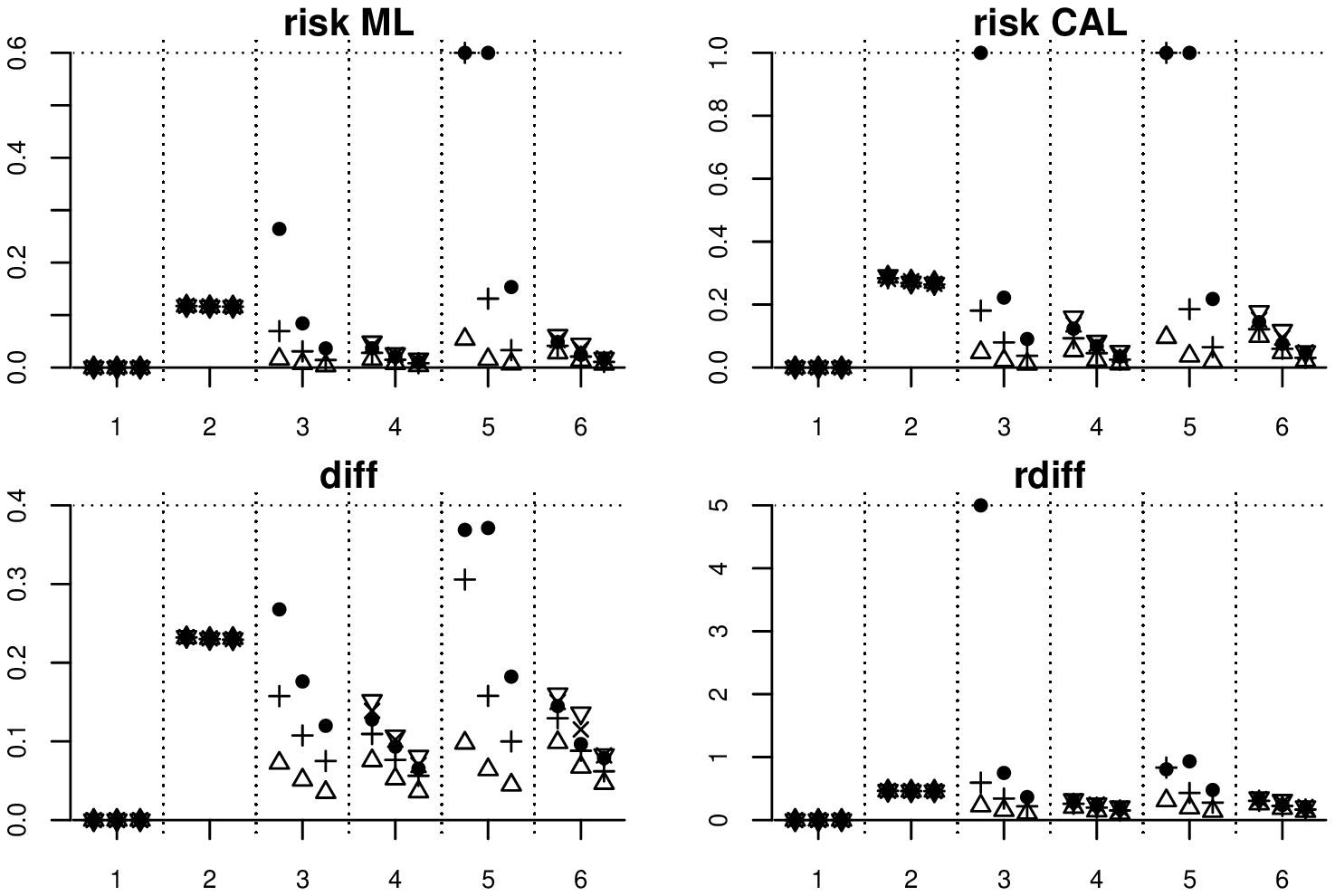} 
\end{tabular}
\end{figure}

Figure~\ref{fig:div-cor} shows the root mean squared errors of global measures $\tilde\kappa_{\mbox{\tiny ML}} (\hat\myeta)$,
$\tilde\kappa_{\mbox{\tiny CAL}} (\hat\myeta) $, $\mbox{mse}(\hat\pi)$, and $\mbox{msre}(\hat\pi)$,  similarly as in Figure~\ref{fig:div},
but when logistic model (\ref{logit-PS}) is correctly specified.
The two regularized estimators $\hat\gamma_{\mbox{\tiny RML}}$ and $\hat\gamma^1_{\mbox{\tiny RCAL}}$ perform similarly to each other, in contrast
with the case with logistic model (\ref{logit-PS}) misspecified.

\newpage
\begin{table}[t]
\caption{ Numbers of samples with non-convergence from 1000 simulations } \label{table:non-conv} \vspace{-.02in}
\scriptsize
\begin{center}
\begin{tabular*}{.8\textwidth}{@{\extracolsep\fill} c ccc c ccc c ccc}
\hline
    & \multicolumn{3}{c}{$n=200$} && \multicolumn{3}{c}{$n=400$} && \multicolumn{3}{c}{$n=800$} \\ \cline{2-4} \cline{6-8} \cline{10-12}
    & $p=4$  & $p=20$  & $p=50$ &&  $p=4$  & $p=20$  & $p=50$  &&  $p=4$  & $p=20$  & $p=50$  \\ \hline
    & \multicolumn{11}{c}{Logistic model correctly specified} \\
ML  & 0  &  0  & 0   &&  0  &  0  & 0   &&   0  &  0  & 0 \\
CAL & 0  & 315 & 999 &&  0  &  0  & 694 &&   0  &  0  & 0 \\ \hline
    & \multicolumn{11}{c}{Logistic model misspecified} \\
ML  & 0  &  0  & 0   &&  0  &  0  & 0   &&   0  &  0  & 0 \\
CAL & 1  & 315 & 998 &&  0  &  2  & 574 &&   0  &  0  & 0 \\
\hline
\end{tabular*} \\[.1in]
\parbox{.8\textwidth}{\small Note: ML is non-penalized maximum likelihood and CAL is non-penalized calibrated estimation, both
implemented using R package $\texttt{trust}$. Non-convergence is declared by \texttt{trust} when the termination criteria are not satisfied after 1000 iterations.
In the non-convergence cases, the values of $\ell_{\mbox{\tiny CAL}}$ obtained, minus $\kappa_{\mbox{\tiny CAL}}(g^*)$, are found to range from $-10,000$ or smaller to $-10$,
indicating that the loss function $\ell_{\mbox{\tiny CAL}}$ may not have a finite minimum.}
\end{center}
\end{table}

\begin{table}[b!]
\caption{ Average numbers of nonzero coefficients estimated from 1000 simulations } \label{table:non-zero} \vspace{-.02in}
\scriptsize
\begin{center}
\begin{tabular*}{1\textwidth}{@{\extracolsep\fill} c ccccc c ccccc c ccccc} \hline
    & \multicolumn{5}{c}{$n=200$} && \multicolumn{5}{c}{$n=400$} && \multicolumn{5}{c}{$n=800$} \\ \cline{2-6} \cline{8-12} \cline{14-18}
$p$  & $4$  & $20$  & $50$  & $100$  & $200$  &&  $4$  & $20$  & $50$  & $100$  & $200$  &&  $4$  & $20$  & $50$  & $100$  & $200$  \\ \hline
    & \multicolumn{17}{c}{Logistic model correctly specified} \\
RML  & $3.5$  & $6.9$  & $8.6$  &  $10.2$ & $12.2$ && $3.7$ & $8.1$ & $10.6$ & $11.6$ & $11.5$ && $3.9$ & $9.1$ & $10.9$ & $14.0$ & $18.5$ \\
     & $3.5$  & $2.9$  & $2.5$  &  $2.3$  & $2.1$  && $3.7$ & $3.3$ & $3.0$  & $2.8$  & $2.6$  && $3.9$ & $3.6$ & $3.3$  & $3.2$  & $3.1$ \\
RCAL & $2.9$  & $3.6$  & $2.4$  &  $2.6$  & $2.8$  && $3.4$ & $4.8$ & $6.4$  & $6.7$  & $3.0$  && $3.7$ & $7.1$ & $7.1$  & $4.6$  & $6.1$ \\
     & $2.9$  & $2.2$  & $1.8$  &  $1.6$  & $1.4$  && $3.4$ & $2.8$ & $2.7$  & $2.4$  & $1.8$  && $3.7$ & $3.4$ & $3.0$  & $2.8$  & $2.8$ \\ \hline
     & \multicolumn{17}{c}{Logistic model misspecified} \\
RML  & $3.0$  & $6.2$  & $7.7$  &  $9.5$  & $10.7$ && $3.1$ & $6.8$ & $8.7$  & $9.8$  & $11.2$ && $3.1$ & $6.8$ & $9.4$  & $13.0$ & $14.3$ \\
     & $3.0$  & $2.3$  & $2.0$  &  $1.9$  & $1.7$  && $3.1$ & $2.5$ & $2.2$  & $2.2$  & $2.1$  && $3.1$ & $2.4$ & $2.3$  & $2.2$  & $2.1$ \\
RCAL & $2.1$  & $2.3$  & $2.1$  &  $2.5$  & $2.2$  && $2.5$ & $3.6$ & $4.2$  & $2.4$  & $1.7$  && $2.8$ & $4.3$ & $3.9$  & $4.8$  & $6.8$ \\
     & $2.1$  & $1.6$  & $1.4$  &  $1.2$  & $0.9$  && $2.5$ & $2.2$ & $1.9$  & $1.6$  & $1.5$  && $2.8$ & $2.5$ & $2.3$  & $2.2$  & $2.2$ \\ \hline
\end{tabular*} \\[.1in]
\parbox{1\textwidth}{\small Note: RML is regularized maximum likelihood and CAL is regularized calibrated estimation, both with Lasso.
Each cell gives the average number of all nonzero coefficients (upper) and the average number of nonzero coefficients only associated with the first 4 covariates (lower). }
\end{center}
\end{table}

\newpage
\begin{table}[t]
\caption{ Root mean squared errors of $ \hat\mu^1_{\mbox{\tiny rIPW}}(\hat\pi; h) $ with $h(X)=$``lin1" } \label{table:est-lin1} \vspace{-.02in}
\scriptsize
\begin{center}
\begin{tabular*}{1\textwidth}{@{\extracolsep\fill} r rrrrr r rrrrr r rrrrr} \hline
    & \multicolumn{5}{c}{$n=200$} && \multicolumn{5}{c}{$n=400$} && \multicolumn{5}{c}{$n=800$} \\ \cline{2-6} \cline{8-12} \cline{14-18}
$p$  & $4$  & $20$  & $50$  & $100$  & $200$  &&  $4$  & $20$  & $50$  & $100$  & $200$  &&  $4$  & $20$  & $50$  & $100$  & $200$  \\ \hline
    & \multicolumn{17}{c}{Logistic model correctly specified} \\
True & $.18$ & $.18$ & $.18$ & $.17$ & $.18$ && $.13$ & $.13$ & $.13$ & $.13$ & $.13$ && $.09$ & $.09$ & $.09$ & $.09$ & $.09$ \\
Const& $.39$ & $.39$ & $.39$ & $.39$ & $.39$ && $.38$ & $.38$ & $.38$ & $.38$ & $.38$ && $.37$ & $.38$ & $.37$ & $.37$ & $.37$ \\
ML   & $.14$ & $.19$ & $.29$ & ---   & ---   && $.10$ & $.11$ & $.15$ & ---   & ---   && $.07$ & $.07$ & $.08$ & ---   & --- \\
RML  & $.14$ & $.18$ & $.20$ & $.22$ & $.23$ && $.10$ & $.13$ & $.15$ & $.17$ & $.18$ && $.07$ & $.10$ & $.11$ & $.12$ & $.14$ \\
CAL  & $.09$ & $.09$ & $.15$ & ---   & ---   && $.07$ & $.06$ & $.07$ & ---   & ---   && $.05$ & $.05$ & $.05$ & ---   & --- \\
RCAL & $.14$ & $.19$ & $.22$ & $.22$ & $.24$ && $.09$ & $.13$ & $.14$ & $.16$ & $.19$ && $.06$ & $.09$ & $.11$ & $.13$ & $.13$ \\
     & \multicolumn{17}{c}{Logistic model misspecified} \\
True & $.18$ & $.18$ & $.18$ & $.17$ & $.18$ && $.13$ & $.13$ & $.13$ & $.13$ & $.13$ && $.09$ & $.09$ & $.09$ & $.09$ & $.09$ \\
Const& $.39$ & $.39$ & $.39$ & $.39$ & $.39$ && $.38$ & $.38$ & $.38$ & $.38$ & $.38$ && $.37$ & $.38$ & $.37$ & $.37$ & $.37$ \\
ML   & $.33$ & $.37$ & $.44$ & ---   & ---   && $.39$ & $.40$ & $.47$ & ---   & ---   && $.44$ & $.46$ & $.43$ & ---   & --- \\
RML  & $.23$ & $.22$ & $.23$ & $.25$ & $.26$ && $.30$ & $.24$ & $.21$ & $.21$ & $.19$ && $.37$ & $.29$ & $.18$ & $.17$ & $.17$ \\
CAL  & $.14$ & $.14$ & $.21$ & ---   & ---   && $.12$ & $.12$ & $.13$ & ---   & ---   && $.12$ & $.11$ & $.10$ & ---   & --- \\
RCAL & $.21$ & $.22$ & $.24$ & $.25$ & $.29$ && $.16$ & $.17$ & $.18$ & $.20$ & $.20$ && $.14$ & $.14$ & $.15$ & $.15$ & $.15$ \\ \hline
\end{tabular*}
\end{center}
\end{table}

\begin{table}[h]
\caption{ Root mean squared errors of $ \hat\mu^1_{\mbox{\tiny rIPW}}(\hat\pi; h) $ with $h(X)=$``lin2" }  \label{table:est-lin2} \vspace{-.02in}
\scriptsize
\begin{center}
\begin{tabular*}{1\textwidth}{@{\extracolsep\fill} r rrrrr r rrrrr r rrrrr} \hline
    & \multicolumn{5}{c}{$n=200$} && \multicolumn{5}{c}{$n=400$} && \multicolumn{5}{c}{$n=800$} \\ \cline{2-6} \cline{8-12} \cline{14-18}
$p$  & $4$  & $20$  & $50$  & $100$  & $200$  &&  $4$  & $20$  & $50$  & $100$  & $200$  &&  $4$  & $20$  & $50$  & $100$  & $200$  \\ \hline
    & \multicolumn{17}{c}{Logistic model correctly specified} \\
True & $.45$ & $.44$ & $.45$ & $.42$ & $.43$ && $.31$ & $.31$ & $.33$ & $.31$ & $.31$ && $.22$ & $.22$ & $.22$ & $.22$ & $.22$ \\
Const& $.45$ & $.46$ & $.47$ & $.45$ & $.46$ && $.38$ & $.38$ & $.39$ & $.38$ & $.38$ && $.33$ & $.34$ & $.33$ & $.33$ & $.33$ \\
ML   & $.36$ & $.46$ & $.72$ & ---   & ---   && $.24$ & $.28$ & $.32$ & ---   & ---   && $.16$ & $.17$ & $.19$ & ---   & --- \\
RML  & $.33$ & $.35$ & $.37$ & $.37$ & $.37$ && $.23$ & $.25$ & $.27$ & $.27$ & $.28$ && $.16$ & $.19$ & $.21$ & $.21$ & $.23$ \\
CAL  & $.25$ & $.27$ & $.30$ & ---   & ---   && $.18$ & $.18$ & $.18$ & ---   & ---   && $.13$ & $.13$ & $.13$ & ---   & --- \\
RCAL & $.31$ & $.36$ & $.38$ & $.37$ & $.38$ && $.21$ & $.25$ & $.27$ & $.26$ & $.27$ && $.15$ & $.18$ & $.21$ & $.22$ & $.23$ \\
     & \multicolumn{17}{c}{Logistic model misspecified} \\
True & $.45$ & $.44$ & $.45$ & $.42$ & $.43$ && $.31$ & $.31$ & $.33$ & $.31$ & $.31$ && $.22$ & $.22$ & $.22$ & $.22$ & $.22$ \\
Const& $.45$ & $.46$ & $.47$ & $.45$ & $.46$ && $.38$ & $.38$ & $.39$ & $.38$ & $.38$ && $.33$ & $.34$ & $.33$ & $.33$ & $.33$ \\
ML   & $.61$ & $.74$ & $.86$ & ---   & ---   && $.66$ & $.62$ & $.70$ & ---   & ---   && $.66$ & $.70$ & $.56$ & ---   & --- \\
RML  & $.48$ & $.44$ & $.42$ & $.41$ & $.40$ && $.55$ & $.40$ & $.37$ & $.36$ & $.31$ && $.58$ & $.46$ & $.30$ & $.31$ & $.29$ \\
CAL  & $.41$ & $.44$ & $.49$ & ---   & ---   && $.34$ & $.35$ & $.38$ & ---   & ---   && $.32$ & $.31$ & $.31$ & ---   & --- \\
RCAL & $.39$ & $.38$ & $.39$ & $.38$ & $.41$ && $.31$ & $.29$ & $.29$ & $.27$ & $.27$ && $.28$ & $.26$ & $.24$ & $.24$ & $.23$ \\ \hline
\end{tabular*}
\end{center}
\end{table}

\newpage
\begin{table}[t]
\caption{ Root mean squared errors of $ \hat\mu^1_{\mbox{\tiny rIPW}}(\hat\pi; h) $ with $h(X)=$``quad1" }  \label{table:est-quad1} \vspace{-.02in}
\scriptsize
\begin{center}
\begin{tabular*}{1\textwidth}{@{\extracolsep\fill} r rrrrr r rrrrr r rrrrr} \hline
    & \multicolumn{5}{c}{$n=200$} && \multicolumn{5}{c}{$n=400$} && \multicolumn{5}{c}{$n=800$} \\ \cline{2-6} \cline{8-12} \cline{14-18}
$p$  & $4$  & $20$  & $50$  & $100$  & $200$  &&  $4$  & $20$  & $50$  & $100$  & $200$  &&  $4$  & $20$  & $50$  & $100$  & $200$  \\ \hline
    & \multicolumn{17}{c}{Logistic model correctly specified} \\
True & $.37$ & $.35$ & $.35$ & $.33$ & $.36$ && $.27$ & $.25$ & $.27$ & $.26$ & $.24$ && $.19$ & $.18$ & $.19$ & $.17$ & $.18$ \\
Const& $.33$ & $.33$ & $.34$ & $.32$ & $.34$ && $.30$ & $.29$ & $.29$ & $.29$ & $.30$ && $.27$ & $.28$ & $.28$ & $.28$ & $.28$ \\
ML   & $.34$ & $.45$ & $.56$ & ---   & ---   && $.27$ & $.27$ & $.32$ & ---   & ---   && $.16$ & $.17$ & $.19$ & ---   & --- \\
RML  & $.29$ & $.27$ & $.29$ & $.28$ & $.29$ && $.24$ & $.22$ & $.22$ & $.22$ & $.23$ && $.16$ & $.17$ & $.17$ & $.18$ & $.19$ \\
CAL  & $.24$ & $.27$ & $.30$ & ---   & ---   && $.17$ & $.18$ & $.21$ & ---   & ---   && $.12$ & $.12$ & $.13$ & ---   & --- \\
RCAL & $.25$ & $.27$ & $.29$ & $.28$ & $.29$ && $.18$ & $.21$ & $.21$ & $.21$ & $.23$ && $.13$ & $.16$ & $.17$ & $.19$ & $.19$ \\
     & \multicolumn{17}{c}{Logistic model misspecified} \\
True & $.37$ & $.35$ & $.35$ & $.33$ & $.36$ && $.27$ & $.25$ & $.27$ & $.26$ & $.24$ && $.19$ & $.18$ & $.19$ & $.17$ & $.18$ \\
Const& $.33$ & $.33$ & $.34$ & $.32$ & $.34$ && $.30$ & $.29$ & $.29$ & $.29$ & $.30$ && $.27$ & $.28$ & $.28$ & $.28$ & $.28$ \\
ML   & $.91$ & $.92$ & $.89$ & ---   & ---   && $1.13$& $1.21$& $1.41$& ---   & ---   && $1.39$& $1.50$& $1.32$& ---   & --- \\
RML  & $.56$ & $.38$ & $.31$ & $.30$ & $.29$ && $.90$ & $.70$ & $.62$ & $.47$ & $.31$ && $1.19$& $1.04$& $.58$ & $.48$ & $.41$ \\
CAL  & $.23$ & $.27$ & $.32$ & ---   & ---   && $.16$ & $.17$ & $.20$ & ---   & ---   && $.11$ & $.12$ & $.13$ & ---   & --- \\
RCAL & $.25$ & $.26$ & $.27$ & $.27$ & $.29$ && $.18$ & $.18$ & $.19$ & $.19$ & $.19$ && $.13$ & $.13$ & $.14$ & $.14$ & $.14$ \\ \hline
\end{tabular*}
\end{center}
\end{table}

\begin{table}[h]
\caption{ Root mean squared errors of $ \hat\mu^1_{\mbox{\tiny rIPW}}(\hat\pi; h) $ with $h(X)=$``quad2" } \label{table:est-quad2} \vspace{-.02in}
\scriptsize
\begin{center}
\begin{tabular*}{1\textwidth}{@{\extracolsep\fill} r rrrrr r rrrrr r rrrrr} \hline
    & \multicolumn{5}{c}{$n=200$} && \multicolumn{5}{c}{$n=400$} && \multicolumn{5}{c}{$n=800$} \\ \cline{2-6} \cline{8-12} \cline{14-18}
$p$  & $4$  & $20$  & $50$  & $100$  & $200$  &&  $4$  & $20$  & $50$  & $100$  & $200$  &&  $4$  & $20$  & $50$  & $100$  & $200$  \\ \hline
    & \multicolumn{17}{c}{Logistic model correctly specified} \\
True & $.27$ & $.26$ & $.26$ & $.25$ & $.26$ && $.19$ & $.19$ & $.19$ & $.19$ & $.19$ && $.13$ & $.13$ & $.14$ & $.13$ & $.15$ \\
Const& $.35$ & $.35$ & $.35$ & $.35$ & $.35$ && $.31$ & $.30$ & $.31$ & $.31$ & $.31$ && $.28$ & $.28$ & $.28$ & $.28$ & $.28$ \\
ML   & $.25$ & $.28$ & $.38$ & ---   & ---   && $.17$ & $.20$ & $.22$ & ---   & ---   && $.12$ & $.12$ & $.14$ & ---   & --- \\
RML  & $.23$ & $.22$ & $.23$ & $.23$ & $.24$ && $.16$ & $.16$ & $.17$ & $.17$ & $.18$ && $.12$ & $.12$ & $.12$ & $.12$ & $.14$ \\
CAL  & $.21$ & $.23$ & $.23$ & ---   & ---   && $.15$ & $.16$ & $.18$ & ---   & ---   && $.10$ & $.11$ & $.12$ & ---   & --- \\
RCAL & $.21$ & $.22$ & $.23$ & $.23$ & $.25$ && $.15$ & $.15$ & $.16$ & $.16$ & $.17$ && $.10$ & $.11$ & $.12$ & $.12$ & $.13$ \\
     & \multicolumn{17}{c}{Logistic model misspecified} \\
True & $.27$ & $.26$ & $.26$ & $.25$ & $.26$ && $.19$ & $.19$ & $.19$ & $.19$ & $.19$ && $.13$ & $.13$ & $.14$ & $.13$ & $.15$ \\
Const& $.35$ & $.35$ & $.35$ & $.35$ & $.35$ && $.31$ & $.30$ & $.31$ & $.31$ & $.31$ && $.28$ & $.28$ & $.28$ & $.28$ & $.28$ \\
ML   & $.34$ & $.37$ & $.51$ & ---   & ---   && $.33$ & $.34$ & $.42$ & ---   & ---   && $.29$ & $.29$ & $.30$ & ---   & --- \\
RML  & $.31$ & $.28$ & $.27$ & $.28$ & $.29$ && $.28$ & $.27$ & $.27$ & $.23$ & $.23$ && $.27$ & $.22$ & $.21$ & $.20$ & $.19$ \\
CAL  & $.32$ & $.32$ & $.32$ & ---   & ---   && $.26$ & $.25$ & $.27$ & ---   & ---   && $.23$ & $.22$ & $.22$ & ---   & --- \\
RCAL & $.27$ & $.26$ & $.27$ & $.27$ & $.30$ && $.22$ & $.20$ & $.21$ & $.21$ & $.20$ && $.19$ & $.18$ & $.18$ & $.17$ & $.18$ \\ \hline
\end{tabular*}
\end{center}
\end{table}

\newpage
\begin{table}[t]
\caption{ Root mean squared errors of $ \hat\mu^1_{\mbox{\tiny rIPW}}(\hat\pi; h) $ with $h(X)=$``exp1" }  \label{table:est-exp1} \vspace{-.02in}
\scriptsize
\begin{center}
\begin{tabular*}{1\textwidth}{@{\extracolsep\fill} r rrrrr r rrrrr r rrrrr} \hline
    & \multicolumn{5}{c}{$n=200$} && \multicolumn{5}{c}{$n=400$} && \multicolumn{5}{c}{$n=800$} \\ \cline{2-6} \cline{8-12} \cline{14-18}
$p$  & $4$  & $20$  & $50$  & $100$  & $200$  &&  $4$  & $20$  & $50$  & $100$  & $200$  &&  $4$  & $20$  & $50$  & $100$  & $200$  \\ \hline
    & \multicolumn{17}{c}{Logistic model correctly specified} \\
True & $.37$ & $.36$ & $.33$ & $.36$ & $.36$ && $.27$ & $.23$ & $.26$ & $.24$ & $.27$ && $.21$ & $.17$ & $.17$ & $.17$ & $.19$ \\
Const& $.33$ & $.33$ & $.34$ & $.32$ & $.33$ && $.31$ & $.31$ & $.30$ & $.30$ & $.30$ && $.29$ & $.30$ & $.29$ & $.29$ & $.29$ \\
ML   & $.34$ & $.57$ & $.52$ & ---   & ---   && $.26$ & $.26$ & $.30$ & ---   & ---   && $.19$ & $.17$ & $.17$ & ---   & --- \\
RML  & $.29$ & $.27$ & $.26$ & $.27$ & $.26$ && $.23$ & $.20$ & $.20$ & $.20$ & $.21$ && $.19$ & $.15$ & $.15$ & $.16$ & $.16$ \\
CAL  & $.23$ & $.26$ & $.25$ & ---   & ---   && $.17$ & $.16$ & $.19$ & ---   & ---   && $.12$ & $.12$ & $.12$ & ---   & --- \\
RCAL & $.23$ & $.25$ & $.26$ & $.26$ & $.26$ && $.17$ & $.19$ & $.19$ & $.19$ & $.20$ && $.12$ & $.14$ & $.15$ & $.16$ & $.16$ \\
     & \multicolumn{17}{c}{Logistic model misspecified} \\
True & $.37$ & $.36$ & $.33$ & $.36$ & $.36$ && $.27$ & $.23$ & $.26$ & $.24$ & $.27$ && $.21$ & $.17$ & $.17$ & $.17$ & $.19$ \\
Const& $.33$ & $.33$ & $.34$ & $.32$ & $.33$ && $.31$ & $.31$ & $.30$ & $.30$ & $.30$ && $.29$ & $.30$ & $.29$ & $.29$ & $.29$ \\
ML   & $.74$ & $1.06$& $.81$ & ---   & ---   && $1.19$& $1.03$& $1.22$& ---   & ---   && $1.71$& $1.37$& $.97$ & ---   & --- \\
RML  & $.45$ & $.44$ & $.29$ & $.30$ & $.28$ && $.98$ & $.53$ & $.41$ & $.46$ & $.25$ && $1.58$& $.89$ & $.39$ & $.42$ & $.31$ \\
CAL  & $.24$ & $.27$ & $.28$ & ---   & ---   && $.19$ & $.19$ & $.22$ & ---   & ---   && $.15$ & $.14$ & $.14$ & ---   & --- \\
RCAL & $.26$ & $.26$ & $.27$ & $.27$ & $.28$ && $.21$ & $.20$ & $.21$ & $.20$ & $.20$ && $.16$ & $.16$ & $.16$ & $.17$ & $.16$ \\ \hline
\end{tabular*}
\end{center}
\end{table}

\begin{table}[h]
\caption{ Root mean squared errors of $ \hat\mu^1_{\mbox{\tiny rIPW}}(\hat\pi; \varepsilon) $ with $\varepsilon=$``noise" } \label{table:est-eps} \vspace{-.02in}
\scriptsize
\begin{center}
\begin{tabular*}{1\textwidth}{@{\extracolsep\fill} r rrrrr r rrrrr r rrrrr} \hline
    & \multicolumn{5}{c}{$n=200$} && \multicolumn{5}{c}{$n=400$} && \multicolumn{5}{c}{$n=800$} \\ \cline{2-6} \cline{8-12} \cline{14-18}
$p$  & $4$  & $20$  & $50$  & $100$  & $200$  &&  $4$  & $20$  & $50$  & $100$  & $200$  &&  $4$  & $20$  & $50$  & $100$  & $200$  \\ \hline
    & \multicolumn{17}{c}{Logistic model correctly specified} \\
True & $.12$ & $.12$ & $.12$ & $.12$ & $.12$ && $.08$ & $.08$ & $.09$ & $.09$ & $.08$ && $.06$ & $.06$ & $.06$ & $.06$ & $.06$ \\
Const& $.10$ & $.10$ & $.10$ & $.10$ & $.10$ && $.07$ & $.07$ & $.07$ & $.07$ & $.07$ && $.05$ & $.05$ & $.05$ & $.05$ & $.05$ \\
ML   & $.12$ & $.15$ & $.20$ & ---   & ---   && $.09$ & $.09$ & $.11$ & ---   & ---   && $.06$ & $.06$ & $.07$ & ---   & --- \\
RML  & $.11$ & $.11$ & $.11$ & $.10$ & $.11$ && $.08$ & $.07$ & $.08$ & $.08$ & $.07$ && $.06$ & $.06$ & $.06$ & $.06$ & $.05$ \\
CAL  & $.12$ & $.15$ & $.14$ & ---   & ---   && $.08$ & $.09$ & $.11$ & ---   & ---   && $.06$ & $.06$ & $.07$ & ---   & --- \\
RCAL & $.11$ & $.10$ & $.11$ & $.10$ & $.11$ && $.08$ & $.07$ & $.08$ & $.08$ & $.07$ && $.06$ & $.06$ & $.06$ & $.06$ & $.05$ \\
     & \multicolumn{17}{c}{Logistic model misspecified} \\
True & $.12$ & $.12$ & $.12$ & $.12$ & $.12$ && $.08$ & $.08$ & $.09$ & $.09$ & $.08$ && $.06$ & $.06$ & $.06$ & $.06$ & $.06$ \\
Const& $.10$ & $.10$ & $.10$ & $.10$ & $.10$ && $.07$ & $.07$ & $.07$ & $.07$ & $.07$ && $.05$ & $.05$ & $.05$ & $.05$ & $.05$ \\
ML   & $.18$ & $.20$ & $.23$ & ---   & ---   && $.17$ & $.15$ & $.17$ & ---   & ---   && $.15$ & $.14$ & $.16$ & ---   & --- \\
RML  & $.14$ & $.12$ & $.11$ & $.10$ & $.11$ && $.13$ & $.10$ & $.09$ & $.09$ & $.08$ && $.13$ & $.10$ & $.08$ & $.07$ & $.06$ \\
CAL  & $.13$ & $.14$ & $.14$ & ---   & ---   && $.09$ & $.09$ & $.11$ & ---   & ---   && $.06$ & $.06$ & $.07$ & ---   & --- \\
RCAL & $.11$ & $.10$ & $.11$ & $.10$ & $.11$ && $.08$ & $.07$ & $.08$ & $.08$ & $.07$ && $.06$ & $.06$ & $.06$ & $.06$ & $.05$ \\ \hline
\end{tabular*}
\end{center}
\end{table}

\clearpage
\section{Additional results from data analysis}

We present additional results from the data analysis in Section~\ref{sec:application}.

\begin{figure}
\caption{\small Maximum absolute standardized differences against the numbers of
nonzero estimates of $(\gamma_1,\ldots,\gamma_p)$ (left) and the relative variances of the inverse probability weights (right),
similarly as in Figure~\ref{fig:tune1} but in
the untreated sample, as the tuning parameter $\lambda$ varies.} \label{fig:tune0} \vspace{.15in}
\begin{tabular}{c}
\includegraphics[width=6.2in, height=2in]{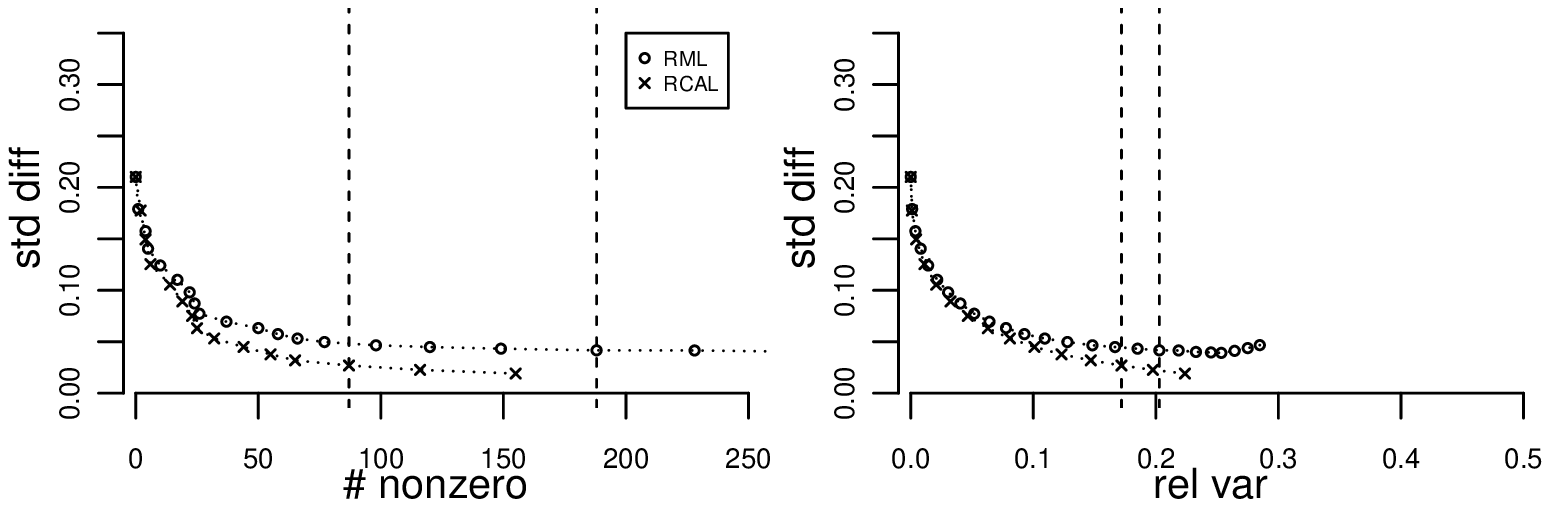} \vspace{-.15in}
\end{tabular}
\end{figure}

Figure~\ref{fig:tune0} shows maximum absolute standardized differences, $\max_j | \mbox{CAL}^0 (\hat\pi^0;f_j)|$, related to
the numbers of nonzero estimates of $\gamma_j$ and the relative variances of the inverse probability weights in the untreated sample
$\{i: T_i=0, i=1,\ldots,n\}$.
For a function $h(X)$, the standardized calibration difference $\mbox{CAL}^0 (\hat\pi^0;h)$ is defined as
\begin{align*}
\mbox{CAL}^0(\hat\pi^0; h) = \frac{\hat\mu^0_{\mbox{\tiny rIPW}}(\hat\pi^0; h) - \tilde E \{h(X)\}}{ \sqrt{ \widetilde \var \{h(X)\}} },
\end{align*}
where $\hat\mu^0_{\mbox{\tiny rIPW}}(\hat\pi^0; h)$ is defined as $\hat\mu^1_{\mbox{\tiny rIPW}}(\hat\pi^1; h)$ with $T$ replaced by $1-T$ and $\hat\pi^1$ replaced by $1-\hat\pi^0$.
The fitted propensity scores $\hat\pi^1$ and $\hat\pi^0$ are the same when maximum likelihood is used,
but separately computed for regularized calibrated estimation, as in the simulation study.

The comparison between regularized maximum likelihood and calibrated estimation is similar as in Figure~\ref{fig:tune1}.
The maximum absolute standardized difference is reduced to $4.2\%$
with $188$ nonzero estimates of coefficients $\gamma_j$ for $\hat\pi_{\mbox{\tiny RML}}$,
but reduced to $2.7\%$ with $87$ nonzero estimates of coefficients $\gamma_j$,
when the tuning parameter $\lambda$ in each case is determined from 5-fold cross validation.
The relative variances of the inverse probability weights, $1/\{1-\hat\pi^0_{\mbox{\tiny RCAL}}(X_i)\}$,
are also consistently smaller than  $1/\{1-\hat\pi^0_{\mbox{\tiny RML}}(X_i)\}$ in the untreated sample,
although the differences are not as substantial as in the treated sample from Figure~\ref{fig:tune1}.

\begin{figure}
\caption{\small Estimates $\hat\mu^1_{\mbox{\tiny rIPW}}(\hat\pi^1)$ and $\hat\mu^0_{\mbox{\tiny rIPW}}(\hat\pi^0)$ (left) and ATE (right) for 30 day survival (i.e., $Y \ge 30$), against the numbers of
nonzero estimates of $(\gamma_1,\ldots,\gamma_p)$ as the tuning parameter $\lambda$ varies. Nominal confidence intervals are also shown for the estimates of ATE at selected $\lambda$.} \label{fig:outcome} \vspace{.15in}
\begin{tabular}{c}
\includegraphics[width=6.2in, height=2in]{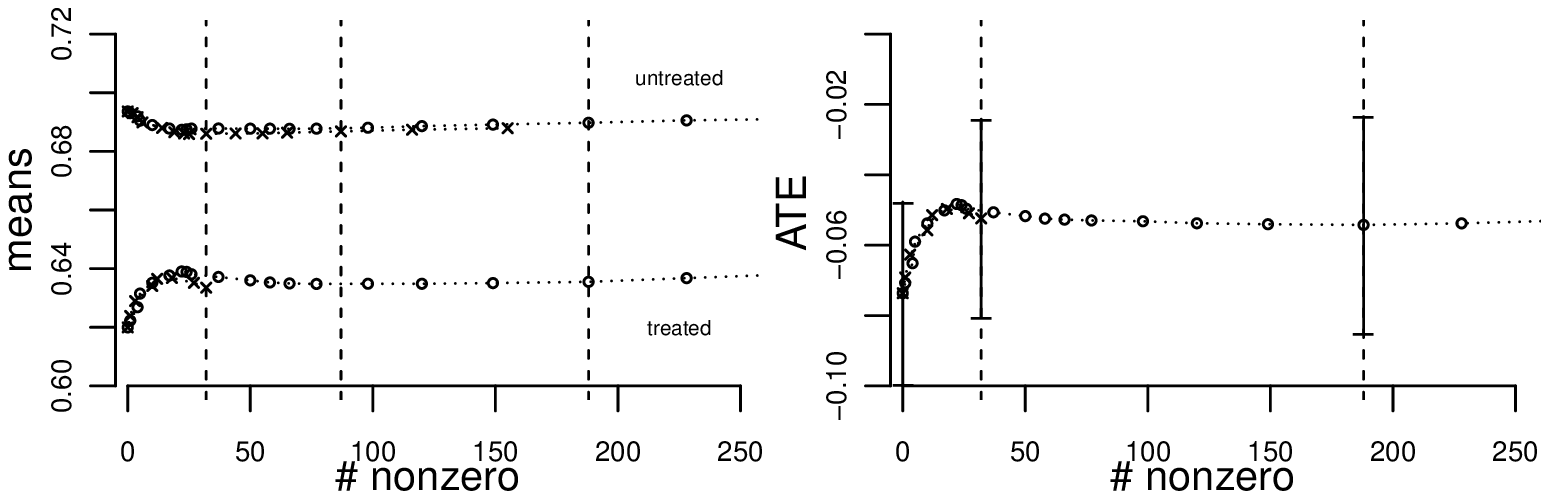} \vspace{-.15in}
\end{tabular}
\end{figure}

Figure~\ref{fig:outcome} shows the estimates $\hat\mu^1_{\mbox{\tiny rIPW}}(\hat\pi^1)$ and $\hat\mu^0_{\mbox{\tiny rIPW}}(\hat\pi^0)$ and ATE for 30 day survival (i.e., $Y \ge 30$)
as the tuning parameter $\lambda$ varies. For regularized calibrated estimation, cross validation separately for
$\hat\pi^1_{\mbox{\tiny RCAL}}$ and $\hat\pi^0_{\mbox{\tiny RCAL}}$ leads to different values of the tuning parameter $\lambda$
and hence also different numbers of nonzero estimates of $\gamma_j$.
For simplicity, the estimates of ATE are computed as $\hat\mu^1_{\mbox{\tiny rIPW}}(\hat\pi^1_{\mbox{\tiny RCAL}}) - \hat\mu^0_{\mbox{\tiny rIPW}}(\hat\pi^0_{\mbox{\tiny RCAL}})$ with the same value of $\lambda$
for both $\hat\pi^1_{\mbox{\tiny RCAL}}$ and $\hat\pi^0_{\mbox{\tiny RCAL}}$, and
a vertical line is placed corresponding to the value of $\lambda$ selected by cross validation for $\hat\pi^1_{\mbox{\tiny RCAL}}$.
In addition, for informal illustration, nominal confidence intervals are computed using twice the nominal standard errors obtained by
ignoring the variation in the fitted propensity scores $\hat\pi_{\mbox{\tiny RML}}$, $\hat\pi^1_{\mbox{\tiny RCAL}}$ and $\hat\pi^0_{\mbox{\tiny RCAL}}$.
The point estimates of ATE are consistently below zero, similar to each other from regularized maximum likelihood and calibrated estimation.
The nominal standard error from the latter method is slightly smaller: the ratio of estimated variances is $(.0154/.0141)^2=1.19$
at the values of $\lambda$ selected by cross validation.

\clearpage

\clearpage
\section{Technical details}

\subsection{Proofs of Propositions 2--5}

\noindent\textbf{Proof of Proposition~\ref{pro2}.}
First, $\kappa_{\mbox{\tiny ML}}(\myeta)$  in (\ref{kappa-ML}) can be rewritten as
\begin{align*}
\kappa_{\mbox{\tiny ML}} (\myeta) = \tilde E \left[ \log\left\{ 1+ \me^{-\myeta(X)} \right\} +(1-T) \myeta(X) \right] = \tilde E \left[ -\log\pi(X) +(1-T) \myeta(X) \right].
\end{align*}
By direct calculation, we have
$\langle \nabla\kappa_{\mbox{\tiny ML}} (\myeta^\prime), \myeta - \myeta^\prime \rangle = \tilde E [\{ -1+ \pi^\prime(X) + (1-T) \}\{\myeta(X)-\myeta^\prime(X)\} ]$ and hence
\begin{align*}
& D_{\mbox{\tiny ML}} (\myeta, \myeta^\prime) = \kappa_{\mbox{\tiny ML}} (\myeta) - \kappa_{\mbox{\tiny ML}} (\myeta^\prime) - \langle \nabla\kappa_{\mbox{\tiny ML}} (\myeta^\prime), \myeta - \myeta^\prime \rangle \\
& = \tilde E \left[ -\log \frac{\pi(X)}{\pi^\prime(X)} + \{1- \pi^\prime(X) \}\{\myeta(X)-\myeta^\prime(X)\} \right]\\
& = \tilde E \left[ -\log \frac{\pi(X)}{\pi^\prime(X)} + \{1- \pi^\prime(X)\} \left\{\log\frac{\pi(X)}{1-\pi(X)}-\log\frac{\pi^\prime(X)}{1-\pi^\prime(X)} \right\}\right] \\
& = \tilde E \left[ \pi^\prime(X) \log \frac{\pi(X)}{\pi^\prime(X)} + \{1- \pi^\prime(X)\} \log\frac{1-\pi(X)}{1-\pi^\prime(X)} \right] ,
\end{align*}
that is, $D_{\mbox{\tiny ML}} (\myeta, \myeta^\prime) =\tilde E [ L\{ \pi(X), \pi^\prime(X) \}]$.

Second, by the definition of $\kappa_{\mbox{\tiny CAL}}(\myeta)$  in (\ref{kappa-CAL}), we have
$\langle \nabla\kappa_{\mbox{\tiny CAL}} (\myeta^\prime), \myeta - \myeta^\prime \rangle = \tilde E [\{- T \me^{-\myeta^\prime(X)} + (1-T) \}\{\myeta(X)-\myeta^\prime(X)\} ]$ and hence
\begin{align*}
& D_{\mbox{\tiny CAL}} (\myeta, \myeta^\prime) = \kappa_{\mbox{\tiny CAL}} (\myeta) - \kappa_{\mbox{\tiny CAL}} (\myeta^\prime) - \langle \nabla\kappa_{\mbox{\tiny CAL}} (\myeta^\prime), \myeta - \myeta^\prime \rangle \\
& = \tilde E \left[ T \me^{-\myeta (X)} - T \me^{-\myeta^\prime(X)} + T \me^{-\myeta^\prime(X)} \{\myeta(X)-\myeta^\prime(X)\}  \right] \\
& = \tilde E \left[ T \frac{1-\pi(X)}{\pi(X)} - T \frac{1-\pi^\prime(X)}{\pi^\prime(X)} + T \frac{1-\pi^\prime(X)}{\pi^\prime(X)} \left\{\log\frac{\pi(X)}{1-\pi(X)}-\log\frac{\pi^\prime(X)}{1-\pi^\prime(X)} \right\} \right] \\
& = \tilde E \left( \frac{T}{\pi^\prime(X)} \left[ \frac{\pi^\prime(X)}{\pi(X)} -1+ \{1- \pi^\prime(X)\} \left\{\log\frac{\pi(X)}{1-\pi(X)}-\log\frac{\pi^\prime(X)}{1-\pi^\prime(X)} \right\} \right] \right) .
\end{align*}
The claimed expression for $D_{\mbox{\tiny CAL}} (\myeta, \myeta^\prime)$ follows by using the decomposition
$ \pi^\prime(X) / \pi(X)-1 = K\{ \pi(X), \pi^\prime (X)\} + \log\{\pi^\prime (X) / \pi(X)\}  $. \hfill $\Box$

\vspace{.1in}
\noindent\textbf{Proof of Proposition~\ref{pro3}.} By $E(\xi^2) = E^2(\xi) + \var(\xi)$ for a random variable $\xi$, we have
\begin{align*}
& E \left[ \left\{ \hat\mu^1_{\mbox{\tiny IPW}}(\gamma) - \mu^1 \right\}^2 \right] \\
& = E^2 \left[ \left\{ \frac{T}{\pi(X;\gamma)} -1 \right\} Y^1 \right] + \frac{1}{n} \var \left\{ \frac{T}{\pi(X;\gamma)}  Y^1 \right\} .
\end{align*}
The first term is no greater than $c \,\mbox{MSEE}(\gamma)$ by (\ref{bias-bound}). The second term can be calculated by conditioning on $(X,Y^1)$ and using Assumption~(A1) as
\begin{align*}
& \frac{1}{n} \var \left\{\frac{\pi^*(X)}{\pi(X;\gamma)} Y^1 \right\} + \frac{1}{n} E \left\{ \frac{\pi^*(X) (1-\pi^*(X))}{\pi^2(X;\gamma)} (Y^1)^2 \right\} \\
& \le \frac{1}{n} c E\left\{ \frac{\pi^{*2}(X)}{\pi^2(X;\gamma)} \right\} + \frac{1}{n} (\delta^{-1}-1) c E\left\{ \frac{\pi^{*2}(X)}{\pi^2(X;\gamma)} \right\} \\
& = \frac{1}{n \delta} c E\left\{ \frac{\pi^{*2}(X)}{\pi^2(X;\gamma)} \right\} \le \frac{2}{n \delta} c \{1+\mbox{MSRE}(\gamma)\} .
\end{align*}
Combining the preceding inequalities completes the proof. $\Box$

\vspace{.1in}
\noindent\textbf{Proof of Proposition~\ref{pro4}.} We show result (i), which directly gives (ii).
The claim about the ratio of $Q(\rho,\rho^\prime)$ and $L(\rho,\rho^\prime)$ follows because $Q(\rho, a \rho) = (a^{-1} -1 )^2$ but $L(\rho, a \rho) \to 0$ as $\rho \to 0+$.
Let $x = \rho^\prime / \rho$. It remains to show that
for $a \in (0,1/2]$, if $0 < x \le a^{-1}$ then
$$
x - 1 - \log(x) \ge (.6 a) (x-1)^2 .
$$
First, if $0 < x \le 1$ then $x - 1 - \log(x) \ge (x-1)^2 /2$.
This follows because $x - 1 - \log(x) - (x-1)^2/2$ has derivative $2-x^{-1}-x \le 0$ and hence it decreases to 0 over $0 < x \le 1$.
Second, the function $h(x) = x - 1 - \log(x) - (.6 a) (x-1)^2$
has derivative $1 + 1.2 a -1.2x - x^{-1}$, which is nonnegative over $1\le x \le (1.2 a)^{-1}$ and then negative when $x > (1.2a)^{-1}$.
That is, $h(x)$ is increasing over $1\le x \le (1.2 a)^{-1}$ and then decreasing when $x > (1.2a)^{-1}$.
Then it suffices to show that $g(a) := h(a^{-1}) = .2 + .4 /a  - .6a + \log(a)\ge 0$ for $0< a \le 1/2$.
The derivative of $g(a)$ is $-.6 -.4/a^2 + 1/a <0$ and hence $g(a)$ is decreasing over $0 < a \le 1/2$.
In addition, $g(1/2) = .7 + \log(.5) \approx .0069 >0$. The proof is then completed. $\Box$

\vspace{.1in}
\noindent\textbf{Proof of Proposition~\ref{pro5}.} Consider the case where
$\ell_{\mbox{\tiny CAL, Q3}}(\gamma; \tilde\gamma)+ \lambda \|\gamma_{1:p}\|_1$ is used as a surrogate function. The proof is similar when
$\ell_{\mbox{\tiny CAL, Q2}}(\gamma; \tilde\gamma)+ \lambda \|\gamma_{1:p}\|_1$ is used.
First, a standard argument from the MM technique shows that
$ \ell_{\mbox{\tiny CAL, Q2}}(\tilde\gamma^{(1)}; \tilde\gamma)+ \lambda \|\tilde\gamma^{(1)}_{1:p}\|_1 \le
\ell_{\mbox{\tiny CAL, Q3}}(\tilde\gamma^{(1)}; \tilde\gamma)+ \lambda \|\tilde\gamma^{(1)}_{1:p}\|_1
< \ell_{\mbox{\tiny CAL, Q3}}(\tilde\gamma ; \tilde\gamma)+ \lambda \|\tilde\gamma _{1:p}\|_1
= \ell_{\mbox{\tiny CAL, Q2}}(\tilde\gamma ; \tilde\gamma)+ \lambda \|\tilde\gamma _{1:p}\|_1$
where the second inequality follows form the definition of $\tilde\gamma^{(1)}$ and
the first inequality holds by the quadratic lower bound principle (Bohning \& Lindsay 1988).
Let $g(t) = \ell_{\mbox{\tiny CAL, Q2}}(\tilde\gamma^{(t)}; \tilde\gamma)+ \lambda \|\gamma^{(t)}_{1:p}\|_1$
and $h(t) = \ell_{\mbox{\tiny CAL}}(\tilde\gamma^{(t)})+ \lambda \|\gamma^{(t)}_{1:p}\|_1$ for $0\le t\le 1$.
Then $g(0) > g(1)$ and hence, by convexity, any subgradient of $g(\cdot)$ at 0 is negative.
But any subgradient of $h(\cdot)$ at 0 is also that of $g(\cdot)$ at 0, because $h(t)-g(t) = ct^2$ for some constant $c$ by construction.
The desired result then holds. $\Box$

\subsection{Proofs of Proposition~\ref{pro-RCAL} and corollaries}

Proposition~\ref{pro-RCAL} and Corollaries~\ref{cor-RCAL-slow} and \ref{cor-RCAL-fast} follow from respectively Proposition~\ref{pro-reg} and
Corollaries~\ref{cor-slow-rate} and \ref{cor-fast-rate}, with $\psi(T, \myeta) = T\me^{-\myeta} + (1-T)\myeta$.
It suffices to verify Assumptions~\ref{ass-mean0}, \ref{ass-hessian2}, and \ref{ass-hessian} used in Proposition~\ref{pro-reg}.
By direct calculation, we have
$\psi_1(T,\myeta) = - T \me^{-\myeta} + (1-T)$ and
$\psi_2(T,\myeta) = T \me^{-\myeta}$.
Then Assumption~\ref{ass-mean0} holds because $|\psi_1\{T, \bar\myeta(X)\} f_j(X) | \le (\me^{-B_0} + 1) C_0$, uniformly bounded for $j=0,1,\ldots,p$.
Assumption~\ref{ass-hessian2} holds because $ \psi_2\{T, \bar\myeta(X)\} \le \me^{-B_0}$.
Assumption~\ref{ass-hessian} holds because $\psi_2(t,u) = \psi_2(t, u^\prime) \me^{u^\prime-u} \le \psi_2(t, u^\prime) \me^{|u^\prime-u|}$.

It remains to show Corollary~\ref{cor-diff}.
By the Cauchy--Schwartz inequality, we have
\begin{align}
& \left| \hat \mu^1_{\mbox{\tiny IPW }} ( \hat\pi^1_{\mbox{\tiny RCAL}}) - \hat \mu^1_{\mbox{\tiny IPW }} ( \bar\pi^1_{\mbox{\tiny CAL}}) \right|^2
= \tilde E^2 \left[ \frac{T}{\bar\pi^1_{\mbox{\tiny CAL}}(X)} \left\{\frac{ \bar \pi^1_{\mbox{\tiny CAL}}(X) }{ \hat \pi^1_{\mbox{\tiny RCAL}}(X) } -1\right\}  \right] \nonumber \\
& \le \tilde E \left[ \frac{T}{\bar\pi^1_{\mbox{\tiny CAL}}(X)} \left\{\frac{ \bar \pi^1_{\mbox{\tiny CAL}}(X) }{ \hat \pi^1_{\mbox{\tiny RCAL}}(X) } -1\right\}^2  \right]
 \tilde E \left\{ \frac{TY^2}{\bar\pi^1_{\mbox{\tiny CAL}}(X)} \right\}.  \label{prf-cor-diff-eq}
\end{align}
By simple manipulation, we have
\begin{align*}
& \frac{ \bar \pi^1_{\mbox{\tiny CAL}}(X) }{ \hat \pi^1_{\mbox{\tiny RCAL}}(X) } \le \exp \left\{ (\bar\gamma^1_{\mbox{\tiny CAL}})^\T f(X) \right\} +
\exp \left\{ (\bar\gamma^1_{\mbox{\tiny CAL}} -\hat\gamma^1_{\mbox{\tiny RCAL}})^\T f(X) \right\} \\
& \le \exp(B_0) + \exp \left\{ \|\bar\gamma^1_{\mbox{\tiny CAL}} -\hat\gamma^1_{\mbox{\tiny RCAL}}\|_1 C_0 \right\},
\end{align*}
under conditions (i) and (iii) in Proposition~\ref{pro-RCAL}.
If $\sum_{j=1}^p |\bar \gamma^1_{\mbox{\tiny CAL},j}| \le M_1$, then
(\ref{cor-RCAL-slow-eq}) implies that
$ \| \bar\gamma^1_{\mbox{\tiny CAL}} -\hat\gamma^1_{\mbox{\tiny RCAL}}\|_1 \le O(1) (A_0-1)^{-1} M_1$, and
hence from the preceding inequality, the ratio $\bar \pi^1_{\mbox{\tiny CAL}}(X) /\hat \pi^1_{\mbox{\tiny RCAL}}(X)$
is bounded from above by a constant, which can be taken as $a^{-1}$ for some $a \in (0, 1/2]$ depending only on $(A_0, B_0,C_0,\eta_0,M_1)$ and $\eta_3$ or $\eta_4$.
Then Propositions \ref{pro4}(i) can be applied
together with Proposition \ref{pro2}(i) to obtain
\begin{align*}
\tilde E \left[ \frac{T}{\bar\pi^1_{\mbox{\tiny CAL}}(X)} \left\{\frac{ \bar \pi^1_{\mbox{\tiny CAL}}(X) }{ \hat \pi^1_{\mbox{\tiny RCAL}}(X) } -1\right\}^2  \right]
\le \frac{5}{3a} D_{\mbox{\tiny CAL}} ( \hat \myeta^1_{\mbox{\tiny RCAL}}, \bar \myeta^1_{\mbox{\tiny CAL}} ) .
\end{align*}
Inequality (\ref{cor-diff-eq1}) then follows from (\ref{prf-cor-diff-eq}) and
$D_{\mbox{\tiny CAL}} ( \hat \myeta^1_{\mbox{\tiny RCAL}}, \bar \myeta^1_{\mbox{\tiny CAL}} ) \le O(1) \lambda_0 \sum_{j=1}^p |\bar \gamma^1_{\mbox{\tiny CAL},j}|$,
which is also implied by (\ref{cor-RCAL-slow-eq}).
Similarly, (\ref{cor-diff-eq2}) can be shown to result from (\ref{cor-RCAL-fast-eq}).
It suffices to note that (\ref{cor-RCAL-fast-eq}) implies that
$ \| \bar\gamma^1_{\mbox{\tiny CAL}} -\hat\gamma^1_{\mbox{\tiny RCAL}}\|_1 \le O(1) (A_0-1)^{-1} |S| \lambda_0$,
which is already bounded from above by a constant under condition (iv) in Proposition~\ref{pro-RCAL}.

\subsection{Proofs of Proposition~\ref{pro-reg} and corollaries}

The proof of Proposition~\ref{pro-reg} is completed by combining Lemmas~\ref{lem-prob}--\ref{lem-orac-ineq}.

\begin{lem} \label{lem-prob}
(i) Denote by $\Omega_1$ the event that
\begin{align}
\sup_{j=0,1,\ldots,p} \left| \tilde E \left[ \psi_1\{T, \bar \myeta(X) \} f_j(X) \right] \right| \le \lambda_0 . \label{lem-prob-eq1}
\end{align}
Under Assumption~\ref{ass-mean0}, if $\lambda_0 \ge \sqrt{ 8 (D_0^2 + D_1^2) }\sqrt{\log(p/\epsilon)/n}$, then $P( \Omega_1 ) \ge 1- 2 \epsilon$.\\
(ii) Denote by $\Omega_2$ the event that
\begin{align}
\sup_{j,k=0,1,\ldots,p} | (\tilde \Sigma_{\bar \gamma})_{jk} - (\Sigma_{\bar\gamma})_{jk} | \le \lambda_0, \label{lem-prob-eq2}
\end{align}
Under Assumptions~\ref{ass-covariate} and \ref{ass-hessian2}, if $\lambda_0 \ge (\sqrt{8} C_0^2 C_1 ) \sqrt{\log(p^2/\epsilon^2)/n}$, then $P( \Omega_2 ) \ge 1- 2 \epsilon^2$.
\end{lem}

\begin{prf}
Lemma~\ref{lem-prob}(i) follows directly from Lemma~\ref{lem-max-subG} in Section~\ref{sec:tech-tool} and the union bound.
Lemma~\ref{lem-prob}(ii) follows from Lemma~\ref{lem-max-bounded} in Section~\ref{sec:tech-tool} and the union bound,
with $ |\psi_2\{T, \bar \myeta(X)\} f_j(X) $ $ f_k(X)| \le C_0^2 C_1$ and hence
$ |\psi_2\{T, \bar \myeta(X)\} f_j(X) f_k(X) - (\Sigma_{\bar\gamma})_{jk} | \le 2 C_0^2 C_1$.
\end{prf}

\begin{lem} \label{lem-basic-ineq}
For any coefficient vector $\gamma$ and $\myeta = \gamma^\T f$, we have
\begin{align}
D( \hat \myeta, \myeta ) + D( \myeta, \hat\myeta) + \langle \nabla \kappa(\myeta) , \hat\myeta-\myeta \rangle + \lambda R(\hat\gamma) \le \lambda R (\gamma) , \label{lem1-ineq1}
\end{align}
or equivalently
\begin{align}
D( \hat \myeta, \myeta^*) + D( \myeta, \hat\myeta) + \langle \nabla \kappa(\myeta^*) , \hat\myeta-\myeta \rangle + \lambda R(\hat\gamma) \le D( \myeta, \myeta^*)+ \lambda R (\gamma). \label{lem1-ineq2}
\end{align}
\end{lem}

\begin{prf}
For any $u \in (0,1]$, the definition of $\hat\gamma$ implies
\begin{align*}
 \ell(\hat\gamma) + \lambda R(\hat\gamma) \le \ell\{ (1-u) \hat\gamma + u \gamma\} + \lambda R\{ (1-u) \hat\gamma + u \gamma\} ,
\end{align*}
which gives
\begin{align*}
\ell(\hat\gamma) - \ell\{ (1-u) \hat\gamma + u \gamma\} + \lambda u R(\hat\gamma) \le \lambda u R (\gamma) ,
\end{align*}
by the convexity of $R()$, that is, $R\{ (1-u) \hat\gamma + u \gamma\} \le (1-u) R (\hat\gamma) + u R(\gamma)$.
Dividing both sides of the preceding inequality by $u$ and letting $u \to 0+$ yields
\begin{align*}
\langle \nabla \kappa(\hat \myeta), \hat \myeta-\myeta \rangle + \lambda R(\hat\gamma) \le \lambda R (\gamma) .
\end{align*}
Inequality (\ref{lem1-ineq1}) follows because
$D( \hat \myeta, \myeta ) + D( \myeta, \hat\myeta)  = \langle \nabla \kappa(\hat\myeta) , \hat\myeta-\myeta \rangle -\langle \nabla \kappa( \myeta) , \hat\myeta-\myeta \rangle$ by direct calculation.
In addition, inequality (\ref{lem1-ineq2}) follows because
$D( \hat \myeta, \myeta^* ) + D( \myeta, \hat\myeta) - D(\myeta, \myeta^*)= \langle \nabla \kappa(\hat\myeta) , \hat\myeta-\myeta \rangle -\langle \nabla \kappa( \myeta^*) , \hat\myeta-\myeta \rangle$
by direct calculation.
\end{prf}

\begin{lem} \label{lem2-basic-ineq}
In the event $\Omega_1$ from Lemma~\ref{lem-prob}, we have
\begin{align}
\left| \langle \nabla \kappa(\bar \myeta ) , \hat\myeta- \bar\myeta \rangle \right| \le \lambda_0 \|\hat\gamma- \bar\gamma \|_1, \label{lem2-ineq1}
\end{align}
and for any subset $S \subset \{0,1,\ldots,p\}$ containing $0$,
\begin{align}
D( \hat \myeta, \myeta ) + D( \myeta, \hat\myeta) +(A_0-1) \lambda_0 \|\hat\gamma -\gamma \|_1
\le  2 A_0 \lambda_0  \left\{ \sum_{j\in S} |\hat\gamma_j - \gamma_j| + \sum_{j\not\in S} |\gamma_j| \right\}.  \label{lem2-ineq2}
\end{align}
\end{lem}

\begin{prf}
Inequality (\ref{lem2-ineq1}) follows directly from (\ref{lem-prob-eq1}) and the fact that
 $\langle \nabla \kappa(\bar \myeta ) , \hat\myeta- \bar \myeta \rangle = (\hat\gamma-\bar\gamma)^\T \tilde E[ \psi_1\{T,  \bar \myeta(X)\} f(X)]$.
Combining (\ref{lem1-ineq1}) and (\ref{lem2-ineq1}) yields
\begin{align*}
D( \hat \myeta, \myeta ) + D( \myeta, \hat\myeta) + A_0 \lambda_0 R(\hat\gamma) \le \lambda_0\{ |\hat\gamma_0-\gamma_0| + R(\hat\gamma-\gamma)\}+ A_0 \lambda_0 R (\gamma).
\end{align*}
Applying to the preceding inequality the triangle inequalities
\begin{align*}
| \hat\gamma_j | & \ge |\hat\gamma_j - \gamma_j | - |\gamma_j| , \quad j \not \in S, \\
| \hat\gamma_j | & \ge |\gamma_j| - |\hat\gamma_j - \gamma_j |  , \quad j  \in S \backslash \{0\},
\end{align*}
and rearranging the result gives
\begin{align*}
D( \hat \myeta, \myeta ) + D( \myeta, \hat\myeta) +(A_0-1) \lambda_0 R(\hat\gamma -\gamma)
\le \lambda_0 |\hat\gamma_0-\gamma_0| + 2 A_0 \lambda_0  \left\{ \sum_{j\in S \backslash\{0\}} |\hat\gamma_j - \gamma_j| + \sum_{j\not\in S} |\gamma_j| \right\}.
\end{align*}
The conclusion follows by adding $(A_0-1)\lambda_0 |\hat\gamma_0-\gamma_0|$ to both sides above.
\end{prf}

\begin{lem} \label{lem-hessian}
Suppose that Assumptions~\ref{ass-covariate} and \ref{ass-hessian} hold. Then for any $\myeta=\gamma^T f$ and $\myeta^\prime = {\gamma^\prime}^\T f$,
\begin{align*}
D (\myeta, \myeta^\prime) + D(\myeta^\prime, \myeta) \ge \frac{1-\me^{-C_3 \|b\|_1 } }{C_3 \|b\|_1} \left( b^\T \tilde \Sigma_\gamma b\right),
\end{align*}
where $b=\gamma^\prime- \gamma$ and $C_3=C_0C_2$.
\end{lem}

\begin{prf}
By direct calculation, we have
\begin{align*}
& D (\myeta, \myeta^\prime) + D(\myeta^\prime, \myeta)  = \tilde E \left( \left[ \psi_1\{T,\myeta^\prime(X)\} - \psi_1 \{ T,\myeta(X)\} \right] \left\{ \myeta^\prime(X) - \myeta(X) \right\} \right) \\
& = \tilde E \left[ \left( \int_0^1 \psi_2 \left[ T, \myeta(X) + u \left\{ \myeta^\prime(X)-\myeta(X) \right\} \right]  \dif u \right)  \left\{ \myeta^\prime(X) - \myeta(X) \right\}^2 \right] .
\end{align*}
By Assumption \ref{ass-hessian} and the fact that $ | \myeta^\prime(X) - \myeta(X) | \le \{ \sup_{j=0,1,\ldots,p} |f_j(X) | \}\, \|\gamma^\prime-\gamma\|_1 \le C_0 \|\gamma^\prime-\gamma\|_1$ by Assumption~\ref{ass-covariate},
it follows that
\begin{align*}
& D (\myeta, \myeta^\prime) + D(\myeta^\prime, \myeta)
\ge \tilde E \left[ \left( \int_0^1  \psi_2 \left\{ T, \myeta(X) \right\} \me^{-C_2 u | \myeta^\prime(X)-\myeta(X)|} \dif u \right) \left\{ \myeta^\prime(X) - \myeta(X) \right\}^2 \right] \\
& \ge \tilde E \left[ \psi_2 \left\{ T, \myeta(X) \right\}  \left\{ \myeta^\prime(X) - \myeta(X) \right\}^2 \right] \left( \int_0^1  \me^{-C_3 u \|\gamma^\prime-\gamma\|_1} \dif u \right),
\end{align*}
which gives the desired result.
\end{prf}

\begin{lem} \label{lem-compat}
Suppose that Assumption \ref{ass-rate}(i) holds. In the event $\Omega_2$ from Lemma~\ref{lem-prob}, Assumption~\ref{ass-compat} (theoretical compatibility condition)
implies an empirical compatibility condition:
for any vector $b=(b_0,b_1,\ldots,b_p)^\T \in \mathbb R^{1+p} $ satisfying (\ref{ass-compat-eq2}),
\begin{align}
(1-\eta_1) \nu_0^2  \left(\sum_{j\in S} |b_j| \right)^2 \le |S| \left( b^\T \tilde \Sigma_\gamma b  \right) . \label{lem-compat-eq}
\end{align}
\end{lem}

\begin{prf}
By (\ref{lem-prob-eq2}), we have $ |b^\T (\tilde \Sigma_\gamma - \Sigma_\gamma) b | \le \sum_{j,k=0,1,\ldots,p} \lambda_0 |b_jb_k| = \lambda_0 \|b\|_1^2$.
Then Assumption~\ref{ass-compat} implies that for any real vector $b=(b_0,b_1,\ldots,b_p)^\T $ satisfying (\ref{ass-compat-eq2}),
\begin{align*}
&\nu_0^2 \|b_S \|_1^2 \le |S| (b^\T \Sigma_\gamma b) \le |S| \left(b^\T \tilde \Sigma_\gamma b + \lambda_0 \|b\|_1^2 \right) \\
& \le |S| (b^\T \tilde \Sigma_\gamma b ) + |S| \lambda_0 (1+\xi_0)^2 \|b_S \|_1^2 ,
\end{align*}
where $\| b_S \|_1 = \sum_{j\in S} |b_j|$. The last inequality is due to $\|b\|_1 \le (1+\xi_0) \|b_S\|_1$ by (\ref{ass-compat-eq2}).
Then (\ref{lem-compat-eq}) follows because $(1+\xi_0)^2 \nu_0^{-2} |S| \lambda_0 \le \eta_1 \,(<1)$ by Assumption \ref{ass-rate}(i).
\end{prf}

\begin{lem} \label{lem-orac-ineq}
Suppose that Assumptions \ref{ass-compat}, \ref{ass-covariate}, \ref{ass-hessian}, and \ref{ass-rate}(i)--(ii) hold, and $A_0 > (\xi_0+1)/(\xi_0-1)$.
In the event $\Omega_1 \cap \Omega_2$, (\ref{pro-reg-eq}) holds as in Proposition~\ref{pro-reg}.
\end{lem}

\begin{prf}
Denote $b=\hat\gamma-\bar\gamma$ and $D^\dag(\hat\myeta, \bar\myeta) = D( \hat \myeta, \bar\myeta ) + D( \bar\myeta, \hat\myeta) +(A_0-1) \lambda_0 \|b\|_1$, that is, the left hand side of (\ref{lem2-ineq2}).
By Lemma~\ref{lem2-basic-ineq} under (\ref{lem2-ineq1}), inequality (\ref{lem2-ineq2}) with the subset $S$ from Assumption~\ref{ass-compat} leads to two possible cases: either
\begin{align}
\xi_1 D^\dag (\hat \myeta, \bar\myeta) \le 2 A_0  \lambda_0\sum_{j\not\in S} |\bar\gamma_j| , \label{lem3-ineq1}
\end{align}
or $(1-\xi_1) D^\dag(\hat\myeta, \bar\myeta) \le 2 A_0 \lambda_0 \sum_{j\in S} |b_j|$, that is,
\begin{align}
D^\dag (\hat\myeta, \bar\myeta)
\le (\xi_0+1) (A_0-1) \lambda_0 \sum_{j\in S} |b_j| = \xi_2 \lambda_0 \sum_{j\in S} |b_j| , \label{lem3-ineq2}
\end{align}
where $\xi_1 = 1-2A_0 /\{ (\xi_0+1)(A_0-1)\} \in (0,1]$ because $A_0 > (\xi_0+1)/(\xi_0-1)$.
If (\ref{lem3-ineq2}) holds, then $\sum_{j \not\in S} |b_j| \le \xi_0 \sum_{j\in S} |b_j|$,
which, by Lemma~\ref{lem-compat} under (\ref{lem-prob-eq2}) and Assumptions~\ref{ass-compat} and \ref{ass-rate}(i), implies (\ref{lem-compat-eq}), that is,
\begin{align}
\sum_{j\in S} |b_j| \le (1-\eta_1)^{-1/2}\nu_0^{-1} |S|^{1/2} \left( b^\T \tilde \Sigma_{\bar \gamma} b  \right)^{1/2}. \label{lem3-ineq3}
\end{align}
By Lemma~\ref{lem-hessian} under Assumptions \ref{ass-covariate} and \ref{ass-hessian}, we have
\begin{align}
D( \hat \myeta, \bar\myeta ) + D( \bar\myeta, \hat\myeta)  \ge \frac{1-\me^{-C_3 \|b\|_1 } }{C_3 \|b\|_1} \left( b^\T \tilde \Sigma_{\bar \gamma} b\right) . \label{lem3-ineq4}
\end{align}
Combining (\ref{lem3-ineq2}), (\ref{lem3-ineq3}), and (\ref{lem3-ineq4}) and simple manipulation yields
\begin{align}
D^\dag ( \hat \myeta, \bar\myeta ) \le \xi_2 \lambda_0 \sum_{j\in S} |b_j|  \le \xi_2^2(1-\eta_1)^{-1} \nu_0^{-2} \lambda_0^2 |S| \frac{C_3 \|b\|_1}{1-\me^{-C_3 \|b\|_1 } }. \label{lem3-ineq5}
\end{align}
The second inequality in (\ref{lem3-ineq5}) along with Assumption \ref{ass-rate}(ii) implies that $1 - \me^{- C_3 \|b\|_1} \le C_3 \xi_2 (1-\eta_1)^{-1} \nu_0^{-2} \lambda_0 |S| \le \eta_2 \, (<1)$.
As a result, $C_3 \|b\|_1 \le - \log(1- \eta_2)$ and hence
\begin{align*}
\frac{1-\me^{-C_3 \|b\|_1 } }{C_3 \|b\|_1} = \int_0^1 \me^{-C_3 \|b\|_1 u} \dif u \ge \me^{-C_3 \|b\|_1} \ge 1-\eta_2.
\end{align*}
From this bound, inequality (\ref{lem3-ineq5}) then leads to $D^\dag( \hat \myeta, \bar\myeta )  \le \xi_2^2 \nu_1^{-2} \lambda_0^2 |S| $.
Therefore, (\ref{pro-reg-eq}) holds through (\ref{lem3-ineq1}) and (\ref{lem3-ineq2}) in the event $\Omega_1 \cap \Omega_2$.
\end{prf}

\vspace{.1in}
\noindent\textbf{Proof of Corollary~\ref{cor-slow-rate}.} Denote $b_{1:p}=(b_1,\ldots,b_p)^\T$
and $f_{1:p} = (f_1,\ldots, f_p)^\T$.
First, (\ref{cor-slow-rate-eq1}) amounts to saying that for any vector $b_{1:p} \in \mathbb R^p$,
$$
E^2 [ \psi_2\{T,\bar \myeta(X)\} b_{1:p}^\T f_{1:p}(X) ] \le \eta_3^2 E [\psi_2\{T,\bar \myeta(X)\} ] E[\psi_2\{T,\bar \myeta(X)\} \{b_{1:p}^\T f_{1:p}(X) \}^2 ] ,
$$
which implies that the following quadratic function in $b_0 \in \mathbb R$ is always nonnegative:
$$
 \eta_3^2 E [\psi_2\{T,\bar \myeta(X)\} ] b_0^2 + E[\psi_2\{T,\bar \myeta(X)\} \{b_{1:p}^\T f_{1:p}(X) \}^2 ] + 2b_0 E[ \psi_2\{T,\bar \myeta(X)\} b_{1:p}^\T f_{1:p}(X) ] \ge 0.
$$
That is, (\ref{ass-compat-eq1}) holds for any $(b_0,b_1,\ldots,b_p)$, possibly violating (\ref{ass-compat-eq2}), with $S=\{0\}$ and $\nu_0^2 = (1-\eta_3^2) E[ \psi_2\{T,\bar \myeta(X)\} ]$.
It remains to show that (\ref{cor-slow-rate-eq2}) also implies  Assumption~\ref{ass-compat}.
Under (\ref{cor-slow-rate-eq2}), we have by the triangle and Cauchy--Schwartz inequalities,
\begin{align*}
& E[ \psi_2\{T,\bar \myeta(X)\} \{b_{1:p}^\T f_{1:p}(X)\}^2 ] \le
\sum_{j,k=1,\ldots,p} \big|  b_j b_k E [ \psi_2\{T,\bar \myeta(X)\} f_j(X) f_k(X) ] \big| \\
& \le  \eta_4^2 E [ \psi_2\{T,\bar \myeta(X)\} ] \|b_{1:p}\|_1^2  .
\end{align*}
By the Cauchy--Schwartz inequality again, we have
\begin{align*}
(b^\T \Sigma_{\bar \gamma} b)^{1/2} \ge |b_0| E^{1/2}[ \psi_2\{T,\bar \myeta(X)\} ] - E^{1/2}[ \psi_2\{T,\bar \myeta(X)\} \{b_{1:p}^\T f_{1:p}(X)\}^2 ] .
\end{align*}
Combining the preceding inequalities shows that if $\|b_{1:p}\|_1 \le \xi_0 |b_0|$, then
\begin{align*}
(b^\T \Sigma_{\bar \gamma} b)^{1/2} \ge |b_0| E^{1/2}[ \psi_2\{T,\bar \myeta(X)\} ] - \xi_0 \eta_4 |b_0|E^{1/2}[ \psi_2\{T,\bar \myeta(X)\} ].
\end{align*}
That is, (\ref{ass-compat-eq1}) holds for any $(b_0,b_1,\ldots,b_p)$ satisfying (\ref{ass-compat-eq2}), with $S=\{0\}$, any constant $1< \xi_0<\eta_4^{-1}$, and $\nu_0^2 = (1- \xi_0\eta_4)^2 E[ \psi_2\{T,\bar \myeta(X)\} ]$.\hfill $\Box$

\vspace{.1in}
\noindent\textbf{Proof of Corollary~\ref{cor-fast-rate}.} The result follows immediately from Proposition~\ref{pro-reg}. \hfill $\Box$

\vspace{.1in}
\noindent\textbf{Proof of Corollary~\ref{cor-comp-truth}.}
By the proof of Lemma~\ref{lem-basic-ineq}, we have
\begin{align}
D( \hat \myeta, \myeta^* ) + D( \bar \myeta,  \hat \myeta ) - D( \bar \myeta, \myeta^* )  =
D( \hat \myeta, \bar \myeta ) + D( \bar \myeta, \hat\myeta) + \langle \nabla \kappa( \bar \myeta) , \hat\myeta-\myeta \rangle
- \langle \nabla \kappa(\myeta^*) , \hat\myeta-\myeta \rangle. \label{prf-comp-truth-eq}
\end{align}
Let $\Omega_3$ be the event that $\sup_{j=0,1,\ldots,p} | \tilde E  [ \psi_1\{T, \myeta^*(X) \} f_j(X)  ] | \le \lambda_0$.
Similarly as in Lemma~\ref{lem-prob}, $P(\Omega_3) \ge 1-2 \epsilon$ under Assumption~\ref{ass-mean0} with $Z_j$ replaced by $Z_j^*$.
In the event $\Omega_1 \cap \Omega_3$, we have (\ref{lem2-ineq1}) and similarly
$ | \langle \nabla \kappa(\myeta^* ) , \hat\myeta- \bar\myeta \rangle | \le \lambda_0 \|\hat\gamma- \bar\gamma \|_1 $,
which together with (\ref{prf-comp-truth-eq}) imply
\begin{align*}
D( \hat \myeta, \myeta^* ) + D( \bar \myeta,  \hat \myeta ) - D( \bar \myeta, \myeta^* )  \le
D( \hat \myeta, \bar \myeta ) + D( \bar \myeta, \hat\myeta) +2  \lambda_0 \|\hat\gamma- \bar\gamma \|_1 .
\end{align*}
Denote by $\Delta(\bar\myeta,S)$ the right hand side of (\ref{pro-reg-eq}).
By the proof of Proposition~\ref{pro-reg}, in the event $\Omega_1 \cap \Omega_2$, we have $D( \hat \myeta, \bar \myeta ) + D( \bar \myeta, \hat\myeta) \le \Delta(\bar\myeta,S)$
and $(A_0-1) \lambda_0 \|\hat\gamma-\bar\gamma\|_1 \le \Delta (\bar\myeta, S)$.
Applying these bounds to the preceding inequality in the event $\Omega_1\cap\Omega_2\cap\Omega_3$ yields the desired result. \hfill $\Box$

\subsection{Technical tools} \label{sec:tech-tool}

For completeness, we state the following maximal inequalities, which can be obtained from Buhlmann \& van de Geer (2011), Lemma 14.11 and Lemma 14.16.

\begin{lem} \label{lem-max-bounded}
Let $(Y_1,\ldots,Y_n)$ be independent variables such that $E(Y_i)=0$ for $i=1,\ldots,n$ and
$\max_{i=1,\ldots,n } |Y_i| \le D_0$
for some constant $D_0$. Then for any $u >0$,
\begin{align*}
P \left(  \left| \frac{1}{n} \sum_{i=1}^n Y_i \right| > u \right) \le 2 \exp \left(-\frac{nu^2}{2 D_0^2} \right).
\end{align*}
\end{lem}

\begin{lem} \label{lem-max-subG}
Let $(Y_1,\ldots,Y_n)$ be independent variables such that $E(Y_i)=0$ for $i=1,\ldots,n$ and
$\max_{i=1,\ldots,n } D_1^2  E \{\exp(Y_i^2/D_1^2) - 1 \} \le D_2^2$
for some constants $(D_1,D_2)$. Then for any $u >0$,
\begin{align*}
P \left(  \left| \frac{1}{n} \sum_{i=1}^n Y_i \right| > u \right) \le 2 \exp \left\{-\frac{nu^2}{8(D_1^2 + D_2^2)} \right\}.
\end{align*}
\end{lem}

\vspace{.25in}
\centerline{\bf\Large References}

\begin{description}\addtolength{\itemsep}{-.05in}

\item Bickel, P., Ritov, Y., and Tsybakov, A.B. (2009) ``Simultaneous analysis of Lasso and Dantzig selector," {\em Annals
of Statistics}, 37, 1705--1732.

\item Zhang, C.-H. and Zhang, T. (2012) ``A general framework of dual certificate analysis for structured
sparse recovery problems," arXiv:1201.3302.

\end{description}

\end{document}